\documentclass[a4paper,11pt]{article}
\pdfoutput=1
\usepackage{jheppub}
\usepackage{adjustbox}
\usepackage{multirow}
\usepackage{pbox}
\usepackage{makecell}
\usepackage[T1]{fontenc} % if needed
\newcommand{\eqal}[1]{\begin{align}#1\end{align}}
\newcommand{\nb }{\nonumber\\}
\newcommand{\yong}[1]{{\color{black}#1}}

\newcommand{\sampsa}[1]{{\color{black}#1}}

\title{\boldmath Non-standard interactions in SMEFT confronted with terrestrial neutrino experiments}
\author[a,b]{Yong Du,}
\author[a]{Hao-Lin Li,}
\author[c]{Jian Tang,}
\author[c]{Sampsa Vihonen,}
\author[a,d,e,f]{Jiang-Hao Yu}

\affiliation[a]{CAS Key Laboratory of Theoretical Physics, Institute of Theoretical Physics, Chinese Academy of Sciences, Beijing 100190, P. R. China}
\affiliation[b]{Amherst Center for Fundamental Interactions, Physics Department, University of Massachusetts Amherst, Amherst, MA 01003 USA}
\affiliation[c]{School of Physics, Sun Yat-sen University, Guangzhou 510275, China}
\affiliation[d]{School of Physical Sciences, University of Chinese Academy of Sciences, Beijing 100049, P.R. China}
\affiliation[e]{School of Fundamental Physics and Mathematical Sciences, Hangzhou Institute for Advanced Study, UCAS, Hangzhou 310024, China}
\affiliation[f]{International Centre for Theoretical Physics Asia-Pacific, Beijing/Hangzhou, China}

\emailAdd{yongdu@umass.edu}
\emailAdd{lihaolin@itp.ac.cn}
\emailAdd{tangjian5@mail.sysu.edu.cn}
\emailAdd{sampsa@mail.sysu.edu.cn}
\emailAdd{jhyu@itp.ac.cn}

\abstract{The Standard Model Effective Field Theory (SMEFT) provides a systematic and model-independent framework to study neutrino non-standard interactions (NSIs). We study the constraining power of the on-going neutrino oscillation experiments T2K, NO$\nu$A, Daya Bay, Double Chooz and RENO in the SMEFT framework. A full consideration of matching is provided between different effective field theories and the renormalization group running at different scales, filling the gap between the low-energy neutrino oscillation experiments and SMEFT at the UV scale. We first illustrate our method with a top-down approach in a simplified scalar leptoquark model, showing more stringent constraints from the neutrino oscillation experiments compared to collider studies. We then provide a bottom-up study on individual dimension-6 SMEFT operators and find NSIs in neutrino experiments already sensitive to new physics at $\sim$20\,TeV when the Wilson coefficients are fixed at unity. We also investigate the correlation among multiple operators at the UV scale and find it could change the constraints on SMEFT operators by several orders of magnitude compared with when only one operator is considered. Furthermore, we find that accelerator and reactor neutrino experiments are sensitive to different SMEFT operators, which highlights the complementarity of the two experiment types.

}

\begin{document} 
\preprint{ACFI-T20-15}
\maketitle
\flushbottom

%%%%%%%%%%%%%%%%%%%%%%%%%%%%%%%%%%%%%%%%%%%%%
\section{\label{sec:intro}Introduction}
%%%%%%%%%%%%%%%%%%%%%%%%%%%%%%%%%%%%%%%%%%%%%

The discovery of neutrino oscillations more than two decades ago~\cite{Fukuda:1998mi,Ahmad:2002jz} began an era where the understanding of the neutrino properties has improved at an accelerating pace. Not only have the majority of the neutrino oscillation parameters describing the oscillation of $\nu_e$, $\nu_\mu$ and $\nu_\tau$ been measured to a relatively high precision, but the recent observations from the long-baseline (LBL) accelerator experiments have also hinted that these oscillations may indicate {\sl CP} violation in the leptonic sector (see Ref.~\cite{Esteban:2020cvm} for a review). The standard theory of neutrino oscillations has been tested in a variety of neutrino oscillation experiments, which have provided a wealth of data using different neutrino sources and measurement techniques. The present generation of on-going neutrino oscillation experiments is already starting to show the level of precision at which probes for non-standard neutrino physics can be started using the novel neutrino oscillation data.

%According to the standard three-neutrino paradigm, the oscillation among the three active neutrino states $\nu_e$, $\nu_\mu$ and $\nu_\tau$ is controlled by the three mixing angles $\theta_{12}$, $\theta_{13}$ and $\theta_{23}$, the Dirac {\sl CP} phase $\delta_{CP}$ and the two mass-squared differences $\Delta m_{21}^2 \equiv m_2^2 - m_1^2$ and $\Delta m_{31}^2 \equiv m_3^2 - m_1^2$. The mixing angles and the Dirac {\sl CP} phase make up the well-known Pontecorvo-Maki-Nakagawa-Sakata (PMNS) matrix which describe the mixing of the three active neutrinos~\cite{Pontecorvo:1957cp,Pontecorvo:1957qd,Maki:1960ut,Maki:1962mu,Pontecorvo:1967fh}. As the mixing angles and the magnitudes of the mass-squared differences have been measured to a very high precision, the only remaining unknowns in the three-neutrino picture are the relative ordering of the neutrino masses $m_1$, $m_2$ and $m_3$, the octancy of $\theta_{23}$ and the value of $\delta_{CP}$.\footnote{\color{black}Perhaps a very important question is whether neutrinos are Dirac or Majorana particles. In the latter case neutrinos would be their own antiparticles, and there would be Majorana phases present in the neutrino mixing matrix. Majorana phases do not generally affect neutrino oscillation probabilities, and we therefore neglect them in this work.} The present generation of on-going neutrino oscillation experiments is already starting to show the level of precision at which probes for non-standard neutrino physics can be started using the novel neutrino oscillation data.

There have been several attempts to look for signals of the physics beyond the Standard Model (BSM) in the neutrino sector~\cite{Davidson:2003ha,Barranco:2005ps,Barranco:2007ej,Forero:2011zz,Adhikari:2012vc,Girardi:2014kca,DiIura:2014csa,Tang:2017khg,Babu:2019mfe}. While many studies have interpreted the consequences of UV-complete models in the neutrino oscillations through the mixing with the heavy right-handed neutrinos, a great deal of discussion has been placed on the possibility of non-standard interactions (NSIs) influencing neutrino oscillations~\cite{Ohlsson:2012kf,Dev:2019anc,Esteban:2020cvm}. The NSI of neutrinos is typically discussed in the context of the {\sl ad-hoc} parameterization of effective couplings $\epsilon^s$, $\epsilon^d$\,\cite{Grossman:1995wx,GonzalezGarcia:2001mp,Ohlsson:2008gx,Farzan:2017xzy} and $\epsilon^m$\,\cite{Wolfenstein:1977ue}, which describe the non-standard effects in the source, detection and propagation, respectively. The majority of the literature focused on finding the constraints on each parameter in neutrino oscillation experiments, but new physics has so far been mainly treated as uncorrelated energy-independent parameters. The efforts to connect the knowledge into a full UV-complete model have altogether been fairly limited. There is now a growing demand for a systematic treatment that connects the low-energy BSM phenomenology of the neutrino NSI to the high-scale regime.

  \begin{figure}[!htb]
        \center{\includegraphics[width=0.9\textwidth]
        {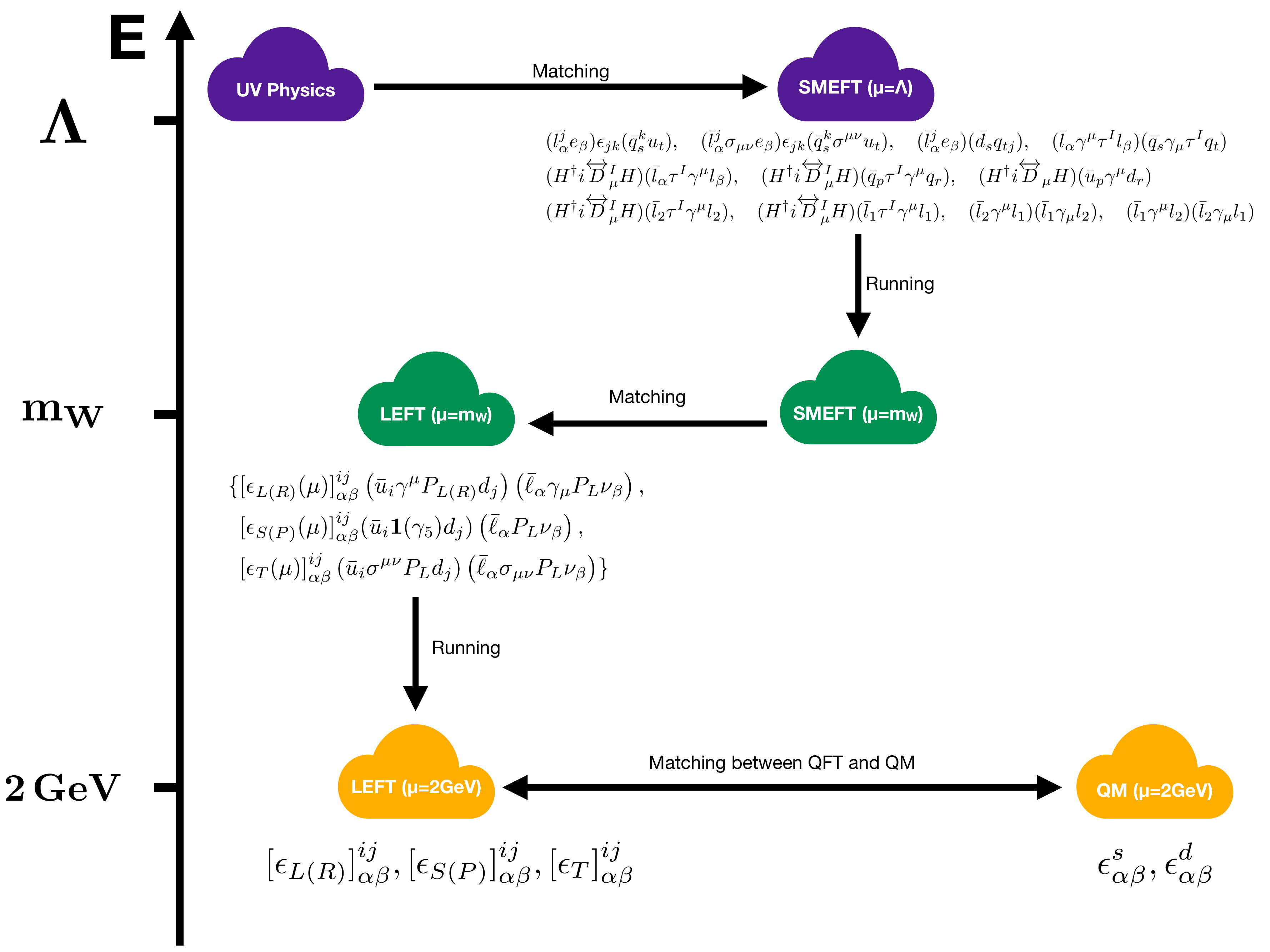}}
        \caption{Schematic description of relating the UV parameters to the neutrino NSI parameters. The vertical axis defines the scale where corresponding EFTs are present. The enumerated operators and/or Wilson coefficients under a certain EFT name at some energy scale $\mu$ are those relevant for our study in this work. See the main text for the details.}
        \label{fig:workflow}
  \end{figure}
  
The Standard Model Effective Fields Theory (SMEFT) provides a systematic \yong{and model-independent} way to parameterize the effects of \yong{some} heavy degrees of freedom in the UV theory into Wilson coefficients of a set of higher dimensional operators. \yong{These operators are constructed by the SM fields and respect the SM gauge symmetries. Therefore, the SMEFT is very robust to study the effects of some UV physics on neutrino NSI parameters.} A schematic workflow to find relations between the parameters in the UV model and the neutrino NSI parameters is shown in figure~\ref{fig:workflow}. Generally, in the top-down point of view, the Wilson coefficients of the SMEFT \yong{operators} can be derived as a function of parameters in the UV theory by integrating out the heavy fields at the UV scale $\Lambda$. \yong{This procedure of deriving the relations between the Wilson coefficients and the couplings in the UV model is called matching, and is indicated by the first row of figure\,\ref{fig:workflow}. Note that the Wilson coefficients after this matching are defined at the UV scale $\Lambda$, while current experiments are performed at a much lower scale. For example, observables from collider experiments are usually defined at the weak scale $m_W$, thus the renormalization group equation (RGE) running of these Wilson coefficients needs to be considered to correctly find contributions to these observables from the UV theory. This step is represented by the arrow connecting the purple and the green in figure\,\ref{fig:workflow}.}

\yong{On the other hand, neutrino oscillation experiments are performed at an even lower scale, and theoretically, the low energy effective field theory (LEFT) around $2\rm\,GeV$ stands as an ideal framework for the study. Obviously, there exists an energy gap between the SMEFT after the aforementioned RGE running and this LEFT. Therefore, extra matching and running need to be appropriately done before making a correct and meaningful prediction for the neutrino NSI parameters from the UV theory. These extra steps are indicated by the last two rows of figure\,\ref{fig:workflow}, where one firstly translates the SMEFT at the weak scale into the LEFT through matching as shown by the arrow connecting the two greens. This second matching is achieved by breaking the electroweak symmetry and integrating out the the top quark, the Higgs boson and the massive gauge bosons. Note that all the Wilson coefficients in the LEFT after this step are defined at the weak scale $m_W$. Then secondly, one applies the RGE and runs these Wilson coefficients down to the 2\,GeV scale for neutrino oscillation experiments. This is indicated by the arrow connecting the green and the yellow in figure\,\ref{fig:workflow}. Now the conventional neutrino NSI parameters in the quantum field theory (QFT) description are encoded in the Wilson coefficients $\epsilon_{L,R,S,P,T}$ of the several charged current operators in LEFT as shown in figure\,\ref{fig:workflow} and eq.\,\eqref{eq:CC} at the 2\,GeV scale. From there on, we leave out the scale parameter $\mu$ and unless otherwise specified, $\epsilon_{L,R,S,P,T}$ shall all be understood as parameters defined at the 2\,GeV scale. Finally, these Wilson coefficients can be matched to the NSI parameters $\epsilon^{s,d}$ in the quantum mechanic (QM) formalism. A faithful matching of these two types of neutrino NSI parameters only exist when the event rates calculated in the QFT formalism are expanded to linear order in the NSI parameters. This point was recently presented in Ref.\,\cite{Falkowski:2019kfn} and is represented by the arrow connecting the two yellows in our schematic plot figure\,\ref{fig:workflow}.}

In the present work, we make an effort to \sampsa{systematically study} the connection between terrestrial neutrino oscillation experiments and the high-energy phenomenology. We study the \yong{physical} potential of the neutrino oscillation experiments by reflecting on the recent findings of the LBL accelerator experiments Tokai-to-Kamioka (T2K)~\cite{Abe:2019vii,Abe:2011ks} and NuMI Off-axis neutrino Appearance (NO$\nu$A)~\cite{Ayres:2007tu,Acero:2019ksn} and the reactor experiments Daya Bay~\cite{An:2012bu,An:2013zwz,Adey:2018zwh}, Double Chooz~\cite{Ardellier:2006mn,DoubleChooz:2019qbj} and Reactor Experiment for Neutrino Oscillation (RENO)~\cite{Ahn:2012nd,Bak:2018ydk}. Practically, we use the \texttt{Wilson} package\,\cite{Aebischer:2018bkb} to take care \yong{of the running from the scale $\Lambda$ down to 2\,GeV, as well as the second matching between SMEFT and the LEFT at $\mu=m_W$ shown in figure\,\ref{fig:workflow}.\footnote{For the matching and running between SMEFT and LEFT for charge-current interactions, see Refs.\,\cite{Cirigliano:2009wk,Gonzalez-Alonso:2017iyc}.} The first matching is model dependent and is thus usually worked out manually model by model. We will discuss more about it in section\,\ref{sec:NSIparam} and provide an example in section\,\ref{sec:leptoquark} for illustration. The last matching is universal and has been studied in Ref.~\cite{Falkowski:2019kfn} recently}. The NSI parameters acquired this way serve as our input to the General Long-Baseline Experiment Simulator ({\tt GLoBES})~\cite{Huber:2004ka,Huber:2007ji} through its new physics extension~\cite{NuPhys}.

This article is decomposed into the following sections: in section\,\ref{sec:NSIparam} we review the standard parameterization of the NSIs in neutrino oscillations and \yong{summarize the matching between the QFT and the QM formalisms relevant for our study}. Section\,\ref{sec:SMEFTtoNSI} focuses on the connection between the SMEFT and the neutrino NSI parameters. We describe the neutrino oscillation data and the related analysis methods used in this work in section\,\ref{sec:nudata}. \yong{Then we present a UV example, i.e., the simplified leptoquark model in section\,\ref{sec:leptoquark} to illustrate our approach following the workflow of figure\,\ref{fig:workflow}}. We finally present a systematical analysis on the \yong{dominant} dimension-6 \yong{SMEFT} operators by providing constraints on the UV scale $\Lambda$ and the Wilson coefficients from neutrino experiments in section\,\ref{sec:pheno}. {The implications on the {\sl CP} violation in the leptonic sector are briefly discussed in section\,\ref{sec:NSIphase}.} The results are then summarized in section\,\ref{sec:concl}.

%%%%%%%%%%%%%%%%%%%%%%%%%%%%%%%%%%%%%%%%%%%%%
\section{\label{sec:NSIparam}Parameterization of Neutrino NSIs}
%%%%%%%%%%%%%%%%%%%%%%%%%%%%%%%%%%%%%%%%%%%%%

\begin{figure}[t]
        \center{\includegraphics[width=1.0\textwidth]
        {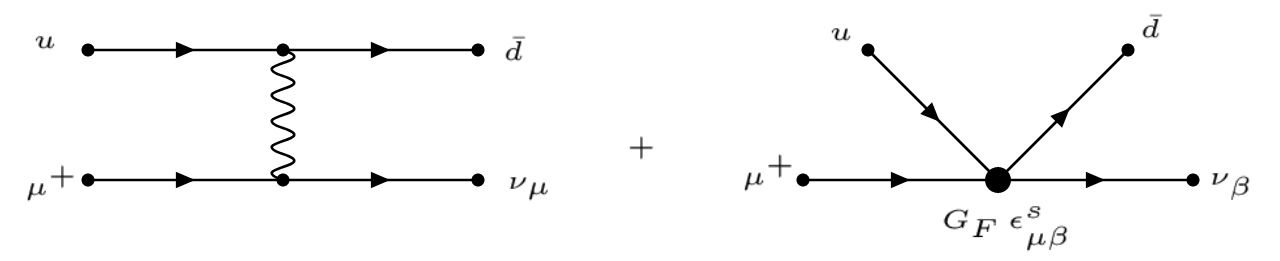}}
        \caption{An illustration of the non-standard interactions in the neutrino source. A positively charged pion decays incoherently into a positive muon and a muon neutrino or a different flavor neutrino.}
        \label{fig:NSIdiagram}
  \end{figure}

Since neutrino oscillations are low-energy experiments, the Quantum Mechanical (QM) formalism is often used to compute the oscillation probability $P_{\alpha\beta}=|\langle \nu_\alpha^d|e^{-i HL}|\nu_\beta^s\rangle|^2$, which describes the probability of a source neutrino of flavor $\beta$ being detected as a neutrino of flavor $\alpha$. The meaning of ``flavor'' here is modified according to the inclusion of the non-standard interactions and is different from the original definition of the flavor that is associated with the lepton doublets diagonalizing the charged lepton mass matrix. The relation between these two flavor eigenstates are related by the QM production and detection NSI parameters $\epsilon^s$ and $\epsilon^d$\,\cite{Grossman:1995wx,GonzalezGarcia:2001mp,Ohlsson:2008gx,Farzan:2017xzy}:
\begin{eqnarray}\label{eq:NSIQM}
|\nu_\alpha^s\rangle = \frac{(1+\epsilon^s)_{\alpha\gamma}}{N^s_\alpha}|\nu_\gamma\rangle,\ \langle\nu_\beta^d| = \langle\nu_\gamma| \frac{(1+\epsilon^d)_{\gamma\beta}}{N^d_\beta},
\end{eqnarray}
where $N_{\alpha}^s=\sqrt{[(1+\epsilon^s)(1+\epsilon^{s\dagger})]_{\alpha\alpha}}$ and $N_{\beta}^d=\sqrt{[(1+\epsilon^{d{\dagger}})(1+\epsilon^d)]_{\beta\beta}}$. The repeated indices $\alpha$ and $\beta$ on the right hand side of equations refer to the diagonal elements, and are not to be confused with indices contraction. The information about matter effects affecting the propagation of the neutrino oscillation in media is encoded in the Hamiltonian $H$:
\begin{equation}
\label{eq:HNSI}
H = \frac{1}{2E_{\nu}}\left[U
\left(
\begin{array}{ccc}
0 \,\,\, & 0 & 0 \\
0 \,\,\, & \Delta m_{21}^2 & 0\\
0 \,\,\, & 0 & \Delta m_{31}^2 
\end{array}
\right) U^{\dagger}
+ A
\left(
\begin{array}{ccc}
1+\varepsilon_{ee}^m & \epsilon_{e\mu}^m & \,\,\, \epsilon_{e\tau}^m \\
\epsilon_{e\mu}^{m*} & \epsilon_{\mu\mu}^m & \,\,\, \epsilon_{\mu\tau}^m \\
\epsilon_{e\tau}^{m*} & \epsilon_{\mu\tau}^{m*} & \,\,\, \epsilon_{\tau\tau}^m
\end{array}
\right)
\right],
\end{equation}
where $U$ is the Pontecorvo-Maki-Nakagawa-Sakata matrix~\cite{Pontecorvo:1957cp,Pontecorvo:1957qd,Maki:1960ut,Maki:1962mu,Pontecorvo:1967fh} and $A = \sqrt{2} G_F N_e$, with $G_F$ denoting the Fermi constant and $N_e$ the electron number density in the medium, defines the standard matter potential~\cite{Wolfenstein:1977ue,Mikheev:1986gs}. The matrix elements $\epsilon^m_{\alpha \beta}$ ($\alpha$, $\beta = e$, $\mu$, $\tau$) are the matter NSI parameters. \sampsa{A schematic illustration of the source NSI is presented in figure\,~\ref{fig:NSIdiagram}, where the pion decay process is influenced by non-standard neutrino interactions.}

The NSI parameters defined in the QM formalism have been widely used in the study of neutrino oscillations, and their concepts are rather phenomenological. However, to study the effects originating from some UV physics, it is better to parameterize the NSI parameters in the QFT framework, i.e., in terms of the Wilson coefficients in the Lagrangian of the LEFT \yong{since these Wilson coefficients are directly related to the parameters in the UV theory as depicted in figure\,\ref{fig:workflow}. On the other hand,} it was recently pointed out in Ref.\,\cite{Falkowski:2019kfn} that matching the differential rate $R_{\alpha\beta}$, defined as the number of oscillation events observed per second per neutrino energy\,\cite{Antusch:2006vwa,Falkowski:2019kfn}, in both formalisms to obtain the connection between the NSI parameters and the Wilson coefficients in LEFT is highly non-trivial. Furthermore, this matching procedure is not guaranteed to give a meaningful match of the two sets of parameters unless one works in the linear approximation in terms of $\epsilon^{s,d}$. In what follows, in light of the smallness of the neutrino NSI parameters, we will adopt the matching procedure described in Ref.\,\cite{Falkowski:2019kfn} to the linear order and study effects from these charge-current NSI parameters.\footnote{For the neutral-current neutrino NSIs, see, for example, Ref.\,\cite{Dev:2019anc} for a recent review and Refs.\,\cite{Choudhury:2018xsm,Masud:2018pig,Friedland:2011za,Pandey:2019apj,Babu:2020nna,Liu:2020emq,Dey:2020fbx} for their recent studies. In this work, however, we will only focus on the charged-current NSI parameters.}

To start, we write down the most general {\color{black}charged-current (CC)} neutrino NSIs in LEFT as\cite{Falkowski:2019kfn}:\footnote{For this CC NSIs, we adopt the convention used in Ref.\,\cite{Falkowski:2019kfn}, where the (V+A), scalar, pseudoscalar and tensor-type interactions are also introduced. In the case of general neutrino interactions, we refer the reader to Ref.\,\cite{Bischer:2019ttk}.}
\begin{align}
\mathcal{L}_{\rm CC} \supset &-\frac{2 V_{u d}}{v^{2}}\left\{\left[\mathbf{1}+\epsilon_{L}\right]^{ij}_{\alpha \beta}\left(\bar{u}_i \gamma^{\mu} P_{L} d_j\right)\left(\bar{\ell}_{\alpha} \gamma_{\mu} P_{L} \nu_{\beta}\right)+\left[\epsilon_{R}\right]^{ij}_{\alpha \beta}\left(\bar{u}_i \gamma^{\mu} P_{R} d_j\right)\left(\bar{\ell}_{\alpha} \gamma_{\mu} P_{L} \nu_{\beta}\right)\right.\nonumber \\
& + \frac{1}{2}\left[\epsilon_{S}\right]^{ij}_{\alpha \beta}(\bar{u}_i d_j)\left(\bar{\ell}_{\alpha} P_{L} \nu_{\beta}\right)-\frac{1}{2}\left[\epsilon_{P}\right]^{ij}_{\alpha \beta}\left(\bar{u}_i \gamma_{5} d_j\right)\left(\bar{\ell}_{\alpha} P_{L} \nu_{\beta}\right)\nonumber \\
&+\left.\frac{1}{4}\left[\epsilon_{T}\right]^{ij}_{\alpha \beta}\left(\bar{u}_i \sigma^{\mu \nu} P_{L} d_j\right)\left(\bar{\ell}_{\alpha} \sigma_{\mu \nu} P_{L} \nu_{\beta}\right)+\mathrm{h.c.}\right\},\label{eq:CC}
\end{align}
where from now on we use the Roman letters $i,j = \{1,2,3\}$ to represent the flavor of quarks, and the Greek letters $\alpha, \beta =\{e,\mu,\tau\}$ to that of charged leptons and neutrinos, $P=P_{L,R}$ are the chiral projection operators for the left- and the right-handed state of Dirac spinors. %and $f,f'=(e,u,d)$. %In this work, we will mainly focus on the first generation quarks. 
Since the energy scale of neutrino oscillation experiments are oftentimes very small, the first generation of quarks is to be understood when the $i,j$ indices are omitted for the NSI parameters $\epsilon_{L,R,S,P,T}$.

Starting from eq.\,\eqref{eq:CC} and following the matching procedure of Ref.\,\cite{Falkowski:2019kfn}, we reproduce the following matching formulas originally derived in Ref.\,\cite{Falkowski:2019kfn} to the linear order of $\epsilon^{s,d}_{\alpha\beta}$'s:\footnote{Here, we use the same notation for the nucleon tensor form factor $g_T$ as in Ref.\,\cite{Falkowski:2019kfn} {in eqs.\,(\ref{eq:sNSI1}-\ref{inverseBetaDecayFor})}.}
\begin{align}
\epsilon_{e \beta}^{s}=&\left[\epsilon_{L}-\epsilon_{R}-\frac{g_{T}}{g_{A}} \frac{m_{e}}{f_{T}\left(E_{\nu}\right)} \epsilon_{T}\right]_{e \beta}^{*},\quad{\text{~($\beta$ decay)}} \label{eq:sNSI1}\\
\epsilon_{\beta e}^{d}=&\left[\epsilon_{L}+\frac{1-3 g_{A}^{2}}{1+3 g_{A}^{2}} \epsilon_{R}-\frac{m_{e}}{E_{\nu}-\Delta}\left(\frac{g_{S}}{1+3 g_{A}^{2}} \epsilon_{S}-\frac{3 g_{A} g_{T}}{1+3 g_{A}^{2}} \epsilon_{T}\right)\right]_{e \beta},{\text{~(inverse $\beta$ decay)}}\label{inverseBetaDecayFor} \\
\epsilon_{\mu \beta}^{s}=&\left[\epsilon_{L}-\epsilon_{R}-\frac{m_{\pi}^{2}}{m_{\mu}\left(m_{u}+m_{d}\right)} \epsilon_{P}\right]_{\mu \beta}^{*},\quad{\text{~(pion decay)}}\label{eq:sNSI2}
\end{align}
where $g_S=1.022(10)$, $g_A=1.251(33)$ and $g_T=0.987(55)$ are the scalar, axial-vector and tensor charges of the nucleon~\cite{Chang:2018uxx,Gupta:2018qil,Aoki:2019cca,Gonzalez-Alonso:2013ura}, {\color{black}$\Delta\equiv m_n-m_p\simeq 1.3 \rm ~MeV$, $E_\nu$ is the neutrino energy, and $f_{T}(E_\nu)$ is the nucleon form factor resulting from the tensor-type NSI in eq.\,\eqref{eq:CC}, for which we use the same parameterization as that in Ref.\,\cite{Falkowski:2019xoe}. Note that for the inverse $\beta$ decay, the formula above only includes the Gamow-Teller type interactions since reactor beta transitions like Daya Bay, Double Chooz and RENO discussed above, are mostly of this type\,\cite{Hayes:2016qnu}. On the other hand, one may also expect contributions to $\epsilon^s_{e\beta}$ from muon decay, however, as we will discuss in the next section, NSI effects from muon decay will be absorbed into the definition of the Fermi constant in our setup.}

Before discussing the effects from the CC NSIs, we would like to emphasize that whenever matching is involved, one shall note that all the parameters in eq.\,\eqref{eq:CC} are to be understood as parameters defined at the $2\rm\,GeV$ scale. Therefore, when one studies the effects of some new physics that lives at a much higher scale, the running effects\yong{, as illustrated in figure\,\ref{fig:workflow},} shall be included as they could contribute significantly to these parameters due to the large QCD coupling. A complete study of the running effects is provided in Ref.\,\cite{Jenkins:2017dyc}. In our numerical analysis, we count on the \texttt{Wilson} package~\cite{Aebischer:2018bkb} to take care of these effects.

%%%%%%%%%%%%%%%%%%%%%%%%%%%%%%%%%%%%%%%%%%%%%
\section{\label{sec:SMEFTtoNSI}SMEFT Contributions to neutrino NSI parameters}
%%%%%%%%%%%%%%%%%%%%%%%%%%%%%%%%%%%%%%%%%%%%%

Since no new particles have been observed at the LHC since 2012 after the discovery of the Higgs, SMEFT has become a popular toolset to systematically study potential deviations from SM predictions due to new physics residing in an energy scale $\Lambda$ much higher than the weak scale. %Neutrino oscillation experiments operate at a relatively low scale, from MeV to GeV, and 
Hence, it is thus natural to study the effects on neutrino NSIs \yong{from this new physics} in the framework of SMEFT. To be specific, we shall derive constraints on the SMEFT Wilson coefficients from neutrino oscillation experiments \yong{following the workflow in figure\,\ref{fig:workflow} and using eqs.\,(\ref{eq:sNSI1}-\ref{eq:sNSI2}) for the matching that essentially relates the UV parameters to $\epsilon^{s,d}$.} Similar work focusing on the neutral-current source NSI can be found in Ref.~\cite{Terol-Calvo:2019vck}, though no source and detection NSI were taken into account. Ref.~\cite{Falkowski:2017pss} studied charge-current NSIs, but they only considered the lepton flavor conserving part of the NSI parameters in eq.\,\eqref{eq:CC}. In this section, we will describe the matching framework we use to derive our results. Our study follows the results in Ref.~\cite{Jenkins:2017jig} and is implemented in the Python package \texttt{Wilson}.

In general, the SMEFT can be organized by the dimension of the operators:
\begin{equation}
    {\cal L}_\text{SMEFT} = \sum_D \sum_{i_D} \frac{c_{i_D}}{\Lambda^{D-4}} {\cal O}_i^{D},
\end{equation}
where $D$ denotes the dimensions of the operators starting from 5, $i_D$ labels the effective operators for a given dimension $D$, $c_{i_D}$ is the dimensionless Wilson coefficient, $\Lambda$ represents the characteristic UV scale. In this work, we will focus on the dimension-6 operators only and assume that neutrino masses are generated by the dimension-5 Weinberg operator\,\cite{Weinberg:1979sa}.

\yong{Starting from a certain UV theory around scale $\Lambda$, its contributions to SMEFT operators can be readily obtained by integrating out the heavy fields with the covariant derivative expansion (CDE) techniques\,\cite{Gaillard:1985uh,Cheyette:1987qz,Henning:2014wua}. Alternatively, one can calculate the amplitudes in both the UV theory and the SMEFT, and then obtain their relations through amplitudes matching. See, for example, Ref.\,\cite{Skiba:2010xn} for a detailed discussion.} This is a paradigm of the top-down EFT approach in studying the phenomenology of new physics\yong{, depicted by the first row of figure\,\ref{fig:workflow}}. We will show a concrete example of this approach to study {\color{black}constraints from neutrino oscillations on a simplified scalar} leptoquark model in section\,\ref{sec:leptoquark}. On the other hand, in the bottom-up EFT approach, one can view the SMEFT at an arbitrary new physics scale as the starting point and treat all the Wilson coefficients as independent parameters. This approach is more general while the correlation among the Wilson coefficients is lost. \yong{In section\,\ref{sec:pheno}, through a careful study on the correlation of dimension-6 SMEFT operators for neutrino oscillation experiments, we will demonstrate this correlation is in general important.}

In both the EFT approaches described above, the Wilson coefficients {\color{black}of the SMEFT operators are all defined at the new physics scale $\Lambda$, while a natural matching between SMEFT and LEFT is around $m_W$, where one integrates out $t$ quark, the Higgs boson, and the $W,Z$ bosons in the SM as implied by the second row of figure\,\ref{fig:workflow}. Therefore, before the matching procedure is done, one needs to run the SMEFT Wilson coefficients from the scale $\Lambda$ down to $m_W$, i.e., from the first to the second row of figure\,\ref{fig:workflow}. This running of the SMEFT Wilson coefficients were thoroughly studied in Ref.~\cite{Alonso:2013hga,Jenkins:2013zja,Jenkins:2013wua}.} {Due to the running effects, neutral current interactions in the UV model can finally generate charged current NSIs. Taking the minimal $Z'$ model as an example, we assume that the SM gauge group is extended by a $U(1)_z$ associated with the new gauge boson $Z'$ and a complex scalar $\phi$ which is responsible for the spontaneous breaking of the $U(1)_z$ giving mass to $Z'$. With no new fermion fields introduced, the gauge anomaly cancellation restricts the $U(1)_z$ charges of the SM fields to be proportional to their $U(1)_Y$ hypercharges or trivially to equal to zero. In this case we find that a tree level matching generates flavor conserving neutral current $C_{\substack{ll \\ ee\mu\mu}}$, which generates $C_{\substack{ll \\ e\mu\mu e}}$ through the RGE running  and further induces the $\epsilon_L$ via the calibration of $G_F$ as we will discuss below.}

The matching between SMEFT and the LEFT is studied in detail in Ref.~\cite{Jenkins:2017jig}. At tree level, %this is done by setting the Higgs field to its vev $v_T$ and integrating out the top quark and the $W,Z$ gauge bosons. 
we classify the SMEFT operators \yong{that contribute to the neutrino NSI operators in the LEFT after matching into two types as listed below, where the type-A ones generates the operators in the LEFT directly via expansions of the left-handed lepton and the quark doublets, and the type-B ones contribute by modifying the interactions between the gauge bosons and fermions after setting the Higgs field to its vacuum expectation value $v_T$ and integrating out the $W$ bosons and setting.}

\begin{center}
{\renewcommand\arraystretch{1.5}
\begin{tabular}{ |c|c| } 
 \hline
 \multirow{2}{*}{ Type-A}&${\cal O}^{(1)}_{\substack{lequ}}=(\bar{l}_{\alpha}^{j} e_{\beta}) \epsilon_{j k}(\bar{q}_{s}^{k} u_{t})\quad {\cal O}^{(3)}_{\substack{lequ}}=(\bar{l}_{\alpha}^{j} \sigma_{\mu \nu} e_{\beta}) \epsilon_{j k}(\bar{q}_{s}^{k} \sigma^{\mu \nu} u_{t})$\\
& $ {\cal O}_{\substack{ledq}}=(\bar{l}_{\alpha}^{j} e_{\beta})(\bar{d}_{s} q_{t j})\quad {\cal O}^{(3)}_{\substack{lq}}=(\bar{l}_{\alpha} \gamma^{\mu} \tau^{I} l_{\beta})(\bar{q}_{s} \gamma_{\mu} \tau^{I} q_{t})$\\
 \hline
 \multirow{4}{*}{ Type-B}&${\cal O}^{(3)}_{\substack{Hl}}=(H^\dag i\overleftrightarrow{D}^I_\mu H)(\bar l_\alpha \tau^I \gamma^\mu l_\beta)\quad {\cal O}^{(3)}_{\substack{Hq}}=(H^\dag i\overleftrightarrow{D}^I_\mu H)(\bar q_p \tau^I \gamma^\mu q_r)
 $\\
 &$\quad {\cal O}_{\substack{Hud}}=(H^\dag i\overleftrightarrow{D}_\mu H)(\bar u_p \gamma^\mu d_r) \quad {\cal O}_{\substack{ll \\ 2112}}=(\bar l_2 \gamma^\mu l_1)(\bar l_1 \gamma_\mu l_2)
 $\\
& ${\cal O}_{\substack{ll \\ 1221}}=(\bar l_1 \gamma^\mu l_2)(\bar l_2 \gamma_\mu l_1)\quad {\cal O}^{(3)}_{\substack{Hl \\ 22}}=(H^\dag i\overleftrightarrow{D}^I_\mu H)(\bar l_2 \tau^I \gamma^\mu l_2)$\\
&$ {\cal O}^{(3)}_{\substack{Hl \\ 11}}=(H^\dag i\overleftrightarrow{D}^I_\mu H)(\bar l_1 \tau^I \gamma^\mu l_1)$\\[1pt]
 \hline
\end{tabular}
}
\end{center}

One example of the type-A operators is:
\begin{equation}
    {\cal O}^{(1)}_{lequ} = \bar{l}^j_\alpha e_\beta\epsilon_{jk} \bar{q}^k_s u_t,\label{eq:lequ1}
\end{equation}
where $l_\alpha = (\nu_{L,\alpha}, \ell_{L,\alpha})^T$ and $q_s = (u_{L,s}, d_{L,s})^T$ are the $SU(2)_L$ lepton and quark doublets. Expanding the conjugate of this operator, one can find that it contains the following CC neutrino NSI relevant operators:
\begin{equation}
   \bar{l}^j_\alpha e_\beta\epsilon_{jk} \bar{q}^k_s u_t \supset (\bar{u}_t P_L d_s)( \bar{\ell}_\alpha P_L \nu_\beta) =  \frac{1}{2}(\bar{u}_t d_s)( \bar{\ell}_\alpha P_L \nu_\beta) -\frac{1}{2}(\bar{u}_t \gamma^5 d_s)( \bar{\ell}_\alpha P_L \nu_\beta).
\end{equation}
Hence, if the corresponding dimensionless Wilson coefficients of the operators in eq.\,\eqref{eq:lequ1} is $C^{(1)}_{\substack{lequ \\ prst}}$, it can then be related to $\epsilon_S$ and $\epsilon_P$ defined in eq.\,\eqref{eq:CC} as:
\begin{equation}
    (\epsilon_S)^{ts}_{\alpha\beta} = (\epsilon_P)^{ts}_{\alpha\beta} = (C^{(1)}_{\substack{lequ \\ \alpha\beta st}})^*.
\end{equation}

For the Type-B operators, one can parameterize their effects in the following Lagrangian~\cite{Jenkins:2017jig}:
\begin{eqnarray}
{\cal L}_{\rm gauge}  = -\frac{\bar{g}_2}{\sqrt{2}}\{\mathcal{W}^+_\mu j^\mu_{\mathcal{W}} +h.c.\}-\bar{g}_Z Z_\mu j^\mu_Z,
\end{eqnarray}
where once again we only focus on the CC $j^\mu_W$:
\begin{eqnarray}
j_{\mathcal{W}}^{\mu}=\left[W_{l}\right]_{\alpha\beta} \bar{\nu}_{L \alpha} \gamma^{\mu} e_{L \beta}+\left[W_{q}\right]_{p r} \bar{u}_{L p} \gamma^{\mu} d_{L r}+\left[W_{R}\right]_{p r} \bar{u}_{R p} \gamma^{\mu} d_{R r},
\end{eqnarray}
with $W_{l,q,R}$ parameterizing the modification of the $W$ boson coupling to the corresponding fermion currents
\begin{eqnarray}
\left[W_{l}\right]_{\alpha\beta}=\left[\delta_{\alpha\beta}+v_{T}^{2} C_{\substack{H l\\ \alpha\beta}}^{(3)}\right], \quad\left[W_{q}\right]_{p r}=\left[\delta_{p r}+v_{T}^{2} C_{\substack{H q\\ pr}}^{(3)}\right], \quad\left[W_{R}\right]_{p r}=\left[\frac{1}{2} v_{T}^{2} C_{\substack{H u d \\ pr}}\right]
\end{eqnarray}
The $W$ boson mass is $M_{\mathcal{W}}=\bar{g}_2^2v_T^2/4$, with $\bar{g}_2$ the modified gauge coupling of the SU(2) group {\color{black}when normalization of the gauge fields due to the presence of ${\cal O}_{HW}=|H|^2W^I_{\mu\nu}W^{I\mu\nu}$~\cite{Jenkins:2017jig}. However, as we shall see immediately, ${\cal O}_{HW}$ is not relevant for our tree-level matching}. After integrating out the $W$ boson at tree level, the effective Lagrangian can be written as:
\begin{eqnarray}
{\cal L}_{\rm C} = -\frac{\bar{g}^2_2}{2M_{\mathcal{W}}}j^\mu_{\mathcal{W}}j_{\mathcal{W}{\mu}}=-\frac{2}{v_T^2}j^\mu_{\mathcal{W}}j_{\mathcal{W}{\mu}},
\end{eqnarray}
note that $\bar{g}_2$ cancels in the numerator and the denominator.
Upon expanding $j^\mu_{\mathcal{W}}$, the above equation will then generate all the terms in eq.~\eqref{eq:CC}. However, one needs to pay particular attention to the fact that $v_T$ above is different from $v$ in eq.\,\eqref{eq:CC}. The latter is defined through the Fermi constant as $G_F=1/(\sqrt{2}v^2)$, and the relation between $v_T$ and $G_F$, the Fermi constant $G_\mu$ obtained from muon decay \yong{in our setup}, is as follows:
\begin{eqnarray}
\frac{4G_F}{\sqrt{2}} = \frac{4G_\mu}{\sqrt{2}}=\frac{2}{v_T^2}-C_{\substack{ll \\ \mu e e\mu}}-C_{\substack{ll \\ e\mu \mu e}}+C^{(3)}_{\substack{Hl \\ \mu \mu}}+C^{(3)}_{\substack{Hl \\ e e}}
\end{eqnarray}
The term on the right hand side is essentially the Wilson coefficients of the LEFT operators $\left(\bar{\nu}_{L \mu} \gamma^{\mu} \nu_{L e}\right)\left(\bar{e}_{L} \gamma_{\mu} \mu_{L}\right)$ after the tree level matching. In principle, there could be a non-vanishing Wilson coefficients of $\left(\bar{\nu}_{L \mu} \gamma^{\mu} \nu_{L e}\right)\left(\bar{e}_{R} \gamma_{\mu} \mu_{R}\right)$ from the new physics effects contributing to the muon decay process, while its contribution is higher order or suppressed by the $m_e/m_\mu$ in the interference terms due to the opposite chirality of the electron in the two operators~\cite{Jenkins:2017jig}. {Since there are two neutrinos involved in muon decay, the matching between the QFT and the QM formalisms are less trivial compared with pion decay, beta and inverse beta decay discussed in section\,\ref{sec:NSIparam}. On the other hand, for different setups with respect to $G_F$ as discussed above, it would become necessary to have the matching formulae for muon decay. For this reason, we present these matching formulae in Appendix~\ref{sec:muonLEFT}. Our results agree with those in Ref.\,\cite{Falkowski:2019kfn} for both neutrinos in the final states after Taylor expansion around small electron mass $m_e$. A comparison between our results and those in Ref.\,\cite{Falkowski:2019kfn} is also discussed in Appendix~\ref{sec:muonLEFT}.}

The last subtlety of this framework is related to the definition of quark flavors in operators. At tree level, the mass matrices $\left[M_{\psi}\right]_{r s}$ with SMEFT dimension-6 operators are:
\begin{eqnarray}
\left[M_{\psi}\right]_{r s}=\frac{v_{T}}{\sqrt{2}}\left(\left[Y_{\psi}\right]_{r s}-\frac{1}{2} v^{2} C_{\psi H}^{*}\right), \quad \psi=u, d.
\end{eqnarray}
Depending on whether the up-type quark or the down-type quark mass matrix is diagonal, there are two types of bases at the definition scale, i.e., the ``Warsaw up'' and the ``Warsaw down'' bases:
\begin{eqnarray}
&& \text{Warsaw up}: M_{d} \rightarrow \operatorname{diag}\left(m_{d}, m_{s}, m_{b}\right) V^{\dagger}, \quad M_{u} \rightarrow \operatorname{diag}\left(m_{u}, m_{c}, m_{t}\right), \\
&&\text{Warsaw down}: M_{d} \rightarrow \operatorname{diag}\left(m_{d}, m_{s}, m_{b}\right) , \quad M_{u} \rightarrow V^{\dagger}\operatorname{diag}\left(m_{u}, m_{c}, m_{t}\right).
\end{eqnarray}

For the lepton sector, in {\color{black}either} the ``Warsaw up'' {\color{black}or the} ``Warsaw down'' basis, we always choose the bases of the lepton doublets and the right-handed charged lepton fields such that the mass matrix of the charged leptons is diagonal. We would like to emphasize here that both bases are specifically defined at a fixed scale such that when running effects are included, off-diagonal elements in the mass matrices are likely to reappear at a different scale. Therefore, whenever a running is performed, one needs to rotate the basis of fermion fields to change the operators back into the  ``Warsaw up'' or the ``Warsaw down'' basis.

%%%%%%%%%%%%%%%%%%%%%%%%%%%%%%%%%%%%%%%%%%%%%
%\section{\label{sec:nudata}Description of the neutrino oscillation data}
%\section{\label{sec:nudata}Brief description of the neutrino oscillation data and the resulting constraints}
\section{\label{sec:nudata}A brief description of the neutrino oscillation data and the resulting constraints to NSI parameters}
%%%%%%%%%%%%%%%%%%%%%%%%%%%%%%%%%%%%%%%%%%%%%

The focus of our work is in the present generation of neutrino oscillation experiments \yong{and to} investigate the effects of dimension-6 \yong{SMEFT} operators on neutrino oscillation. To this end, we simulate the neutrino oscillation corresponding to the recently collected data in the LBL experiments T2K and NO$\nu$A and the reactor experiments Daya Bay, Double Chooz and RENO. We consider the data sets published in the Neutrino 2020 conference~\cite{Neutrino2020} and reproduce them with {\tt GLoBES}. In this section, we briefly describe the {\color{black} nature of the simulated} data and numerical methods used in this work.

\subsection{LBL neutrino experiments T2K and NO$\nu$A}
\label{sec:pheno:LBL}

The LBL neutrino experiments selected for this work are T2K~\cite{Abe:2019vii} and NO$\nu$A~\cite{Acero:2019ksn} experiments. T2K and NO$\nu$A are on-going LBL experiments where intensive beams of muon neutrinos and antineutrinos are created through pion decay. Muon neutrinos and antineutrinos produced via this method make the majority of the beam composition, about 96\%-98\%, while the rest of the beam consists of the intrinsic beam background of electron neutrinos and antineutrinos.

\subsubsection{The T2K facility}
The T2K beam facility is located at J-PARC in Japan, which produces neutrinos and antineutrinos with a proton accelerator. The proton beam generates neutrinos and antineutrinos at 0.77~kW power output. The beam was originally scheduled to run 2 years in neutrino mode and 6 years in antineutrino mode. The produced muon neutrinos and antineutrinos are sent to traverse traverse 295~km underground, upon which they undergo oscillations to other flavor states, mainly electron neutrinos and antineutrinos. The T2K collaboration recently reported oscillation data corresponding to 3.13$\times$10$^{21}$ protons-on-target (POT)~\cite{Abe:2020vdv,Abe:2019vii}. The data consists of the combined run of 1.49$\times$10$^{21}$ POT in neutrino beam and 1.64$\times$10$^{21}$ POT in antineutrino beam modes in T2K beam between 2009 and 2018.

The far detector of T2K is Super-Kamiokande, which is a large ultra-pure water-based neutrino detector where neutrinos and antineutrinos sent from J-PARC are observed via the creation of Cherenkov rings. The neutrinos undergo charged-current interactions in the detector creating charged leptons of the same flavor. It is these charged particles that lead to the observation of the Cherenkov rings. Super-Kamiokande has 22.5~kton fiducial mass and it stands 2.5$^\circ$ off the beam axis. In this position, the neutrino and antineutrino beams from J-PARC peak at about 600~MeV, and the dominant interaction type is therefore the charged-current quasi-elastic (CCQE) interaction with small chance for resonant charged-current pion production (CC1$\pi$). The neutrino and antineutrino beams are also studied at the ND280 near detector complex, which includes a Water Cherenkov part of 1,529~kg fiducial mass at 280~m distance and 2.5$^\circ$ from the source. The near detector data corresponds to 5.8$\times$10$^{20}$ POT and 3.9$\times$10$^{20}$ POT in neutrino and antineutrino beam modes, respectively

The oscillation data from T2K is divided into five distinct samples based on their event topology. The appearance data in T2K probes $\nu_\mu \rightarrow \nu_\mu$ and $\bar{\nu}_\mu \rightarrow \bar{\nu}_\mu$ oscillations, whereas the disappearance data consists of events arising from $\nu_\mu \rightarrow \nu_e$ and $\bar{\nu}_\mu \rightarrow \bar{\nu}_e$ oscillations undergoing CCQE interaction. The appearance data also has a third sample, where the $\nu_e$ are observed together with a positively charged pion in the final state ($\nu_e$CC1$\pi^+$). The $\nu_\mu$ and $\bar{\nu}_\mu$ samples are divided into 28 and 19 even-sized energy bins in the interval [0.2, 3.0]~GeV. The $\nu_e$ and $\bar{\nu}_e$ are collected from 23 equi-sized bins in [0.1, 1.25]~GeV, while the $\nu_e$CC1$\pi^+$ sample has 16 bins in [0.45, 1.25]~GeV.

\subsubsection{The NO$\nu$A facility}
The NO$\nu$A experiment~\cite{Acero:2019ksn} generates muon neutrino and antineutrino beams with the use of pion decay at 0.7~kW average power in the NuMI beam facility based in Fermilab in Illinois, USA. NO$\nu$A has a near detector located at 1~km from the beam facility and a far detector facility at 810~km distance in Ash River, Minnesota. The NO$\nu$A near and far detectors are totally active liquid scintillator detectors with about 193~ton and 14~kton fiducial masses. Both detectors are placed 0.8$^\circ$ off-axis from the source. The neutrinos and antineutrinos produced in NO$\nu$A spread over a wide range of energies around 2~GeV, and they may undergo a variety of charged-current interactions. The neutrinos and antineutrinos are detected through the scintillation light emitted by the charged particles created in the interaction. 

For our work, we simulate the NO$\nu$A data corresponding to 12.33$\times$10$^{20}$ POT exposure in the NuMI source~\cite{Acero:2019ksn}. The data consists of five different samples for electron-like events and two samples for muon-like events. The four electron-like samples are categorized into two categories, which are defined by the purity of the event: low-PID and high-PID\footnote{PID stands for particle identification. The high-PID and low-PID are categories for the convolutional neural network that categorizes neutrino events by their purity.}. The muon-like samples are split into 19 unequally spaced energy bins in the range [0.75, 4.0]~GeV, whereas electron-like samples are separated into 6 bins in [1.0, 4.0]~GeV. The fifth sample is the so-called peripheral sample. In this work, we consider the samples for the electron-like and muon-like events, while the peripheral sample \yong{is} left out of the analysis.

\subsection{Reactor antineutrino experiments Daya Bay, Double Chooz and RENO}
\label{sec:pheno:reactors}

Reactor antineutrino experiments are oscillation experiments where the emission of electron antineutrinos are observed from nuclear reactors. Reactor experiments are sensitive to $\bar{\nu}_e \rightarrow \bar{\nu}_e$ disappearance, which provide access to a set of dimension-6 \yong{SMEFT} operators different \yong{from} what can be studied in the LBL experiments. We consider the presently running reactor neutrino experiments Daya Bay~\cite{An:2012bu}, Double Chooz~\cite{Ardellier:2006mn} and RENO~\cite{Ahn:2012nd} in this work.

Daya Bay is a long-running reactor neutrino experiment based in southern Guangdong, China~\cite{An:2012bu}. The Daya Bay experiment was the first to report simultaneous measurements of reactor antineutrinos at multiple
baselines. The experiment led to the discovery of $\bar{\nu}_e$ oscillations over km-long baseline lengths. The experimental compound consists \yong{of} six reactors and eight antineutrino detectors. Each of the nuclear reactors \yong{is} estimated to have an output of about 2.9~GW thermal power. The antineutrino detectors contain 20~t of Gd-doped liquid scintillator in fiducial mass. The detectors are placed in pairs in three locations, in two near detector halls EH1 and EH2, and in a single far detector hall EH3. The detector compound is equivalent \yong{to} 160~t of liquid scintillator. Daya Bay has collected 1958 days of data since its launch in 2011, and it will continue to run until the end of 2020. In the present work, we consider the oscillation data that corresponds to the 1958 days of data taking in Daya Bay~\cite{Adey:2018zwh}.

Double Chooz is a short-baseline reactor neutrino experiment currently operating in France~\cite{Ardellier:2006mn}. Double Chooz uses two identical Gd-doped liquid scintillators, which are both placed in the vicinity of two 4.2~GW thermal power reactors. The near and far detectors are located approximately 400 and 1050 meters from the reactor cores, respectively. The experiment first started operation in 2005, when nuclear reactors were offline. This allowed the measurement of the backgrounds without the reactor flux. In this work, we consider antineutrino oscillation data corresponding to 1276 days of running with the Double Chooz far detector, and 587 days with the near detector, respectively. Both detectors are assumed to contain 10.6~t of liquid scintillator. The results of the Double Chooz experiment were reported in Ref.~\cite{DoubleChooz:2019qbj}.

The third presently-running reactor neutrino experiment considered in this work is the RENO in South Korea~\cite{Ahn:2012nd}. RENO is based on the same working principle as Daya Bay and Double Chooz, and it has been taking data with two identical detectors from 2011. The two detectors are placed in near and far locations of 300~m and 1400~m from the reactor compound, and they host a total of 40~t of liquid scintillator in fiducial mass. The experiment has observed the disappearance of reactor neutrinos in their interactions with free protons, followed by neutron capture on hydrogen (n-H). RENO has collected 2508 and 2908 days of data in its near and far detectors with 16.8~GW thermal power~\cite{Bak:2018ydk}.

\yong{In this work, we simulate the neutrino oscillation experiments with {\tt GLoBES}, and successfully reproduce the results reported for T2K and NO$\nu$A in Refs.~\cite{Abe:2019vii,Zarnecki:2020yag} and Ref~\cite{Acero:2019ksn}, as well as the results reported for Daya Bay, Double Chooz and RENO experiments in the Neutrino 2020 conference~\cite{Neutrino2020}. We then study constraints on dimension-6 SMEFT operators following the procedure shown in figure\,\ref{fig:workflow} and described in more detail in sections\,\ref{sec:NSIparam} and \ref{sec:SMEFTtoNSI}. The results are presented for a simplified scalar leptoquark model in section \,\ref{sec:leptoquark} and for the dimension-6 SMEFT operators in section \,\ref{sec:pheno}. Before going to the results directly, we first discuss the methods we use for the numerical analysis in the next subsection.}

\subsection{Numerical analysis}
\label{sec:pheno:methods}

\begin{table}[!t]
\caption{\label{LBLExperiments} The long-baseline neutrino experiments simulated in this study.}
\begin{center}
\begin{tabular}{ccc}
\hline\hline
\rule{0pt}{3ex}Experiment & T2K & NO$\nu$A  \\ \hline
\rule{0pt}{3ex}Source location & Japan & USA  \\ %\hline
\rule{0pt}{3ex}Status & operating & operating \\ %\hline
\rule{0pt}{3ex}Beam power & 770~kW & 700~kW \\ %\hline
\rule{0pt}{3ex}Protons-on-target & 1.3$\times$10$^{21}$ & 12.33$\times$10$^{20}$ \\ %\hline
\rule{0pt}{3ex}Fiducial mass (far) & 22.5~kt & 14~kt \\ %\hline
\rule{0pt}{3ex}Fiducial mass (near) & 1.529~t & 193~t \\ %\hline
\rule{0pt}{3ex}Baseline length (far) & 295~km & 810~km \\ %\hline
\rule{0pt}{3ex}Baseline length (near) & 280~m & 1~km \\ %\hline
\rule{0pt}{3ex}Off-axis angle & 2.5$^\circ$ & 0.8$^\circ$ \\ \hline
\rule{0pt}{3ex}References & Ref.~\cite{Abe:2019vii} & Ref.~\cite{Acero:2019ksn} \\ \hline\hline
\end{tabular}
\end{center}
\end{table}

The analysis presented in this work is performed with the {\tt GLoBES} software~\cite{Huber:2004ka,Huber:2007ji} for neutrino oscillation experiments. In order to compute the probabilities with NSIs, the add-on New Physics has been included~\cite{NuPhys}.

In all considered experiments, the simulated neutrino oscillation data is analysed with $\chi^2$ functions, which span over energy bins $i =$ 1, 2, ... and detectors $d$. In the case of LBL experiments, the $\chi^2$ function used in the analysis is
\begin{equation}
\chi^2 = \sum_{d} \left( \sum_{i} 2\left[ T_{i,d} - O_{i,d} \left( 1 + \log\frac{O_{i,d}}{T_{i,d}} \right) \right] + \frac{\zeta_{\text{sg}}^2}{\sigma_{\zeta_{\text{sg}}}^2} + \frac{\zeta_{\text{bg}}^2}{\sigma_{\zeta_{\text{bg}}}^2} \right) + {\rm priors},
\label{LBLChi2}
\end{equation}
where $O_{i,d}$ and $T_{i,d}$ stand for the observed and theoretical/predicted events in the near and the far detectors, which are denoted with {\sl d = N} and {\sl F}, respectively. The systematic uncertainties are addressed with the so-called pull-method~\cite{Fogli:2002pt}. We consider normalized errors for signal and background events with nuisance parameters $\zeta_{\text{sg}}$ and $\zeta_{\text{bg}}$, which influence the predicted events $T_{i,d}$ in near and far detectors with a simple shift: $T_{i,d} = (1+\zeta_{\text{sg}})N^{\text{sg}}_{i,d} + (1+\zeta_{\text{bg}})N^{\text{bg}}_{i,d}$ where $N^{\text{sg}}_{i,d}$ and $N^{\text{bg}}_{i,d}$ are signal and background events, respectively. The prior function is defined as the Gaussian distributions of each of the standard neutrino oscillation parameters as well as the non-standard interaction parameters. We adopt the central values and associated errors from the present fit on the world data. 

In the case of the reactor neutrino experiments, it is sufficient to use the $\chi^2$ function
\begin{equation}
\begin{split}
\chi^2 &= \sum_{d} \sum_{i} \frac{(O_{d,i}-T_{d,i} (1 + a_{\rm norm} + \zeta_{d} + \beta_{d} + \xi_{d}))^2}{O_{d,i}}\\ &+ \sum_{d} \left( \frac{\zeta_{d}^2}{\sigma_\zeta^2} + \frac{\xi_d^2}{\sigma_\xi^2} + \frac{\beta_d^2}{\sigma_\beta^2} \right) + \frac{a_{\rm norm}^2}{\sigma_{a}^2} + {\rm priors},
\end{split}
\label{ReactorChi2}
\end{equation}
where {\color{black}$d$ runs through all detectors in the reactor experiment.} The observed and theoretical events are given for detector $d$ and energy bin $i$ by $O_{d,i}$ and $T_{d,i}$, respectively. {\color{black}The nuisance parameters $\zeta_d$, $\xi_d$ and $\beta_d$} and their Gaussian widths are assigned to address the systematic uncertainties regarding {\color{black}the efficiency, energy calibration and scaling in the detectors, respectively.} The nuisance parameter $a_{\rm norm}$ is related to the overall normalization error in the reactor rates. 

As it is apparent from the $\chi^2$ functions used for accelerator and reactor experiments in eqs\,(\ref{LBLChi2}-\ref{ReactorChi2}), the statistical analysis is based on the log-likelihood and likelihood tests, respectively. The $\chi^2$ function for LBL experiments shown in eqs.\,\eqref{LBLChi2} follows Poissonian statistics and is suitable for experiments with relatively low statistics, such as T2K and NO$\nu$A. For reactor experiments Daya Bay, Double Chooz and RENO on the other hand, the Gaussian distribution used in eq.\,\eqref{ReactorChi2} is suitable for the statistical analysis.

\begin{table}[!t]
\caption{\label{ReactorExperiments} Reactor neutrino experiments considered in this study. Each of the reactor experiments have two or more detectors {\color{black}based in different locations. The baseline lengths correspond to the average distance between the reactors and detector compounds.}}
\begin{center}
\begin{tabular}{cccc}
\hline\hline
\rule{0pt}{3ex}Experiment & Double Chooz & Daya Bay & RENO  \\ \hline
\rule{0pt}{3ex}Location & France & China & South Korea  \\ %\hline
\rule{0pt}{3ex}Thermal power & 8.4~GW & 17.4~GW & 16.4~GW \\ %\hline
\rule{0pt}{3ex}Exposure (far) & 1276~days & 1985~days & 2908~days \\ %\hline
\rule{0pt}{3ex}Exposure (near) & 587~days & 1985~days & 2509~days \\ %\hline
\rule{0pt}{3ex}Fiducial mass (far) & 10.6~t & 80~t & 15.4~t \\ %\hline
\rule{0pt}{3ex}Fiducial mass (near) & 10.6~t & 2$\times$40~t & 15.4~t  \\ %\hline
\rule{0pt}{3ex}Baseline length (far) & 1.05~km & 1.579~km & 1.4~km \\ %\hline
\rule{0pt}{3ex}Baseline length (near) & 400~m & 512~m, 561~m & 300~m  \\ \hline
\rule{0pt}{3ex}References & Ref.~\cite{Ardellier:2006mn} & Ref.~\cite{An:2012bu,An:2013zwz} & Ref.~\cite{Ahn:2012nd} \\ \hline\hline
\end{tabular}
\end{center}
\end{table}

The details of the considered LBL experiments can be found in table\,\ref{LBLExperiments}. The reactor experiments are summarized in table\,\ref{ReactorExperiments}. The near and the far units of the Daya Bay experiment refer to the detector compounds in EH1 and EH2, and EH3, respectively. The simulated data is computed from the present best-fit values of the neutrino oscillation parameters from the world data, summarized in table\,\ref{OscillationBestFits}. We adopt priors for the standard oscillation parameters from the best-fit values shown in the table.

\begin{table}[!t]
\caption{\label{OscillationBestFits} The best-fit values and 1$\,\sigma$ confidence level (CL) uncertainties in the standard three-neutrino mixing~\cite{NuFit:5-0,Esteban:2020cvm}. The values are shown for both normal ordering (NO) and inverted ordering (IO), where $\Delta m_{3\ell}^2$ corresponds to $\Delta m_{31}^2$ (NO) and  $\Delta m_{32}^2$ (IO), respectively.}
\begin{center}
%\resizebox{\linewidth}{!}{%
\begin{tabular}{ccccc}\hline\hline
Parameter & Central value $\pm$ 1\,$\sigma$ (NO) \,\,\, & Central value $\pm$ 1\,$\sigma$ (IO) \\ \hline
\rule{0pt}{3ex}$\theta_{12}$ ($^\circ$) & 33.440 $\pm$ 0.755 & 33.450 $\pm$ 0.765 \\ %\hline
\rule{0pt}{3ex}$\theta_{13}$ ($^\circ$) & 8.570 $\pm$ 0.120 & 8.600 $\pm$ 0.120 \\ %\hline
\rule{0pt}{3ex}$\theta_{23}$ ($^\circ$) & 49.200 $\pm$ 1.050 & 49.300 $\pm$ 1.000 \\ %\hline
\rule{0pt}{3ex}$\delta_\text{CP}$ ($^\circ$) & 197.000 $\pm$ 25.500 & 282.000 $\pm$ 28.000 \\ %\hline
\rule{0pt}{3ex}$\Delta m_{21}^2$ (10$^{-5}$ eV$^2$) & 7.420 $\pm$ 0.205 & 7.420 $\pm$ 0.205 \\ %\hline
\rule{0pt}{3ex}$\Delta m_{3l}^2$ (10$^{-3}$ eV$^2$) & 2.517 $\pm$ 0.027 & -2.498 $\pm$ 0.028 \\ \hline\hline
\end{tabular}%}
\end{center}
\end{table}

{The prior values we choose to use in this work are based on the three-neutrino oscillation fit assuming no non-standard interactions. As the numerical analysis we show in this work is about assessing the neutrino NSI that has origin in the new physics occurring at the high scale, the extended framework should in principle be accounted in the priors as well. As we shall see, however, the impact of such extension on our numerical results falls below the precision on our work.}

\subsection{Constraining NSI parameters in neutrino experiments}

We illustrate the power of using the neutrino oscillation experiments in constraining new physics with the following example. By taking only one source or detection NSI parameter $\epsilon^{s/d}_{\alpha \beta}$ ($\alpha$, $\beta = e$, $\mu$, $\tau$) to be non-zero at a time, we obtain the constraints on the magnitude of the NSI parameters. Since the nature of the source and detection NSI depends on the specific neutrino production and detection method used in the experiment, we simulate the LBL and reactor experiments separately. Since no significant deviations from the standard three-neutrino oscillations has been found in the considered experiments, we obtain the upper bounds for each NSI parameter with the following method. 

The $\chi^2$ distribution is computed as $\Delta \chi^2 \equiv \chi^2_{\rm NSI} - \chi^2_{\rm SI}$, where $\chi^2_{\rm NSI}$ is computed with the appropriate $\chi^2$ function assuming one NSI parameter to be non-zero and $\chi^2_{\rm SI}$ with all NSI parameters zero. The resulting $\Delta \chi^2$ distribution follows approximately a $\chi^2$ distribution of one degree of freedom. The $\chi^2_{\rm NSI}$ and $\chi^2_{\rm SI}$ are minimized over the standard neutrino oscillation parameters with related priors. The sensitivity to each NSI parameter is then projected at 95\% CL of statistical significance by requiring $\Delta \chi^2 = 3.8416$.

We present the upper bounds on the source and detection NSI parameters for the LBL and reactor experiments in table\,\ref{NSIbounds}. The results are obtained for the absolute values of the NSI parameters at 95\% CL. We consider only the source NSI parameters $\epsilon_{\mu e}^s$, $\epsilon_{\mu \mu}^s$ and $\epsilon_{\mu \tau}^s$ for the LBL experiments, while the source and detection parameters $\epsilon_{e\beta}^s$ and $\epsilon_{\alpha e}^d$ ($\alpha$, $\beta = e$, $\mu$, $\tau$) are studied in the reactor experiments. As one shall see from the table, the experiments provide the most stringent results on $\left| \epsilon_{\mu e}^s \right|$ in pion decay, $\left| \epsilon_{e e}^s \right|$ in beta decay and $\left| \epsilon_{e e}^d \right|$ in inverse beta decay.\footnote{We note that the results presented in table\,\ref{NSIbounds} are sensitive to how the experimental setups are described in the definition language in {\tt GLoBES}. Whereas the non-oscillation backgrounds played a crucial part in defining the sensitivity to the reactor neutrino experiments, we found the near detector to be particularly important in the long-baseline neutrino experiments.}.

The upper bounds on the source and detection NSI parameters presented here are specified for the case where only one NSI parameter bears a non-zero value. If one should consider a scenario where more than one NSI parameter has a non-negligible value, the constraints on the individual NSI parameters are relaxed. Furthermore, any of these NSI parameters may also have a complex phase. Hence, the results presented in table\,\ref{NSIbounds} describe the most optimistic setup. In the following sections of this work, it is assumed the magnitude and phase of the NSI parameters can be calculated precisely.

\begin{table}[!t]
\caption{\label{NSIbounds} Upper bounds of the NSI parameters associated with the neutrino production and detection. All constraints are given at 95\% {\color{black}CL.}}
\begin{center}
%\resizebox{\linewidth}{!}{%
\begin{tabular}{ccl}\hline\hline
NSI parameter & Upper bound & Experiments \\ \hline
$\left|\epsilon^s_{\mu e}\right|$ & 0.004 & \multirow{3}{*}{T2K\,\cite{Abe:2020vdv,Abe:2019vii,Zarnecki:2020yag}, NO$\nu$A\,\cite{Acero:2019ksn}} \\ %\hline
$\left|\epsilon^s_{\mu \mu}\right|$ & 0.021 & {} \\ %\hline
$\left|\epsilon^s_{\mu \tau}\right|$ & 0.080 &  {}\\ %\hline
\hline\hline
$\left|\epsilon^d_{e e}\right|$ & 0.007 &  \\ %\hline
$\left|\epsilon^d_{\mu e}\right|$ & 0.018 &  \\ %\hline
$\left|\epsilon^d_{\tau e}\right|$ & 0.021 &  \multirow{1}{*}{Daya Bay\,\cite{An:2012bu,Adey:2018zwh}, Double Chooz\,\cite{Ardellier:2006mn,DoubleChooz:2019qbj},}\\%\hline
$\left|\epsilon^s_{ee}\right|$ & 0.007 &  \multirow{1}{*}{and RENO\,\cite{Ahn:2012nd,Bak:2018ydk}} \\ %\hline
$\left|\epsilon^s_{e\mu}\right|$ & 0.018 &   \\ %\hline
$\left|\epsilon^s_{e\tau}\right|$ & 0.021 & \\ %\hline
\hline\hline
\end{tabular}%}
\end{center}
\end{table}

{The presence of the neutrino NSI could also affect the prior values applicable for this work, as noted in the previous section. The effect of the NC NSI, for example, can lead to the emergence of the LMA-D scenario, where the precision to the solar parameters $\theta_{12}$ and $\Delta m_{21}^2$ is affected by the appearing of a local minimum\,\cite{Esteban:2018ppq}. We investigated the effect of the relaxed solar parameter constraints in the NSI sensitivities presented in table\,\ref{NSIbounds}, however, we found no significant change\footnote{The neutrino NSI discussed in Ref.\,\cite{Esteban:2018ppq} is in fact of different nature compared to the NSI we investigate in the present work. Whereas the former work focuses on the NC NSI originating from light mediators, the CC NSI we investigate in this work has origin at the UV scale.}.}

%%%%%%%%%%%%%%%%%%%%%%%%%%%%%%%%%%%%%%%%%%%%%
\section{\label{sec:leptoquark}A UV Example: The simplified scalar leptoquark model}
%%%%%%%%%%%%%%%%%%%%%%%%%%%%%%%%%%%%%%%%%%%%%

\yong{To illustrate how the matching and the running work, as well as how the numerical analysis of neutrino oscillation experiments discussed in last section can be used to impose constraints on the UV models, we adopt the simplified scalar leptoquark model in this section. We will first set up this simple UV model and then illustrate how neutrino experiments are used to constrain this model. As shown in figure\,\ref{fig:workflow}, since this simplified model is defined at a UV scale much above that at which neutrino experiments are carried out, the running effects need to be considered. The {\tt Wilson} package is used for the running and the matching at the weak scale, after which a comparison between theoretical prediction from this simplified UV model and experimental bounds on the NSI parameters is done to obtain constraints on this model. As we will see later in this section, though neutrino experiments are performed at a very low energy scale, they put much more stringent constraints than those from high-energy experiments at colliders.}

\subsection{Model setup and connection to neutrino oscillation experiments}
Our simplified leptoquark model only contains a scalar leptoquark $S$ with the SM quantum number $(\bar{3},1,1/3)$, the renormalizable Lagrangian is given by\,\cite{Gherardi:2020det}
\begin{eqnarray}
\begin{split}
\mathcal{L}_{\text{LQ}} &= |D_\mu S|^2  - M_{1}^2 |S|^2 - \lambda_{H1} |H|^2 |S|^2 {\color{black}- \frac{c}{2}|S|^4} \\
		& + \left( (\lambda^{L})_{i\alpha} \bar q^c _i \epsilon \ell _\alpha
			+ (\lambda^{R})_{i\alpha} \bar u^c _i   e _\alpha  \right) S_1  + \text{h.c.} \label{LQModelLag}  
\end{split}
\end{eqnarray}
{\color{black}with $\epsilon=i\sigma^2$ and $\sigma^2$ the second Pauli matrix.}
In principle, there also exist baryon number violation terms $\bar{q}^c\epsilon q S^*$ and $\bar{u}^c d S^*$, but they are extremely suppressed due to the constraint from proton decay, we thus simply ignore them in our setup. 
{\color{black}Integrating out the heavy leptoquark $S$, one finds this model contributes to the following dimension-6 SMEFT operators in the Warsaw-down basis at tree level\,\cite{Gherardi:2020det}:\footnote{\yong{Contributions at one-loop level are also studied in Ref.\,\cite{Gherardi:2020det}, but since their contributions to the Wilson coefficients during the RGE running are of two-loop order\,\cite{Henning:2014wua}, and can thus be safely ignored. Therefore, we only focus on the tree-level results in this subsection for illustration.}}}
\begin{align}
\mathcal{O}_{\ell q}^{(1)}&\equiv(\bar{\ell}\gamma^\mu\ell)(\bar{q}\gamma_\mu q),\quad\mathcal{O}_{\ell q}^{(3)}\equiv(\bar{\ell}\gamma^\mu\sigma^I\ell)(\bar{q}\gamma_\mu \sigma^Iq),\quad\mathcal{O}_{\ell equ}^{(1)}\equiv(\bar{\ell}^re)\epsilon_{rs}(\bar{q}^su)\label{LQModelSMEFTOpe1}\\
\mathcal{O}_{\ell equ}^{(3)}&\equiv(\bar{\ell}^r\sigma^{\mu\nu}e)\epsilon_{rs}(\bar{q}^s\sigma_{\mu\nu}u),\quad\mathcal{O}_{eu}\equiv(\bar{e}\gamma^\mu e)(\bar{u}\gamma_\mu u)\label{LQModelSMEFTOpe2},
\end{align}
{\color{black}whose Wilson coefficients are given by, respectively\,\cite{Gherardi:2020det},}\footnote{\yong{Note that different from the usual definition for Wilson coefficients, the UV scale $\Lambda=M$ is included here to reflect the dependence of the NSI parameters on both the couplings and the leptoquark mass.}}
\begin{align}
C_{\substack{lq \\ \alpha \beta i j}}^{(1)} &=\frac{\lambda_{i \alpha}^{ L *} \lambda_{j \beta}^{ L}}{4 M^{2}}, \quad
C_{\substack{lq \\ \alpha \beta i j}}^{(3)} =-\frac{\lambda_{i \alpha}^{ L *} \lambda_{j \beta}^{ L}}{4 M^{2}}, \quad 
C_{\substack{lequ \\ \alpha \beta i j}}^{(1)} =\frac{\lambda_{j \beta}^{ R} \lambda_{i \alpha}^{ L *}}{2 M^{2}}, \quad\\
C_{\substack{lequ \\ \alpha \beta i j}}^{(3)} &=-\frac{\lambda_{j \beta}^{R} \lambda_{i \alpha}^{ L *}}{8 M^{2}}, \quad
C_{\substack{eu \\ \alpha \beta i j}} =\frac{\lambda_{i \alpha}^{ R *} \lambda_{j \beta}^{ R}}{2 M^{2}},
\end{align}
{\color{black}where $r,s,$ denote the fundamental representations of the $\rm SU(2)_L$ group, $\alpha,\beta$ stand for lepton flavor indices, and $i,j$ represent quark flavors.}

\yong{Matching this simplified scalar leptoquark model onto the SMEFT is exactly the first step in the first row of figure\,\ref{fig:workflow}, and the Wilson coefficients after this matching are all defined at $\mu=M$ with $M$ here being the UV scale $\Lambda$. To connect this UV model to the low-energy neutrino experiments represented by the last row of figure\,\ref{fig:workflow}, the {\tt Wilson} package is utilized. To be more specific for the purpose of illustrating our approach, we discuss the intermediate steps next.

We first take the dimension-6 SMEFT operators obtained from this simplified scalar leptoquark model and defined in the Warsaw-down basis in eq.\,(\ref{LQModelSMEFTOpe1}-\ref{LQModelSMEFTOpe2}) as the input for the {\tt Wilson} package. Simultaneously, we set all the other dimension-6 SMEFT operators to zero at the input scale $\mu=M$ since we are interested in the prediction specifically from this simplified model. Next, based on the work in Ref.\,\cite{Alonso:2013hga,Jenkins:2013zja,Jenkins:2013wua}, the {\tt Wilson} package runs all the dimension-6 SMEFT operators down to the weak scale $m_W$ as indicated by the arrow connecting the purple and the green in figure\,\ref{fig:workflow}. At this step, the SMEFT operators mix with each other during the running as a result of the entangled RGEs\,\cite{Skiba:2010xn}. Therefore, vanishing SMEFT operators at the input scale do not necessarily imply that they lead to vanishing contributions at a different scale.

After the RGE running above, the {\tt Wilson} package can also automatically match the SMEFT to the LEFT at $\mu=m_W$ based on Refs.\,\cite{Jenkins:2017jig,Aebischer:2015fzz}, as well as a following RGE running from the weak scale $m_W$ down to the 2\,GeV scale for the LEFT based on the theoretical work of Ref.\,\cite{Jenkins:2017dyc}. These two intermediate steps are represented by the second row and the arrow connecting the green and the yellow in figure\,\ref{fig:workflow} respectively. The LEFT obtained at this scale serves as an ideal framework for the description of neutrino oscillation experiments, and the event rates $R_{\alpha\beta}$ can be expressed in terms of the Wilson coefficients $\epsilon_{L,R,S,P,T}$ in the LEFT in the QFT formalism. On the other hand, neutrino observables used to be expressed in terms of the source and detection NSI parameters $\epsilon^{s,d}$ in the QM formalism, and neutrino experimental results are presented as constraints on these $\epsilon^{s,d}$ parameters. Therefore, to connect the UV model directly to neutrino experiments, this last gap needs to be solved. A consistent matching between the QFT and the QM formalisms has recently been discussed in Ref.\,\cite{Falkowski:2019kfn} and is depicted by the last row of figure\,\ref{fig:workflow}.

Following the procedure described above, we present our study on this simplified scalar leptoquark model in the following subsections.}

\begin{figure}[t]
\centering{
  \begin{adjustbox}{max width = \textwidth}
\begin{tabular}{cc}
\includegraphics[scale=0.3]{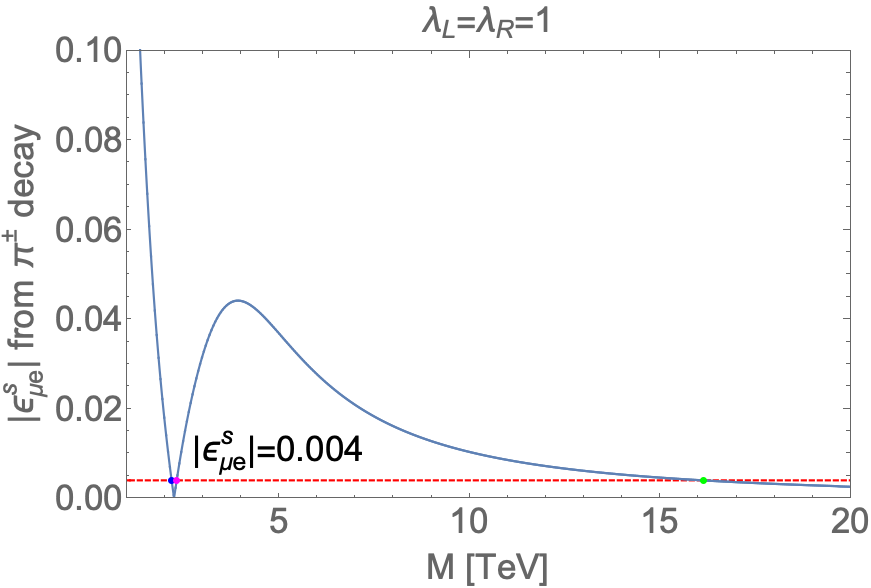}~&~\includegraphics[scale=0.3]{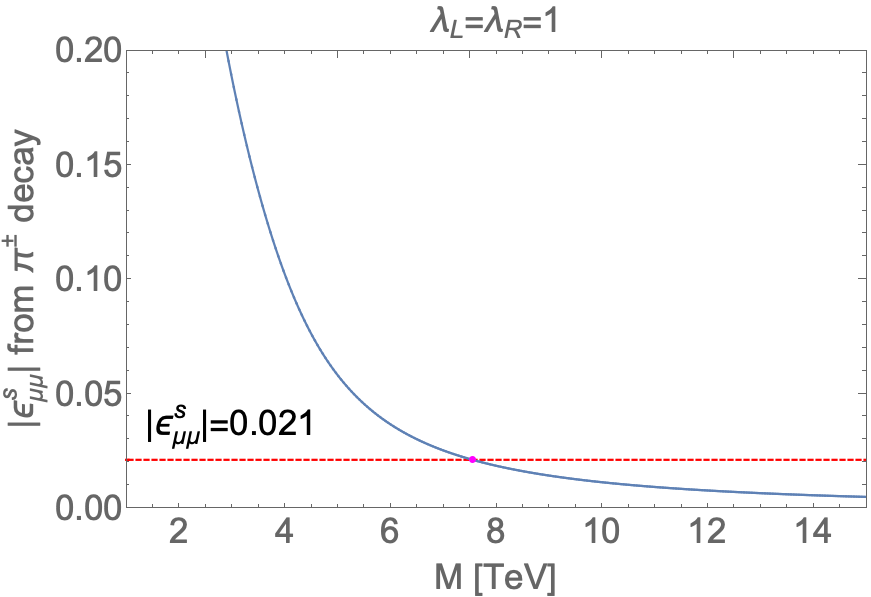}\\
\includegraphics[scale=0.3]{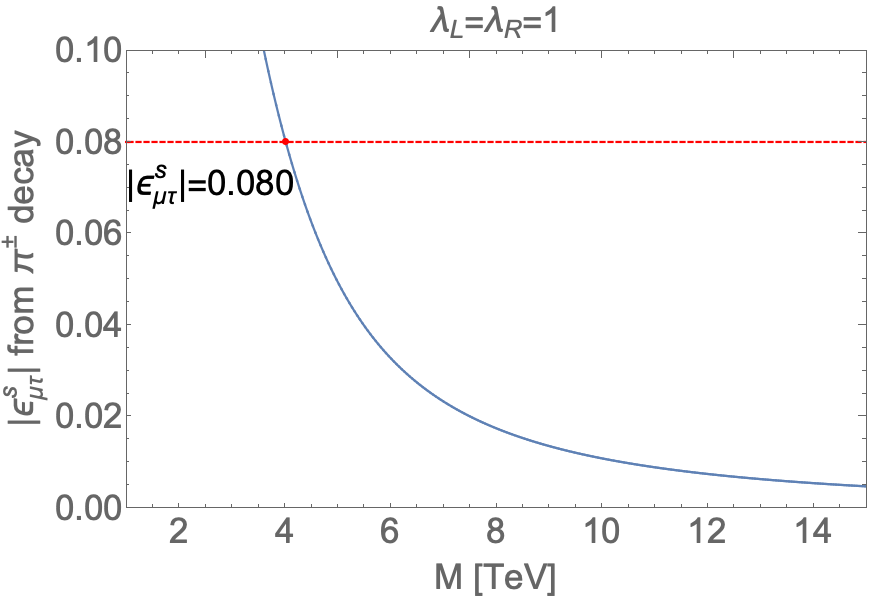}~&~\includegraphics[scale=0.3]{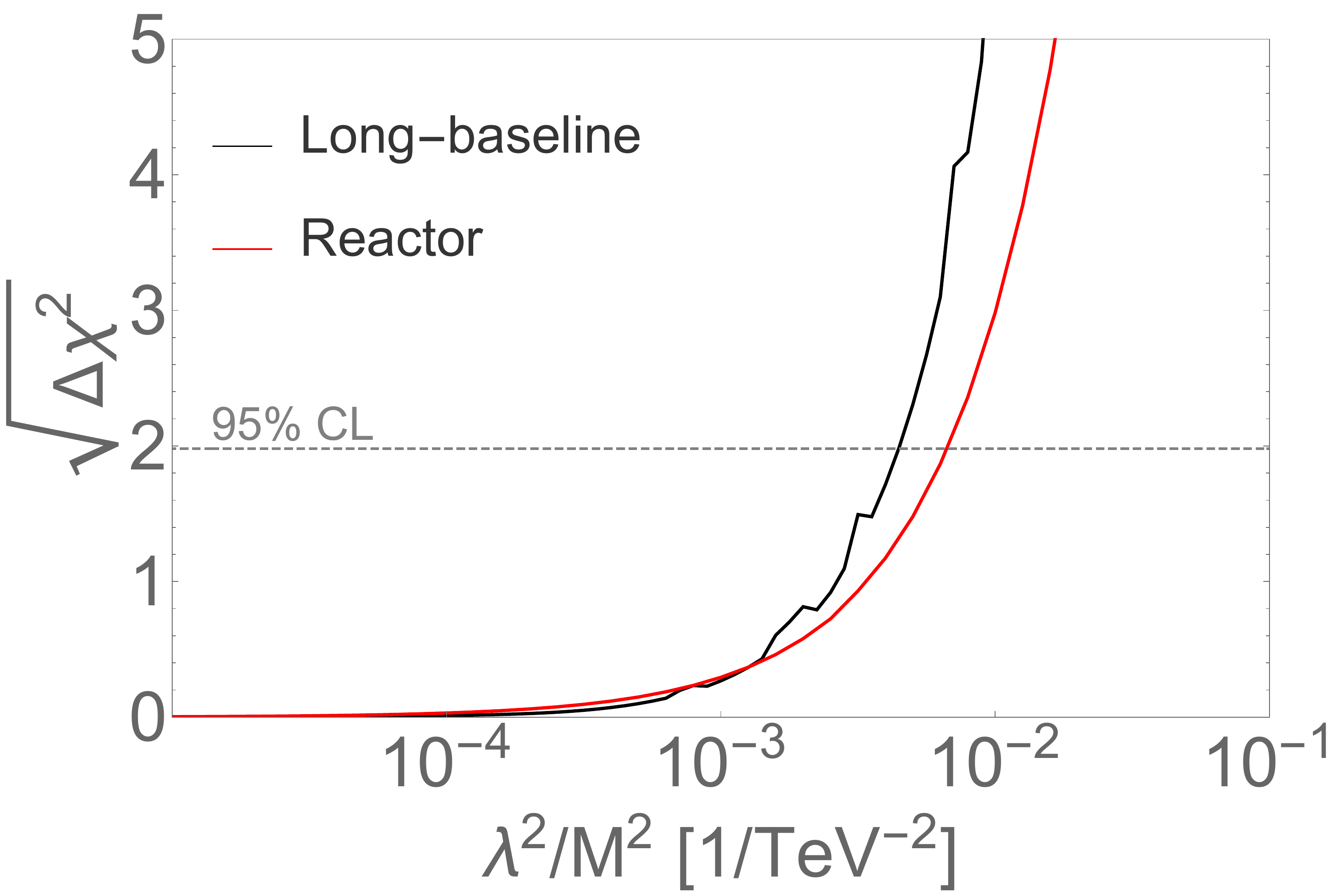}
\end{tabular}
  \end{adjustbox}}
\caption{Constraints on the leptoquark mass from $|\epsilon_{\mu \alpha}^{s}|$ ($\alpha=e,\mu,\tau$) from pion decay. The horizontal red dashed line in each plot corresponds to the upper bounds on the selected NSI parameters summarized in table\,\ref{NSIbounds}. Constraints on the NSI parameters from the beta and the inverse beta decay presently have no constraints on this model as can be seen in section\,\ref{subsec:LQMass}.}\label{LQModelMassScan}
\end{figure}

%%%%%%%%%%%%%%%%%%
\subsection{Constraints on $M$ with $\lambda_{L,R}=1$}\label{subsec:LQMass}
%%%%%%%%%%%%%%%%%%

\yong{As discussed in last subsection, newly introduced interactions between the leptoquark and SM particles in eq.\,\eqref{LQModelLag} generate the tree-level dimension-6 SMEFT operators at the UV scale in eq.\,(\ref{LQModelSMEFTOpe1}-\ref{LQModelSMEFTOpe2}). As a consequence, they contribute to the charged- and neutral-current NSI parameters defined in eq.\,\eqref{eq:CC} at the 2\,GeV scale. Given the current very stringent constraints on these NSI parameters, the low-energy neutrino experiments can thus be applied to explore this simplified model. To that end, we first use the {\tt GLoBES} package to obtain the constraints on each source and detection NSI parameter, summarized in table\,\ref{NSIbounds}. We then use the {\tt Wilson} package to find the prediction on the same set of NSI parameters listed in table\,\ref{NSIbounds} from this UV model.} \sampsa{We also perform this procedure in inverted order, where the output of {\tt Wilson} is transferred to {\tt GLoBES} to directly compute the exclusion limits to the leptoquark model.}

\begin{figure}[t]
\centering{
  \begin{adjustbox}{max width = \textwidth}
\begin{tabular}{cc}
\includegraphics[scale=0.3]{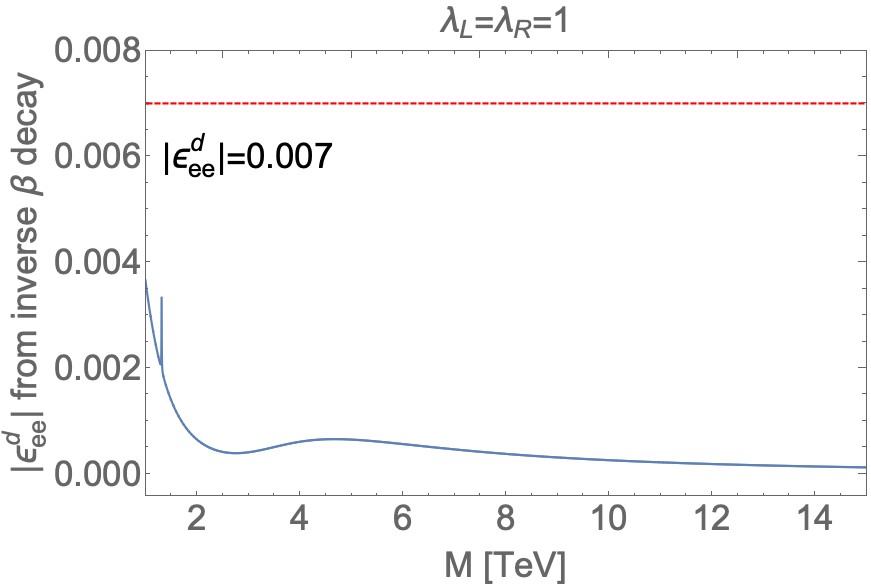} ~& ~ \includegraphics[scale=0.3]{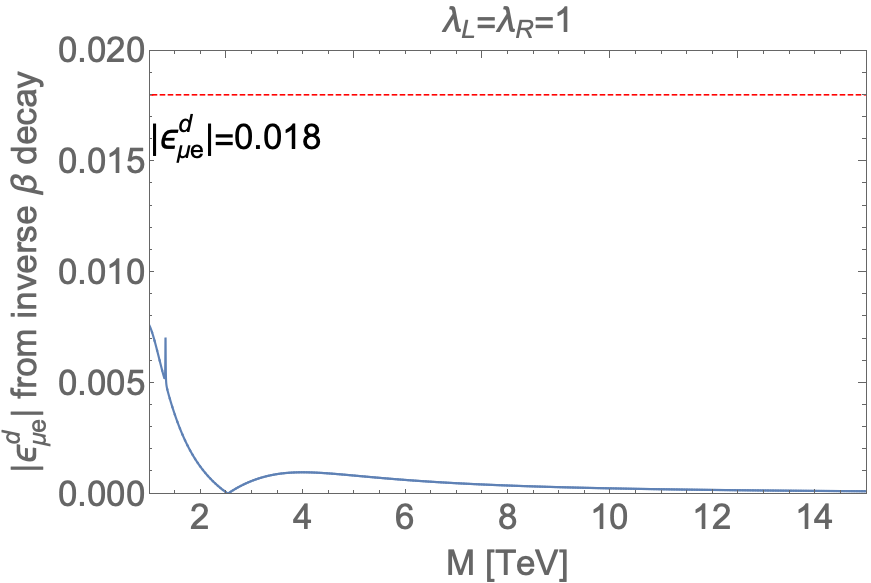}\\
\includegraphics[scale=0.3]{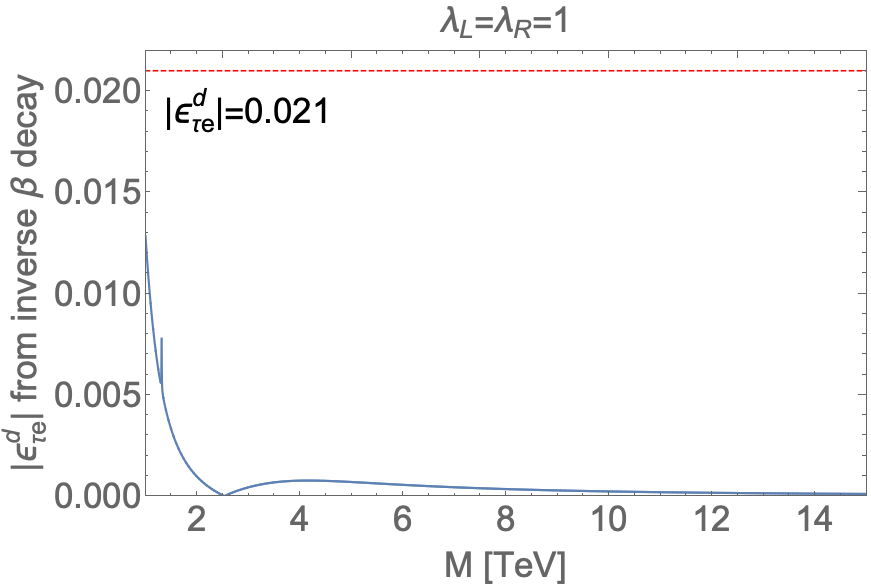}~&~ \includegraphics[scale=0.3]{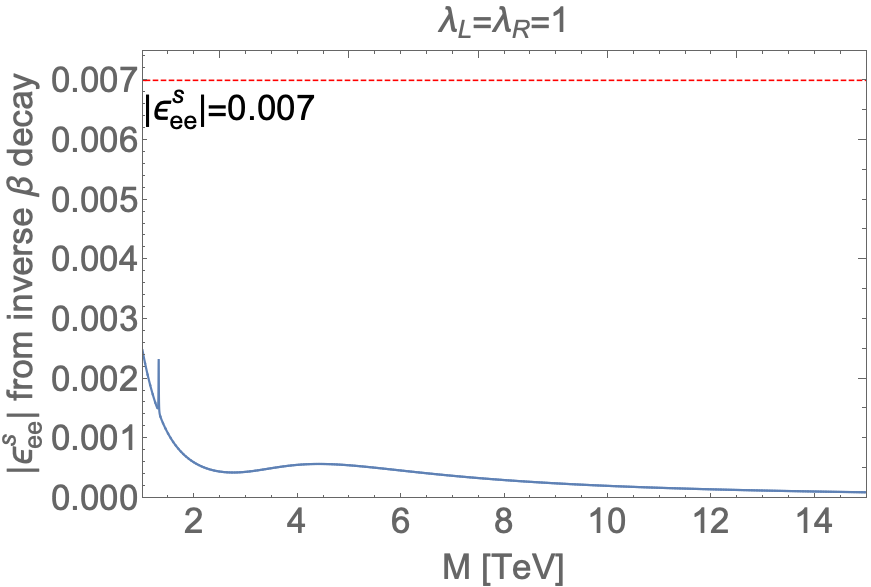}\\
\includegraphics[scale=0.3]{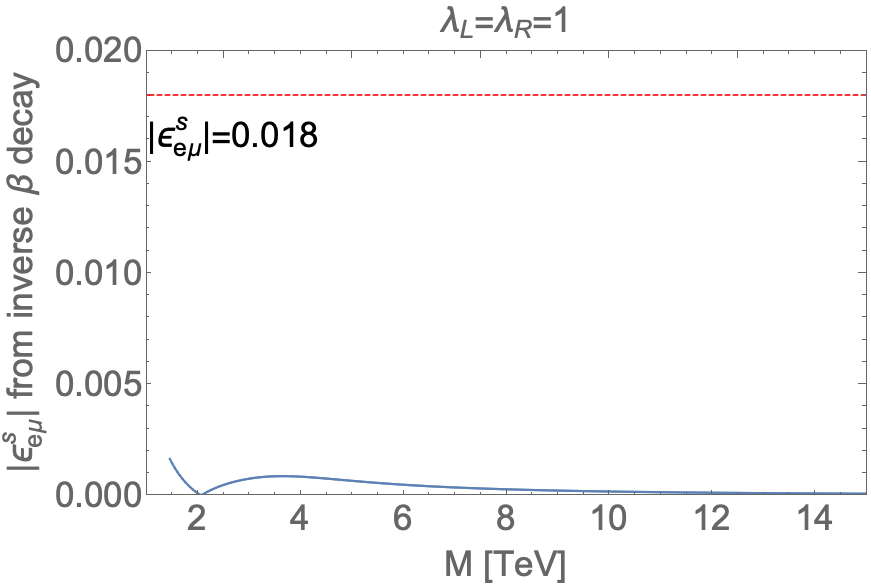}~&~\includegraphics[scale=0.3]{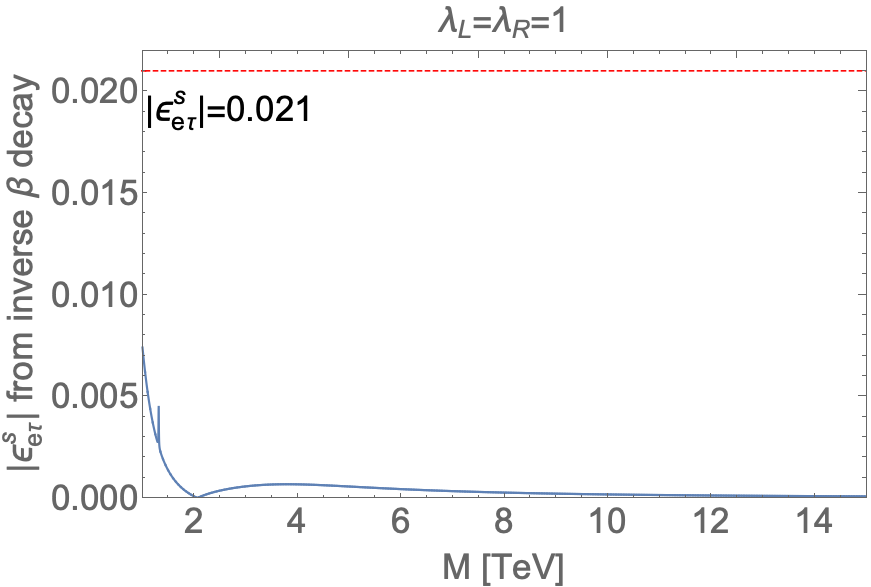}
\end{tabular}
  \end{adjustbox}}
\caption{Constraints on the simplified scalar leptoquark model from the neutrino NSI parameters from beta decay and inverse beta decay. The horizontal red dashed line in each plot corresponds to the current upper bound on the NSI parameter, and the blue curve corresponds to theoretical prediction from this simplified model.}\label{LQModelMassScanMore}
\end{figure}

\yong{In the Warsaw-down basis, since nearly 300 non-vanishing dimension-6 SMEFT operators contribute at the input scale $\Lambda$, for simplicity, we assume $\lambda_{L,R}$ are both real and then present our results in \sampsa{the top-left, top-right and bottom-left panels of} figure\,\ref{LQModelMassScan}. To obtain the plots, we fix $\lambda_{L,R}=1$ and focus on only one NSI parameter at a time. We use a horizontal dashed red line in each subfigure to represent the current upper bound on each NSI parameter summarized in table\,\ref{NSIbounds}. We conclude that, among all the NSI parameters listed in table\,\ref{NSIbounds}, $|\epsilon_{\mu e}^{s}|$ from pion decay leads to the most stringent constraint on the leptoquark model, and it presently excludes a leptoquark lighter than about 16\,TeV as indicated by the upper left plot in figure\,\ref{LQModelMassScan}. Interestingly, we note that there is a tiny window around $M\in[2180,2310]$\,GeV that survives from current constraint on $|\epsilon_{\mu e}^{s}|$, and in general, this tiny region survives even if the experimental sensitivity on $|\epsilon_{\mu e}^{s}|$ is improved. However, if one takes other NSI parameters into account as shown in the upper right and/or lower left plots of figure\,\ref{LQModelMassScan}, this tiny window would have already been excluded.

Another interesting point from figure\,\ref{LQModelMassScan} is that, though the current upper bound on $|\epsilon_{\mu \tau}^{s}|$ only excludes a leptoquark lighter than about 4\,TeV, weakest among the three NSI parameters shown, this exclusion limit is still much stronger than that from collider studies at the ATLAS and the CMS\,\cite{CMS:2020gru,Aaboud:2019bye}, implying the complementarity of high- and low-energy experiments in searching for new physics. A similar observation regarding the complementarity has been obtained for many other low-energy precision experiments like the electric dipole moments in the leptoquark model, see Ref.\,\cite{Fuyuto:2018scm} for example.

Upper bounds on the other NSI parameters from the beta and the inverse beta decay turn out to impose no constraints on this simplified leptoquark model for $\lambda_{L,R}=1$, { as one can see from figure\,\ref{LQModelMassScanMore}.}}

\sampsa{We conclude our investigation with the lower-right panel of figure\,\ref{LQModelMassScan}, where the overall sensitivity to the leptoquark model is shown both in the LBL experiments T2K and NO$\nu$A, and the reactor experiments Daya Bay, Double Chooz and RENO. The two sensitivity plots show the statistical significance at which the model can be excluded in the experiments as function of $\lambda^2/M^2$, whereby $\lambda\equiv\lambda_L = \lambda_R = 1$ is assumed for simplicity. The 95\% CL is showcased with the horizontal dashed line. At this statistical limit, the LBL experiments perform better at $\lambda^2 / M^2 \lesssim$ 4.4$\times$10$^{-3}$~1/TeV$^2$, whereas the reactor experiment set the bound at $\lambda^2 / M^2 \lesssim$ 6.5$\times$10$^{-3}$~1/TeV$^2$. The corresponding masses are approximately 15~TeV and 12.5~TeV respectively. The results from the LBL and reactor experiments are therefore comparable, which are also stronger than the constraints from ATLAS and CMS.}

%%%%%%%%%%%%%%%%%%
\subsection{Constraints on $\lambda_{L,R}$ with fixed $M$}
%%%%%%%%%%%%%%%%%%

\begin{figure}[t]
\centering{
  \begin{adjustbox}{max width = \textwidth}
\begin{tabular}{cccc}
\includegraphics[scale=0.25]{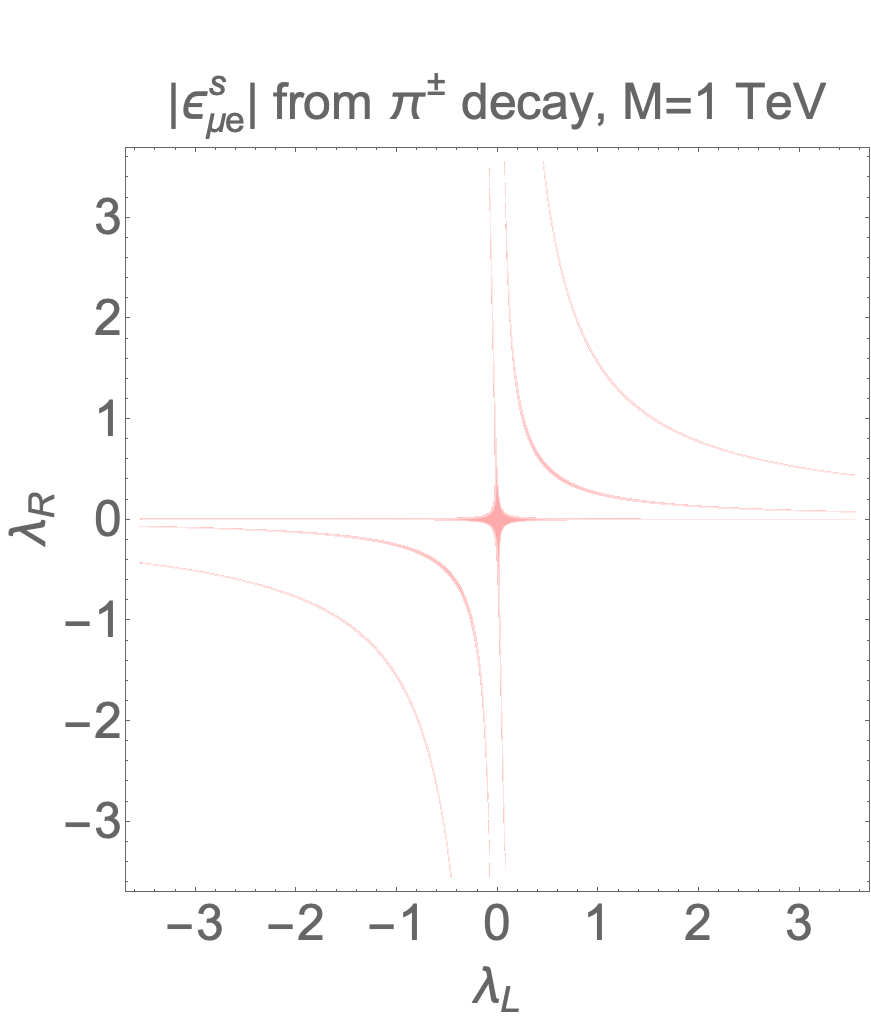} ~& ~ \includegraphics[scale=0.25]{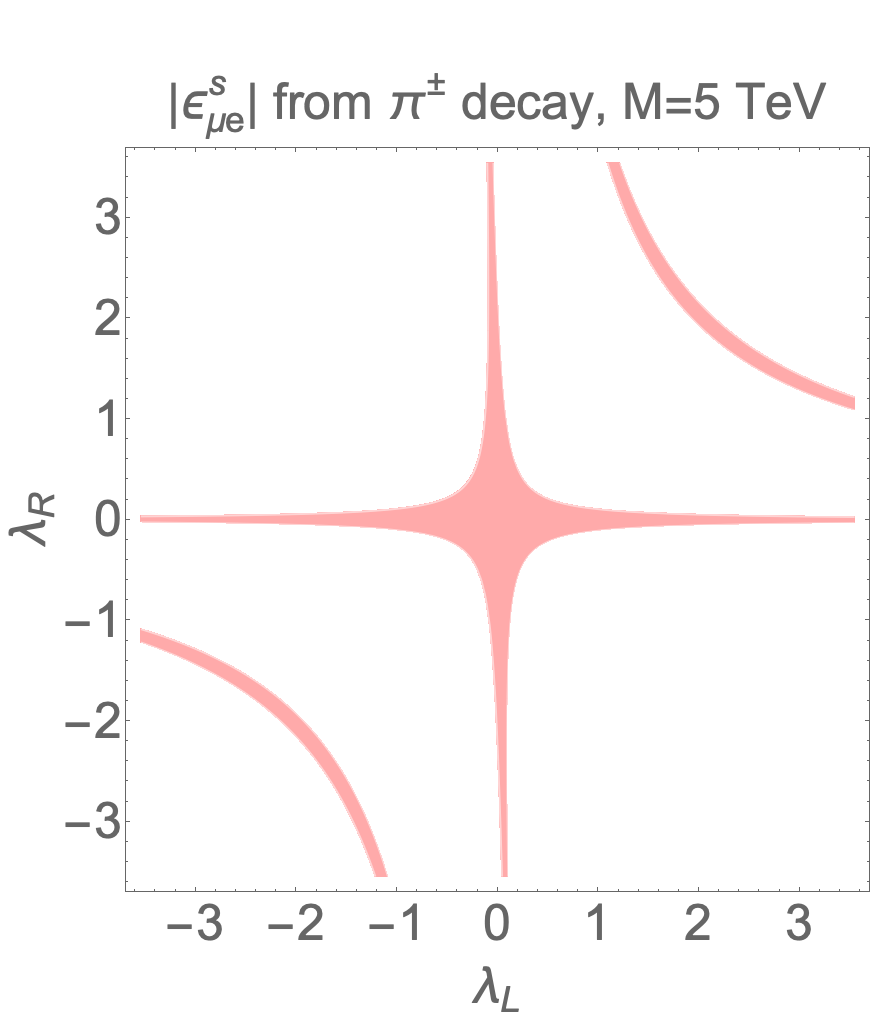}~&~\includegraphics[scale=0.25]{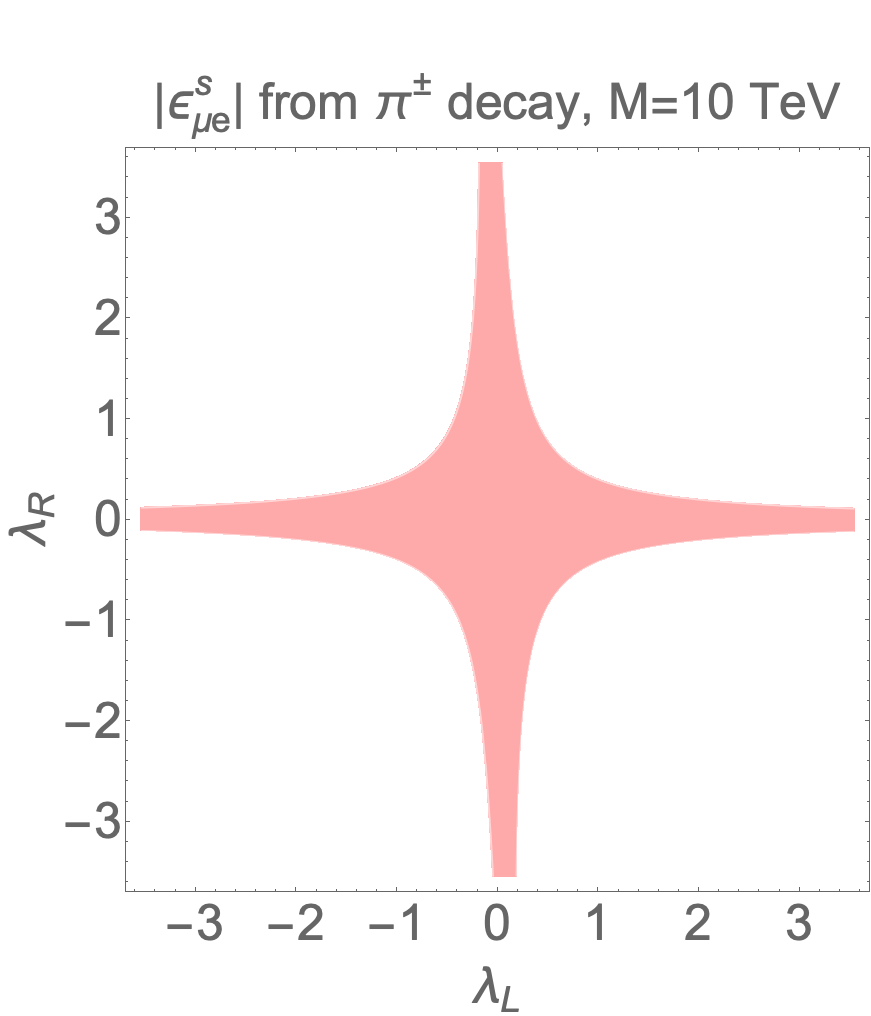}~&~\includegraphics[scale=0.25]{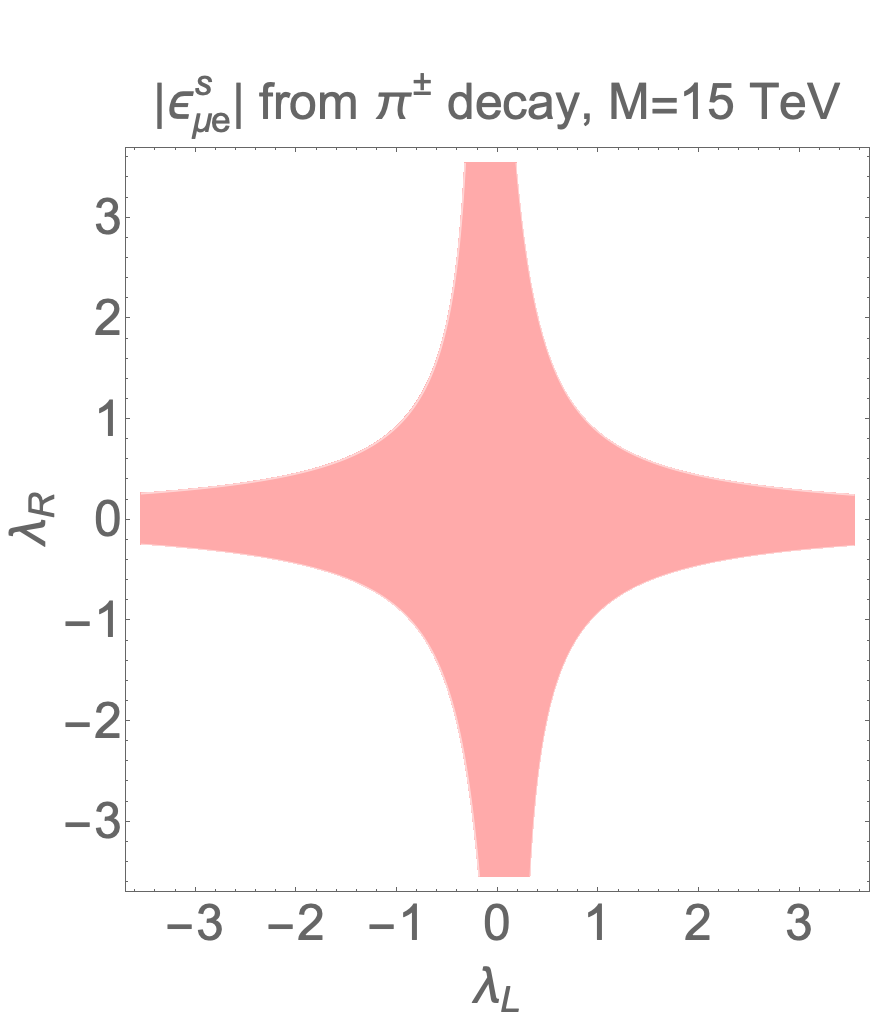}\\
\includegraphics[scale=0.25]{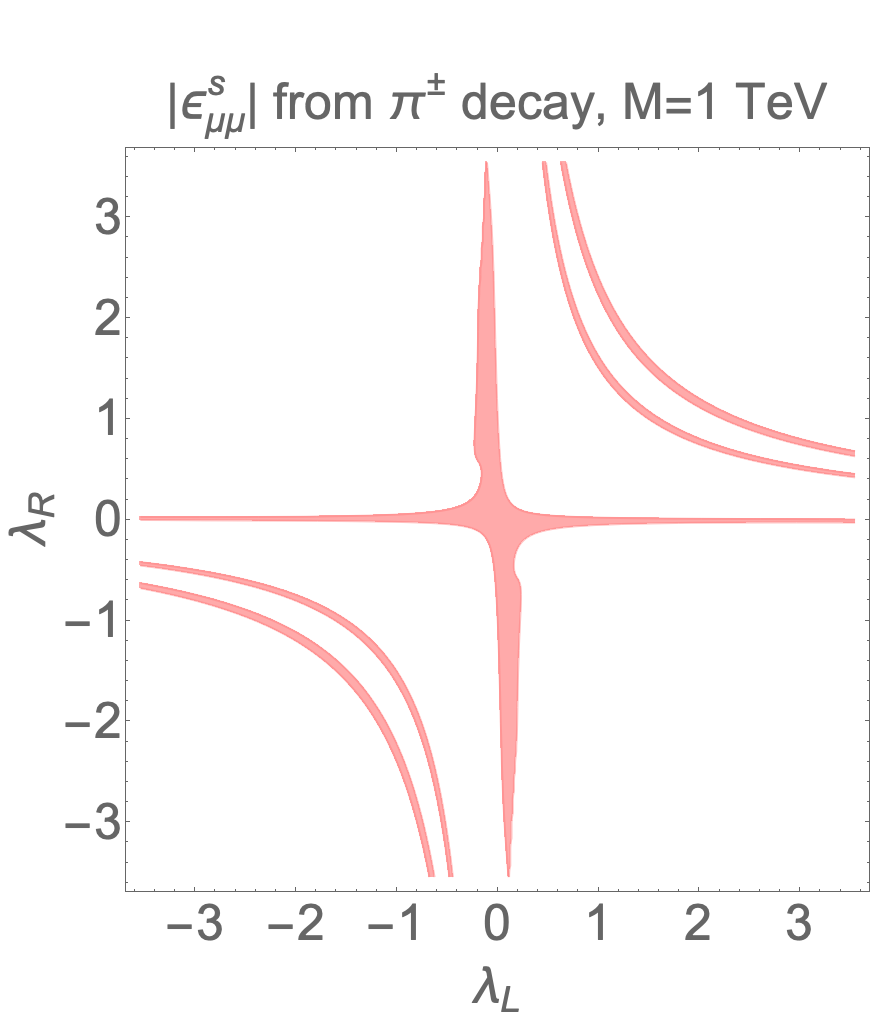} ~& ~ \includegraphics[scale=0.25]{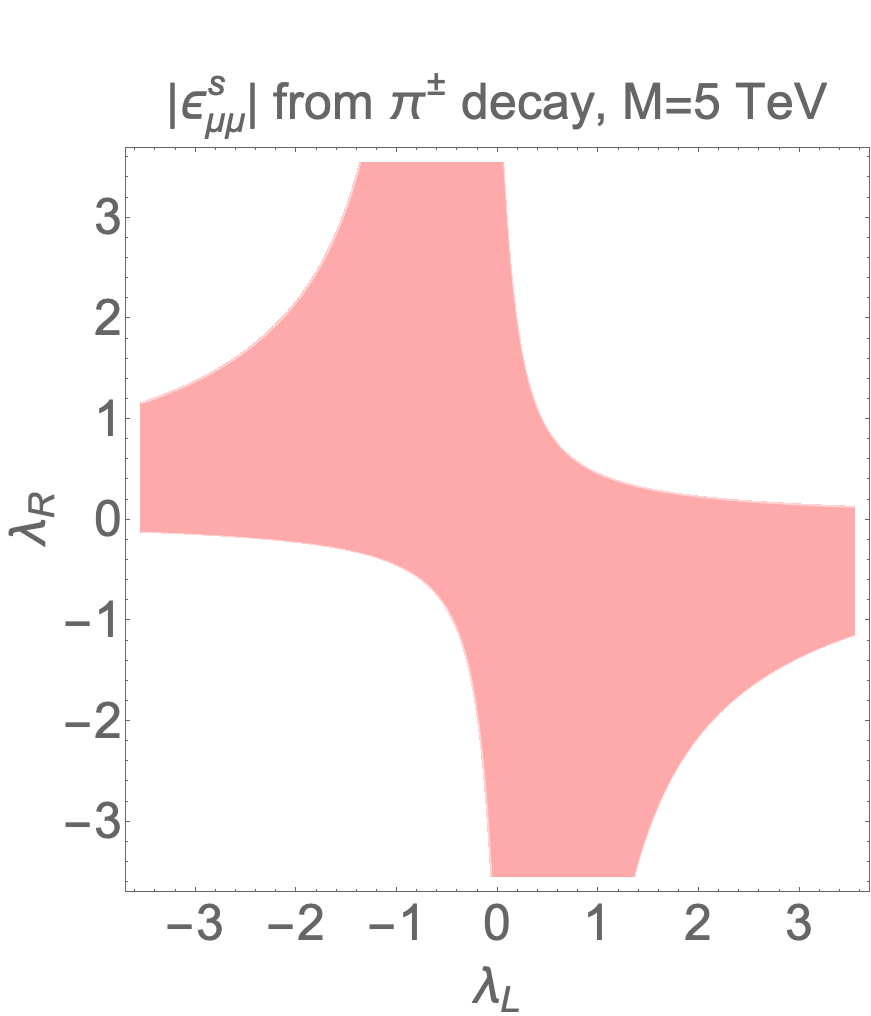}~&~\includegraphics[scale=0.25]{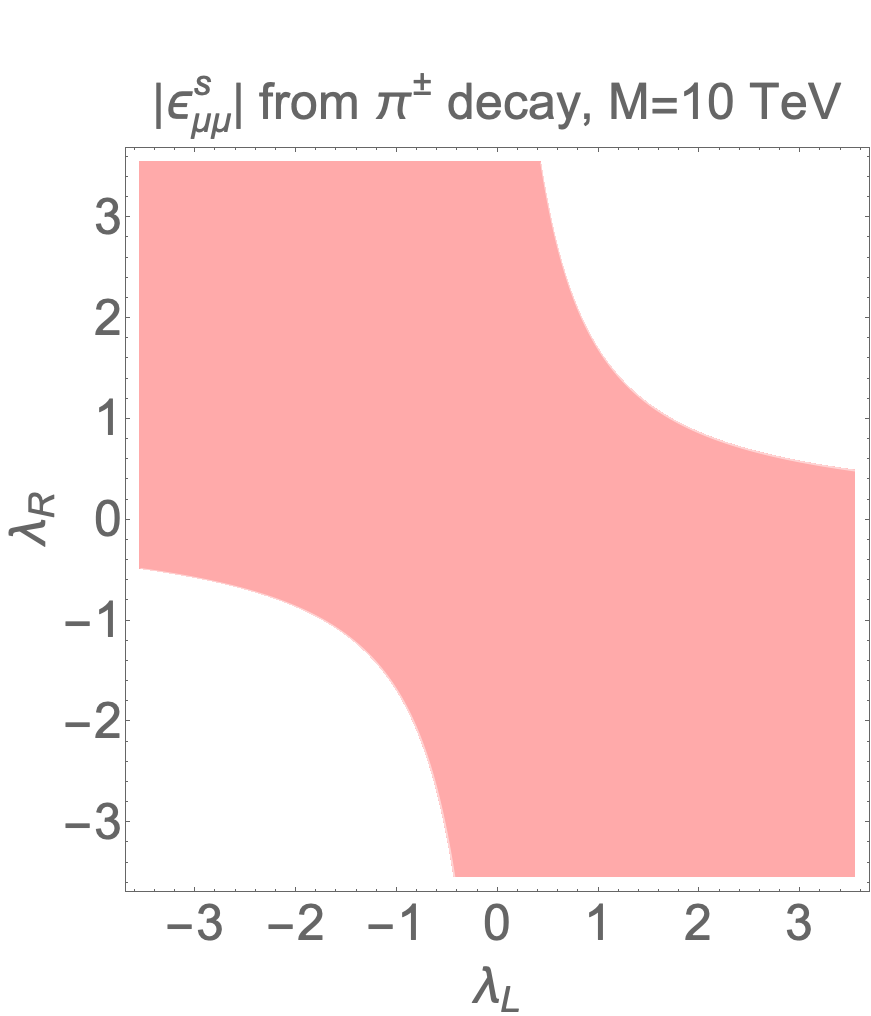}~&~\includegraphics[scale=0.25]{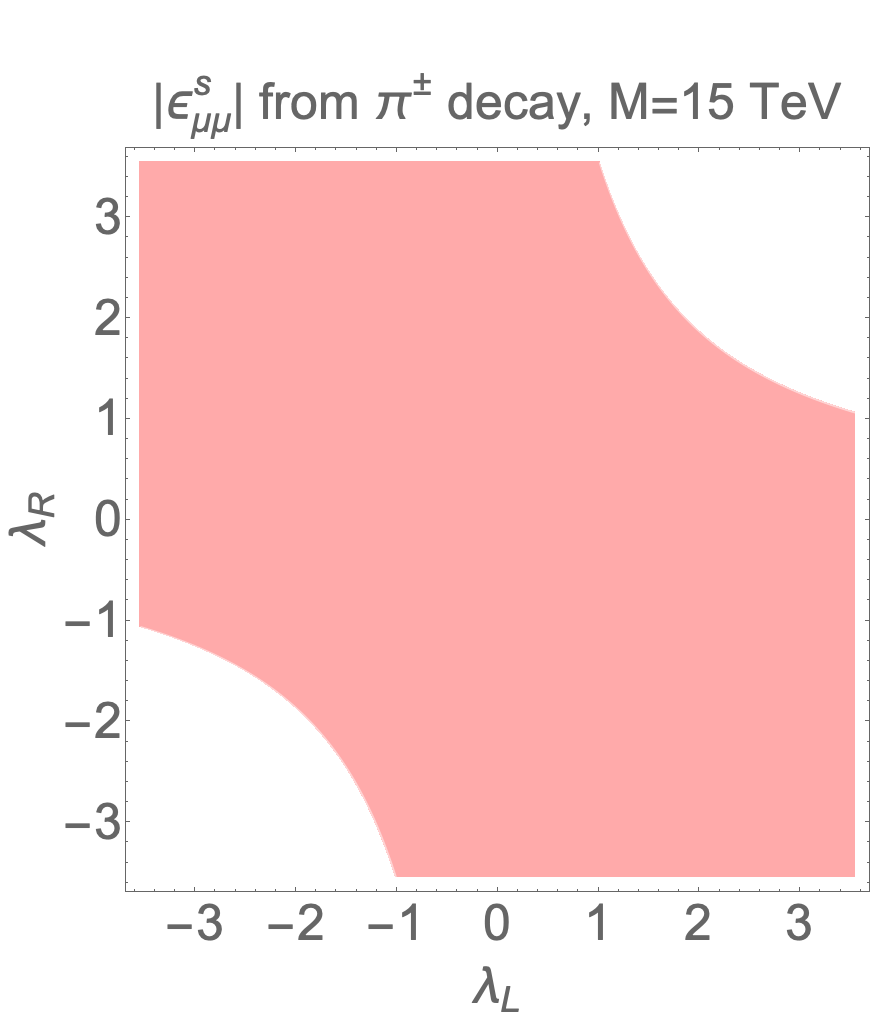}
\\
\includegraphics[scale=0.25]{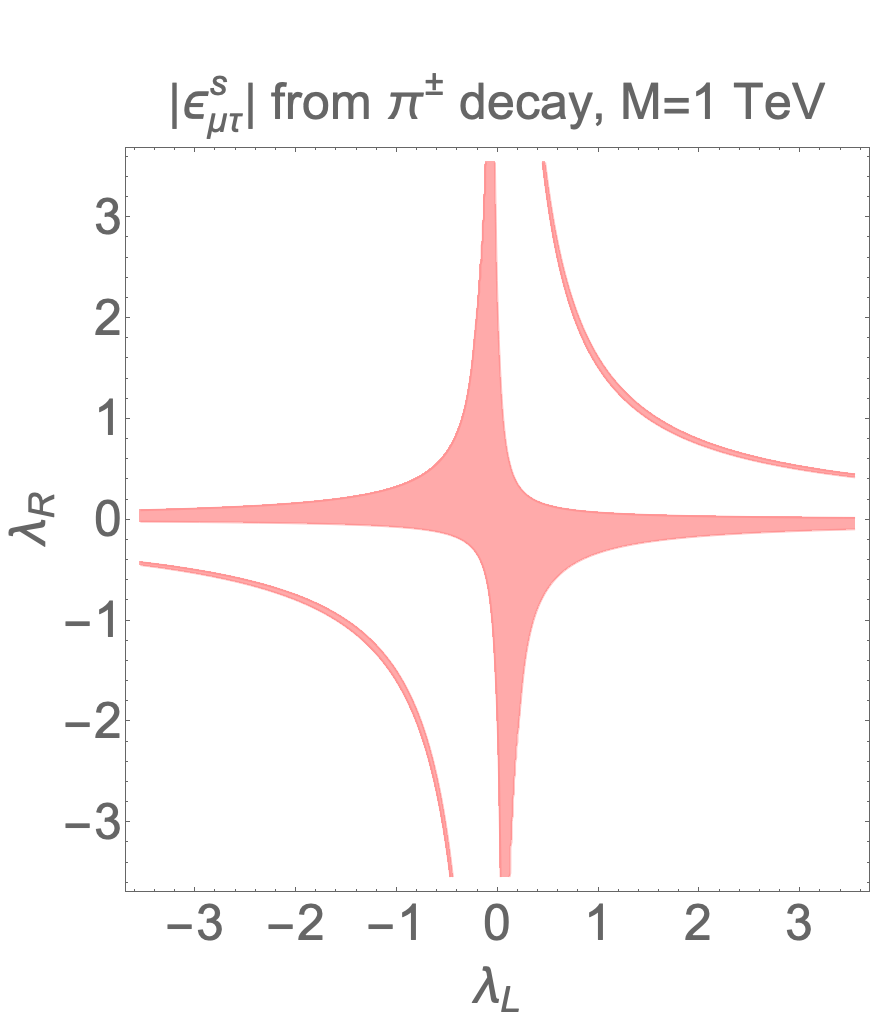} ~& ~ \includegraphics[scale=0.25]{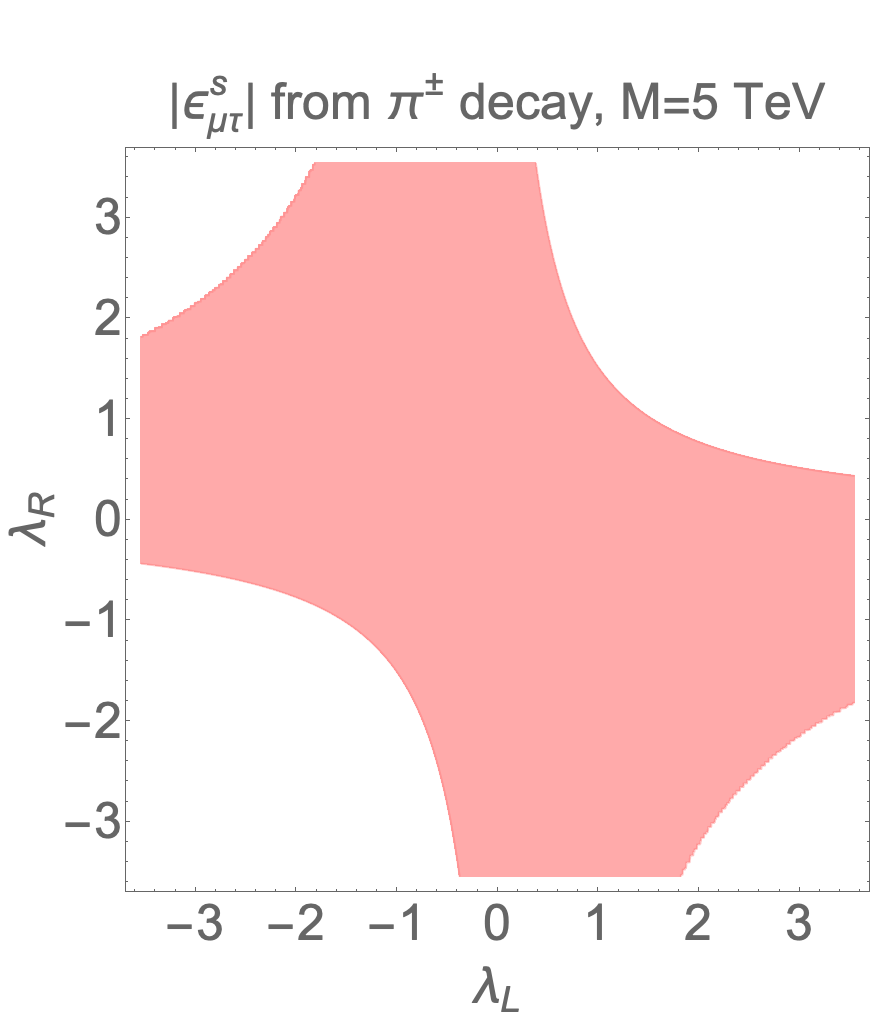}~&~\includegraphics[scale=0.25]{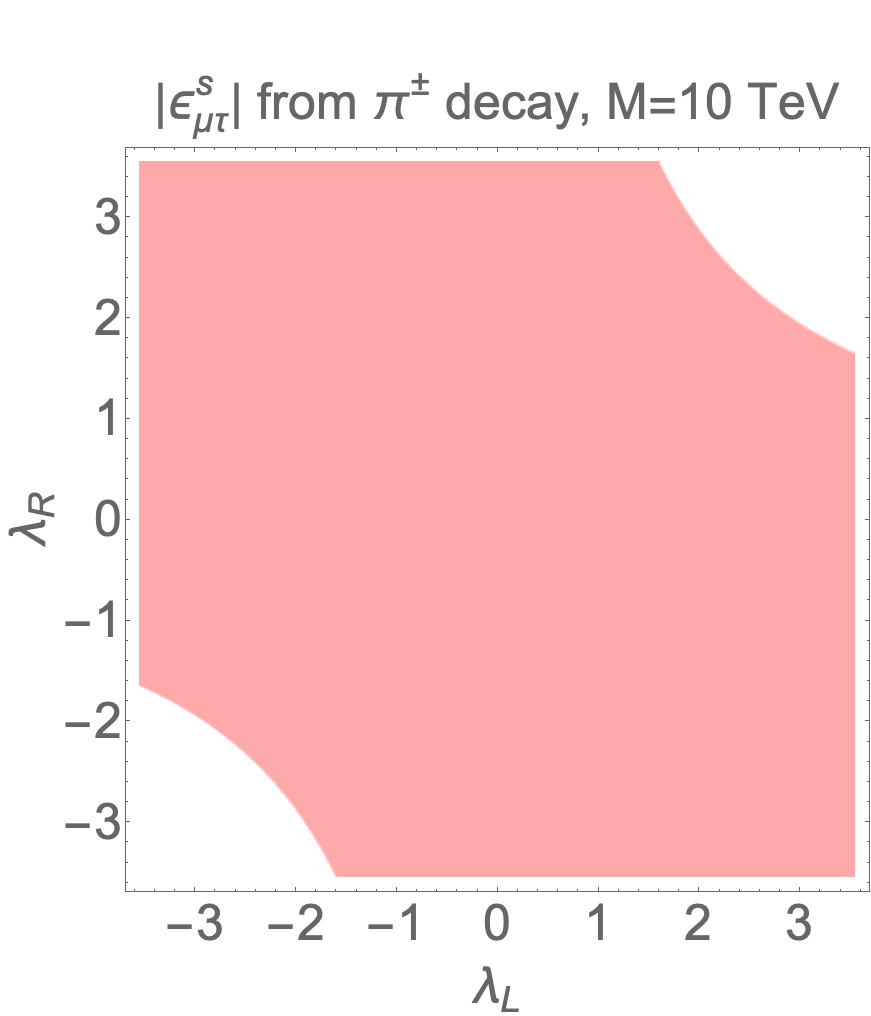}~&~\includegraphics[scale=0.25]{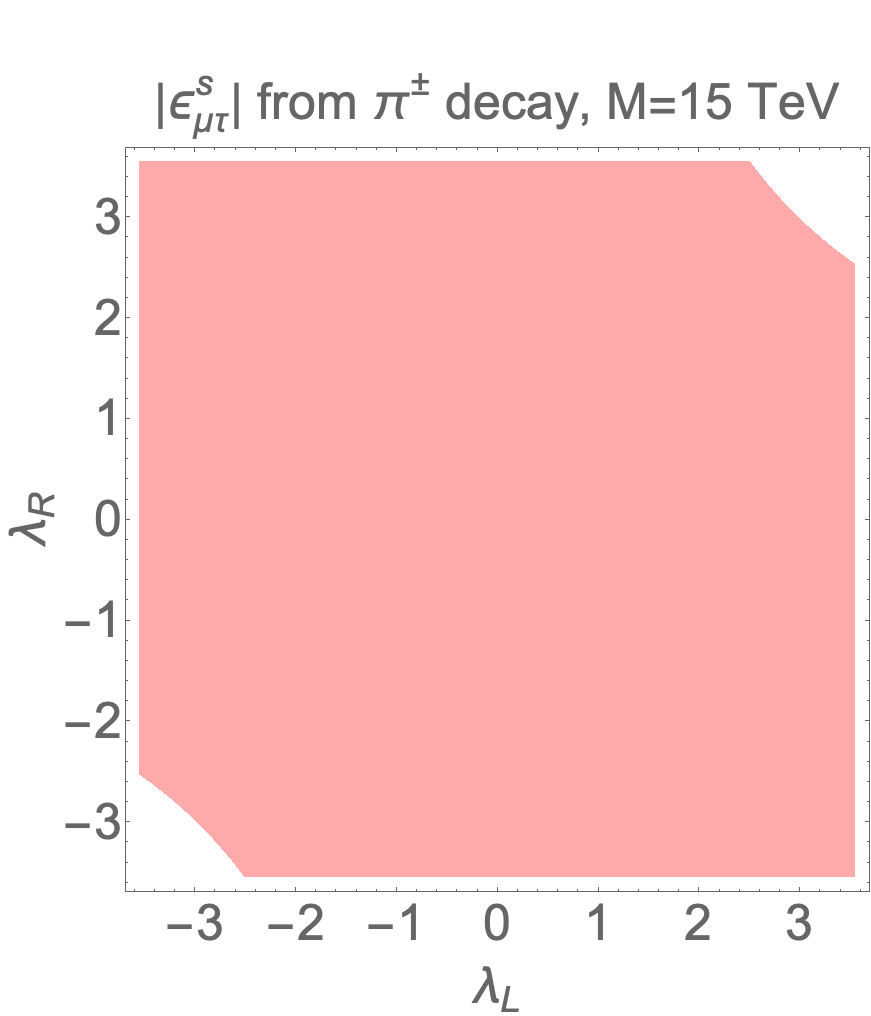}
\end{tabular}
  \end{adjustbox}}
\caption{Constraints on the simplified leptoquark model parameter space from current constraints on $|\epsilon^s_{\mu\ell}|$ ($\ell=e,\mu,\tau$) summarized in table\,\ref{NSIbounds}. The pink region(s) in each figure is (are) still allowed at 95\% CL, obtained by fixing $\lambda_{L}=\lambda_{R}=1$. See the main text for details.}\label{NSIRegionLQ}
\end{figure}

\iffalse
\begin{figure}[t]
\centering{
  \begin{adjustbox}{max width = \textwidth}
\begin{tabular}{ccc}
\includegraphics[scale=0.3333]{plots/CoupScan1TeV1} ~& ~ \includegraphics[scale=0.3333]{plots/CoupScan1TeV2}~&~\includegraphics[scale=0.3333]{plots/CoupScan1TeV3}\\
\includegraphics[scale=0.3333]{plots/CoupScan5TeV1} ~& ~ \includegraphics[scale=0.3333]{plots/CoupScan5TeV2}~&~\includegraphics[scale=0.3333]{plots/CoupScan5TeV3}\\
\includegraphics[scale=0.3333]{plots/CoupScan10TeV1} ~& ~ \includegraphics[scale=0.3333]{plots/CoupScan10TeV2}~&~\includegraphics[scale=0.3333]{plots/CoupScan10TeV3}\\
\includegraphics[scale=0.3333]{plots/CoupScan15TeV1} ~& ~ \includegraphics[scale=0.3333]{plots/CoupScan15TeV2}~&~\includegraphics[scale=0.3333]{plots/CoupScan15TeV3}
\end{tabular}
  \end{adjustbox}}
\caption{Constraints on the simplified leptoquark model parameter space from current constraints on $|\epsilon^s_{\mu\ell}|$ ($\ell=e,\mu,\tau$) summarized in table\,\ref{NSIbounds}. The pink region(s) in each figure is (are) still allowed at 95\% CL, obtained by fixing $\lambda_{L}=\lambda_{R}=1$. See the main text for details.}\label{NSIRegionLQ}
\end{figure}
\fi

\yong{On the other hand, since $\lambda_{L,R}$ are particularly interesting for collider studies, we also present our results with fixed leptoquark masses in figure\,\ref{NSIRegionLQ} upon scanning over $\lambda_{L,R}$. The pink region(s) in each plot of figure\,\ref{NSIRegionLQ} is (are) still allowed at 95\% CL, obtained from a comparison between theoretical prediction from this simplified model and the experimental upper bounds on the neutrino NSI parameters summarized in table\,\ref{NSIbounds}. 

Similar to what we observe in section\,\ref{subsec:LQMass}, the current constraint on $|\epsilon_{\mu e}^{s}|$ from pion decay dominates all the other NSI parameters in table\,\ref{NSIbounds}, which can be seem from the first column of figure\,\ref{NSIRegionLQ}. Particularly, in the case where $M=1$\,TeV, the current constraint on $|\epsilon_{\mu e}^{s}|$ excludes almost the entire parameter space of this simplified model unless $\lambda_{L}$ and/or $\lambda_{R}$ are very tiny to suppress contributions from the UV physics. This point will become clearer if one matches the simplified UV model to the LEFT directly at the UV scale $\mu=M$ without taking into account any subleading running effects, which gives\footnote{Note that we assume both $\lambda_L$ and $\lambda_R$ are real for simplicity.}}

\begin{align}
\epsilon_{\mu \beta}^s (\mu=M) &=  \left(1 + \frac{m_\pi^2}{m_\mu (m_u + m_d)} \right) \frac{v^2 \lambda_L \lambda_R}{4M^2} \left(\frac{\sum\limits_{x=d,s,b}V^*_{ux}}{V_{ud}}\right)^*,\label{WilsonLQModel1}\\
\epsilon_{e \beta}^s (\mu=M) &=  \left(1 + \frac{g_T}{g_A} \frac{m_e}{f_T (E_\nu)} \right) \frac{v^2 \lambda_L \lambda_R}{4M^2} \left(\frac{\sum\limits_{x=d,s,b}V^*_{ux}}{V_{ud}}\right)^*,\label{WilsonLQModel2}\\
\epsilon_{\beta e}^d (\mu=M) &=  \left(1 + \frac{m_e}{E_\nu - \Delta} \frac{g_S - 3 g_A g_T}{1 + 3 g_A^2} \right) \frac{v^2 \lambda_L \lambda_R}{4M^2}\left( \frac{\sum\limits_{x=d,s,b}V^*_{ux}}{V_{ud}} \right)^*.
\end{align}

\begin{figure}[t]
\centering{
  \begin{adjustbox}{max width = \textwidth}
\begin{tabular}{cccc}
\includegraphics[scale=0.25]{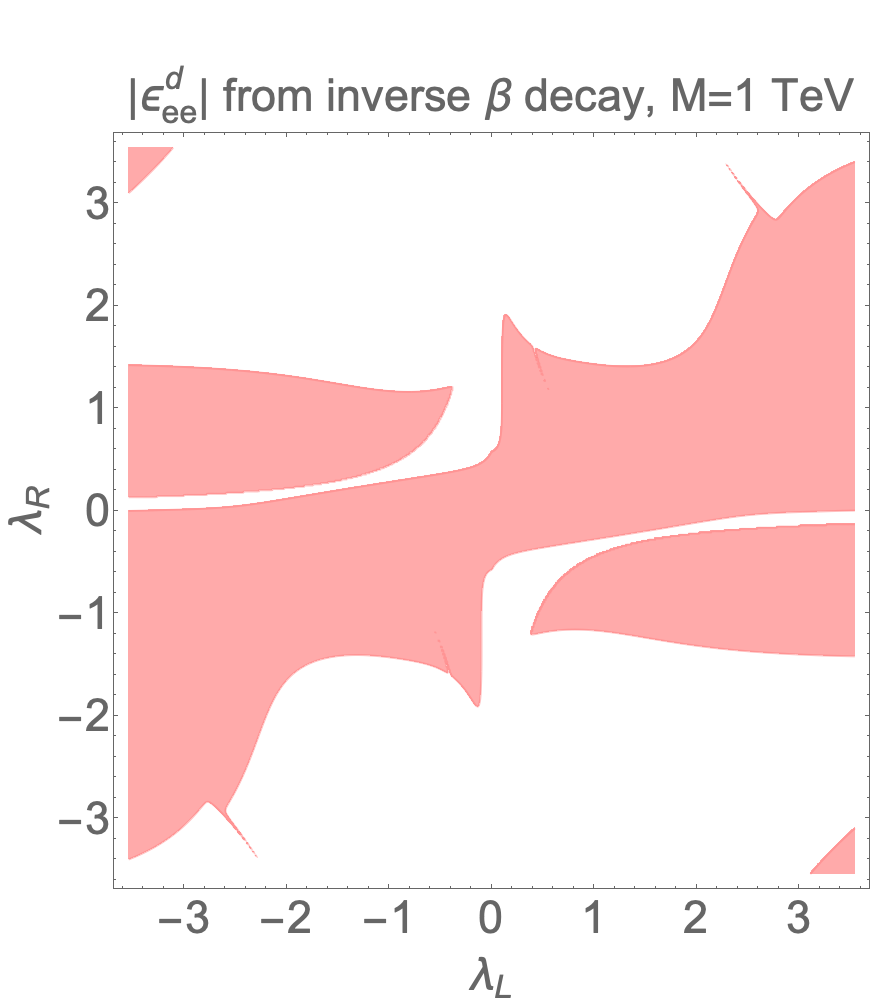} ~& ~ \includegraphics[scale=0.25]{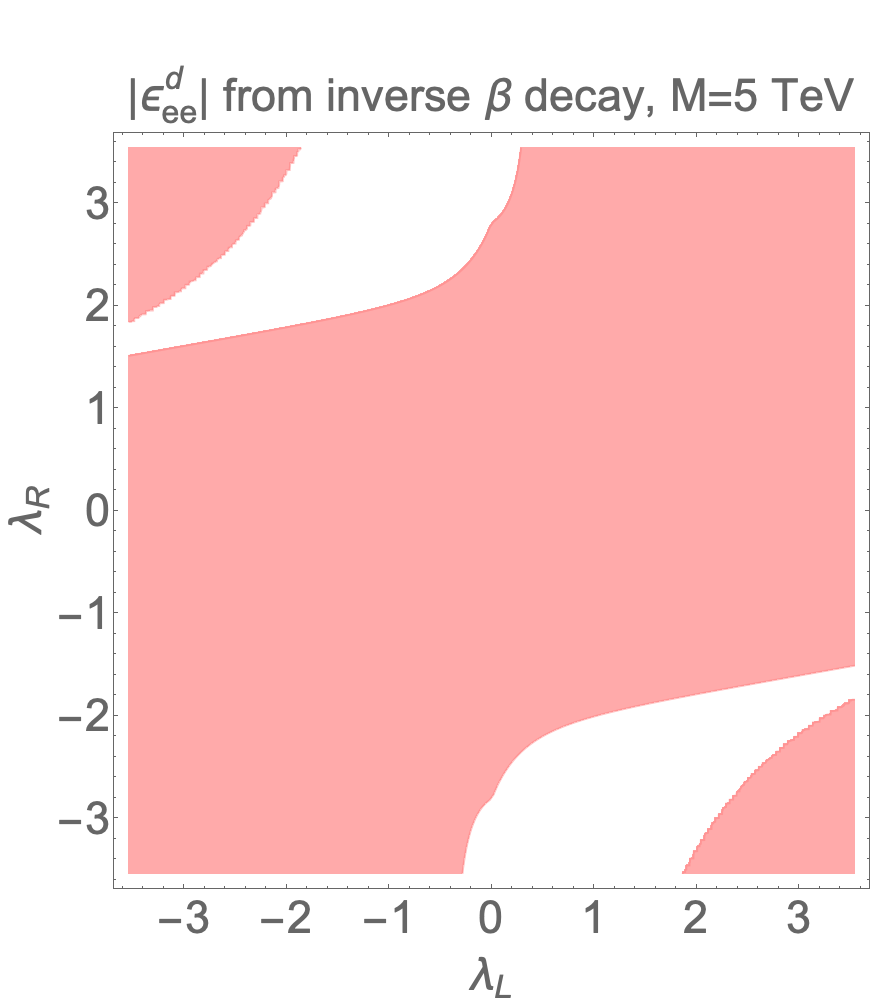}~&~\includegraphics[scale=0.25]{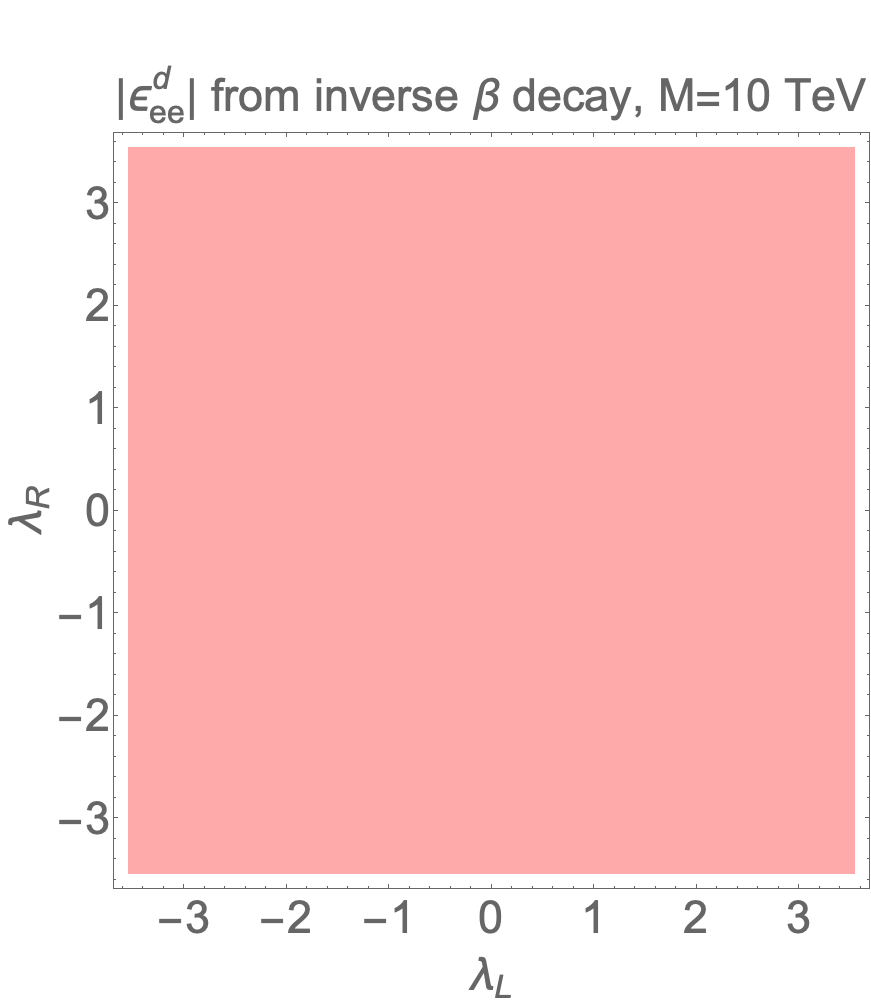}~&~\includegraphics[scale=0.25]{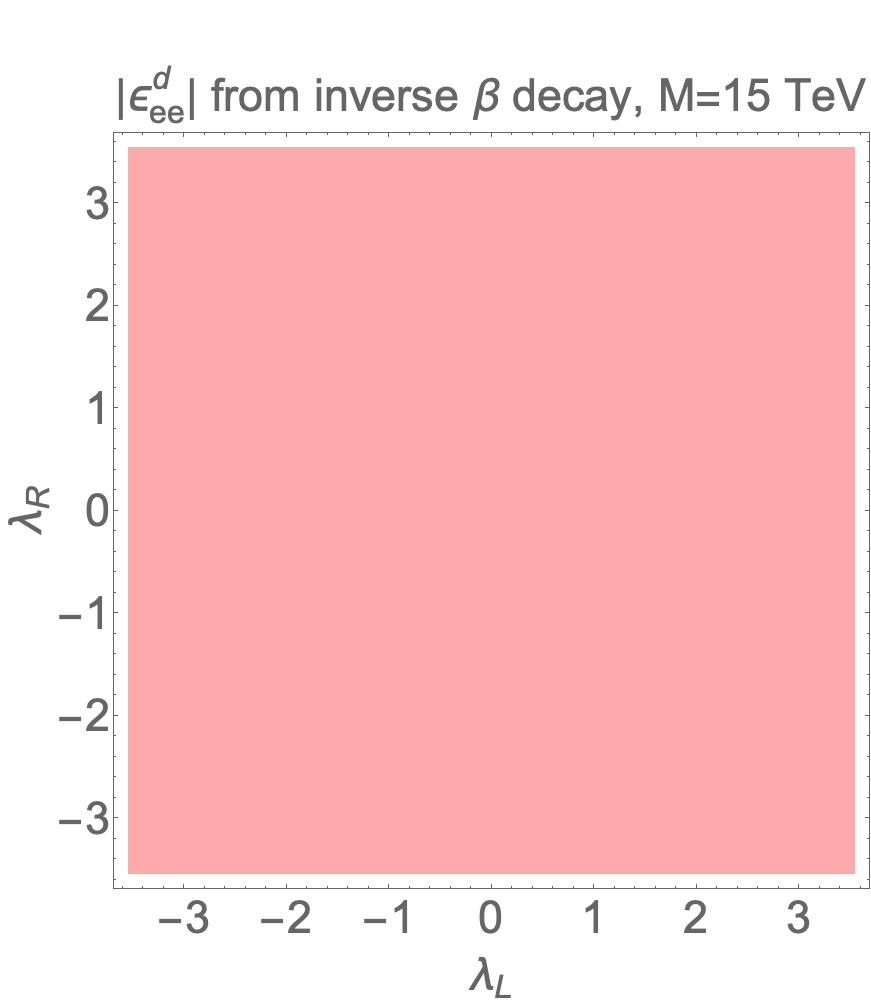}\\
\includegraphics[scale=0.25]{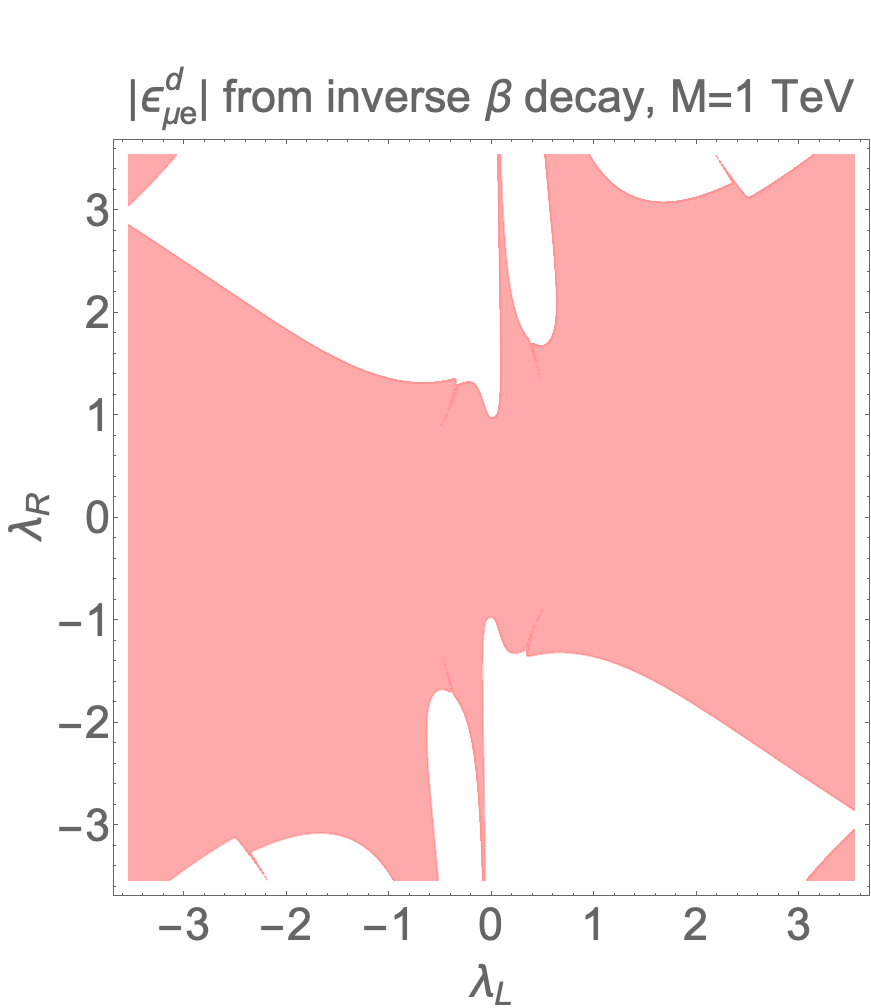} ~& ~ \includegraphics[scale=0.25]{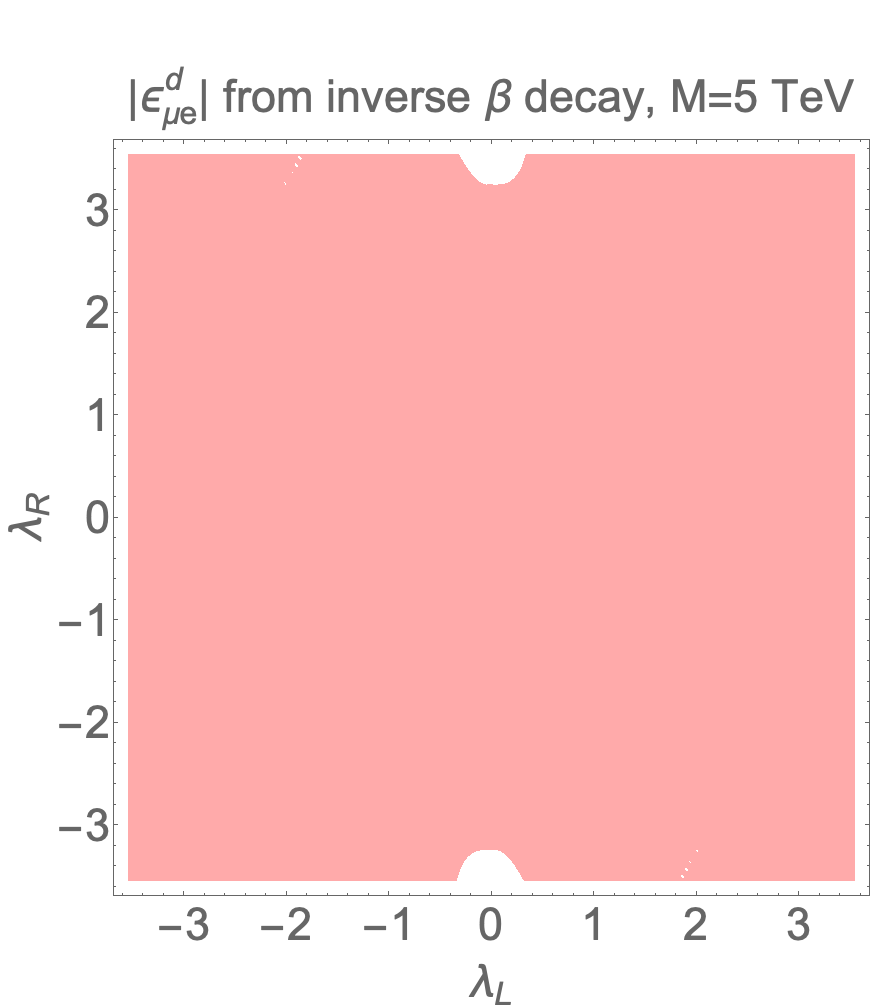}~&~\includegraphics[scale=0.25]{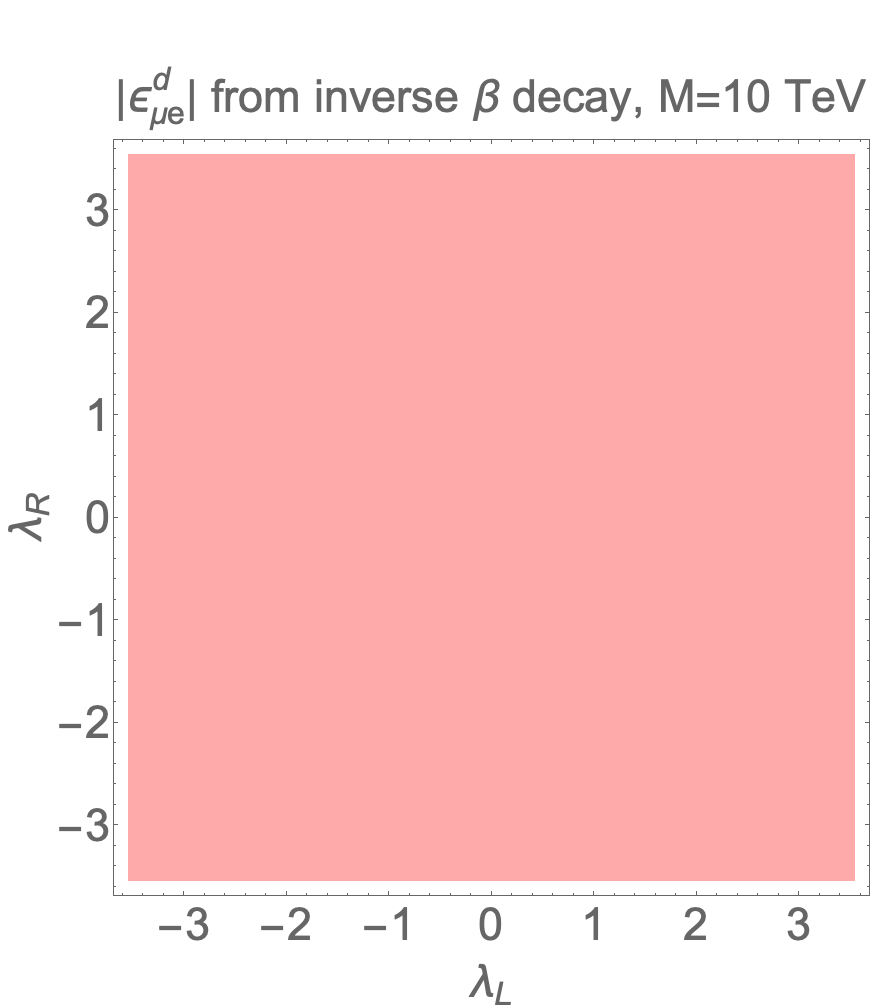}~&~\includegraphics[scale=0.25]{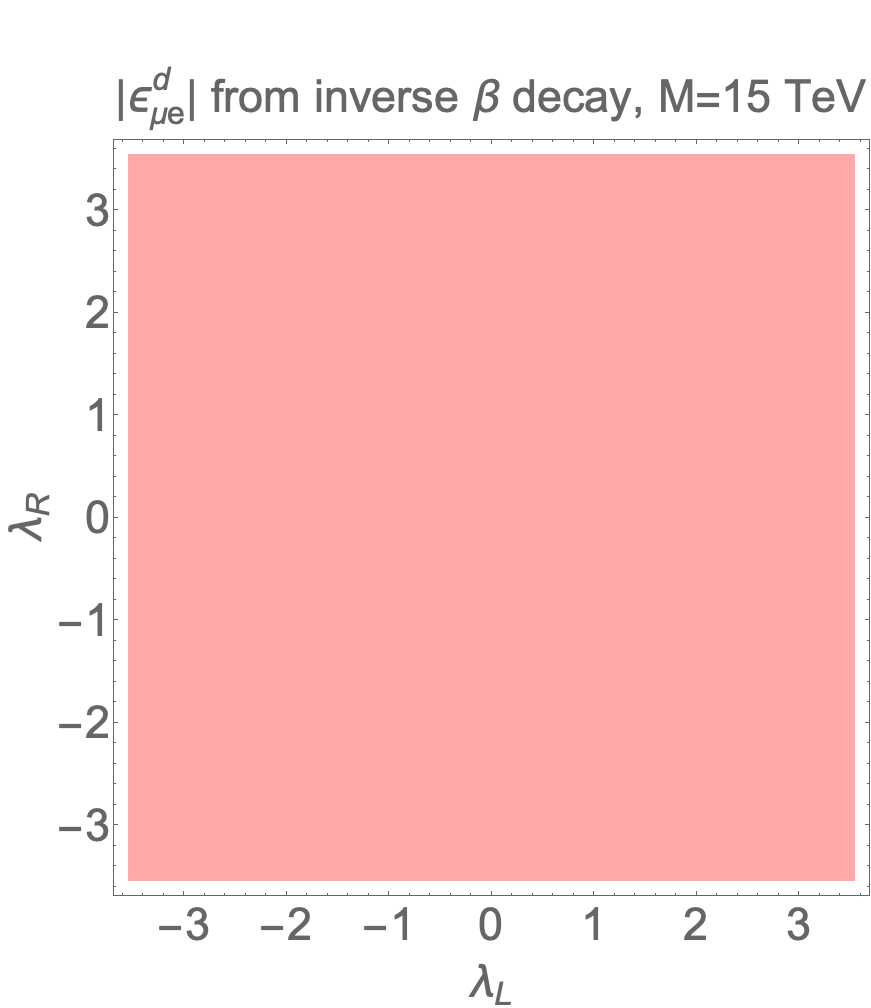}
\\
\includegraphics[scale=0.25]{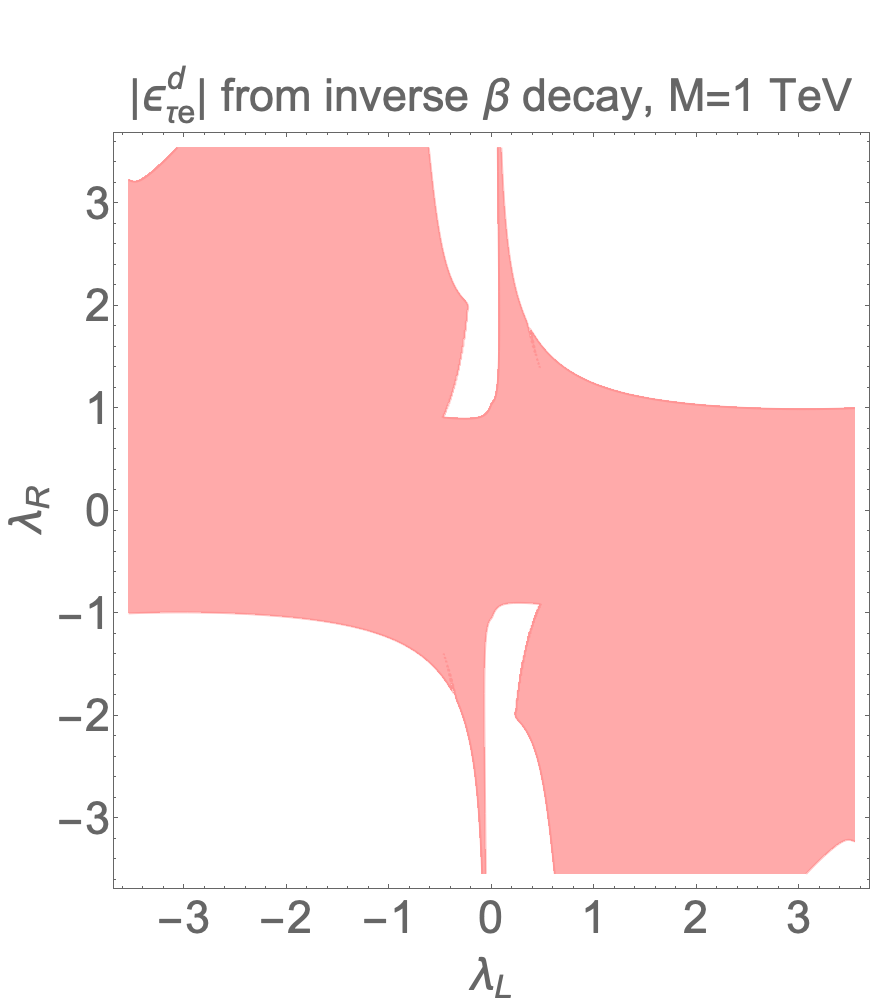} ~& ~ \includegraphics[scale=0.25]{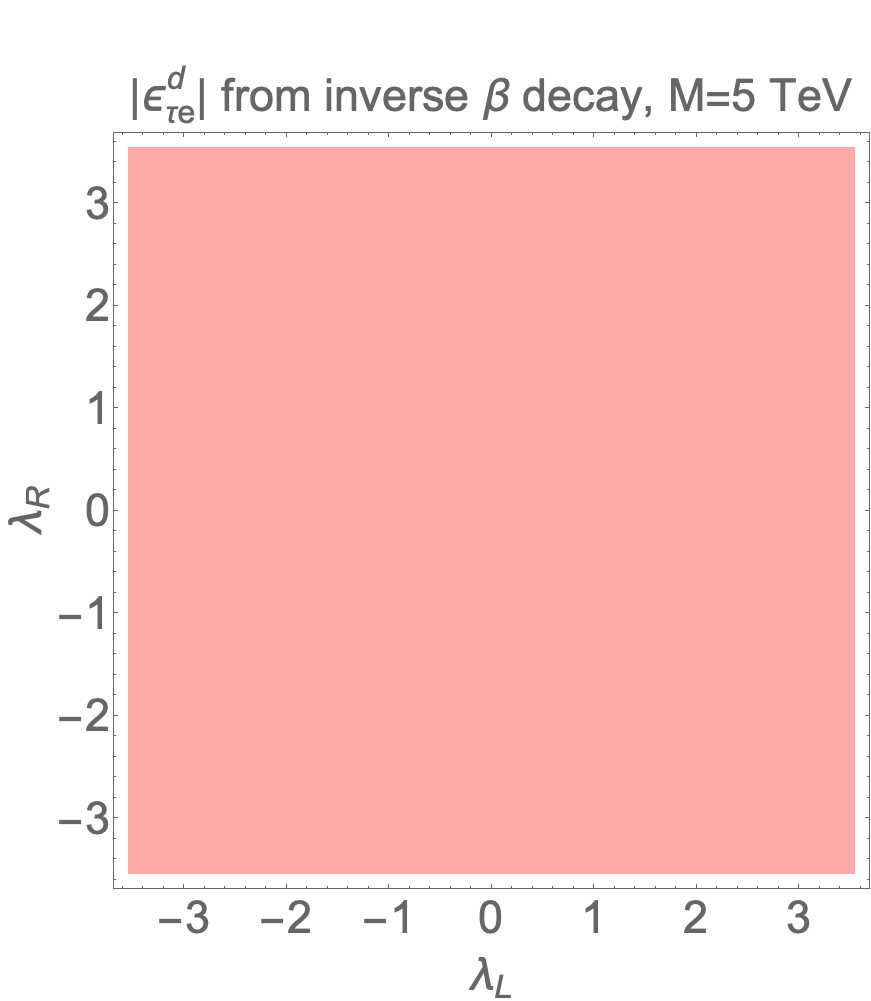}~&~\includegraphics[scale=0.25]{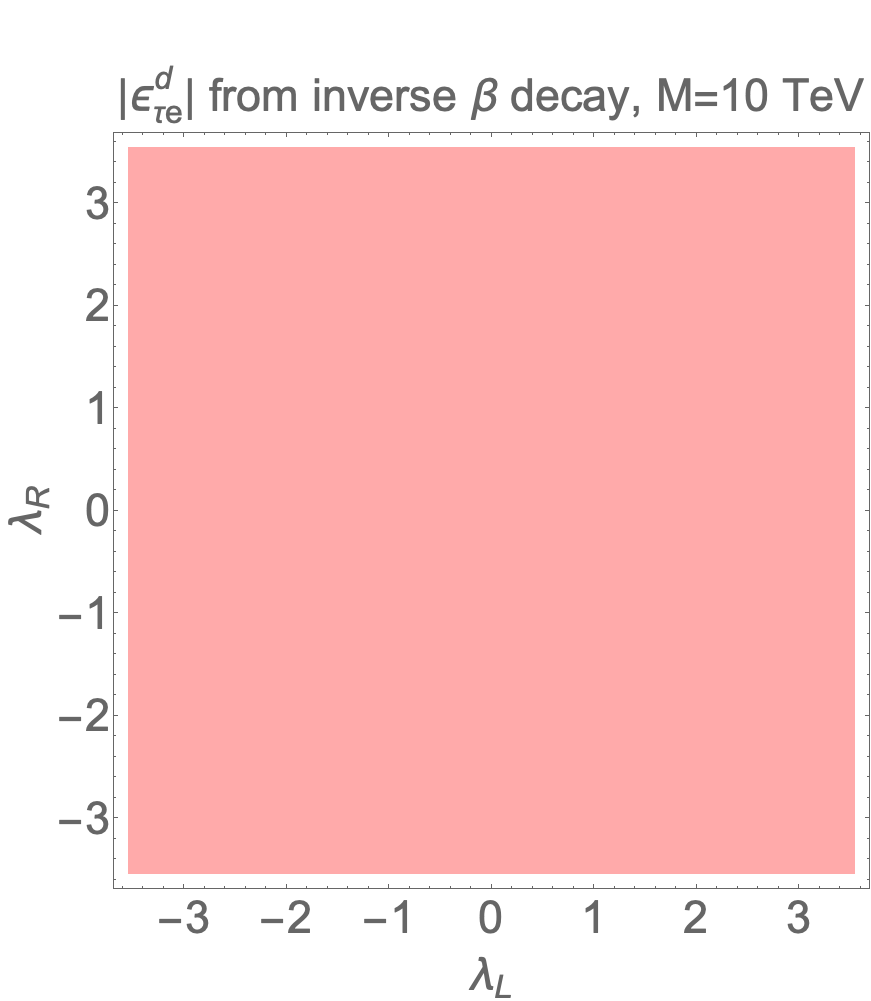}~&~\includegraphics[scale=0.25]{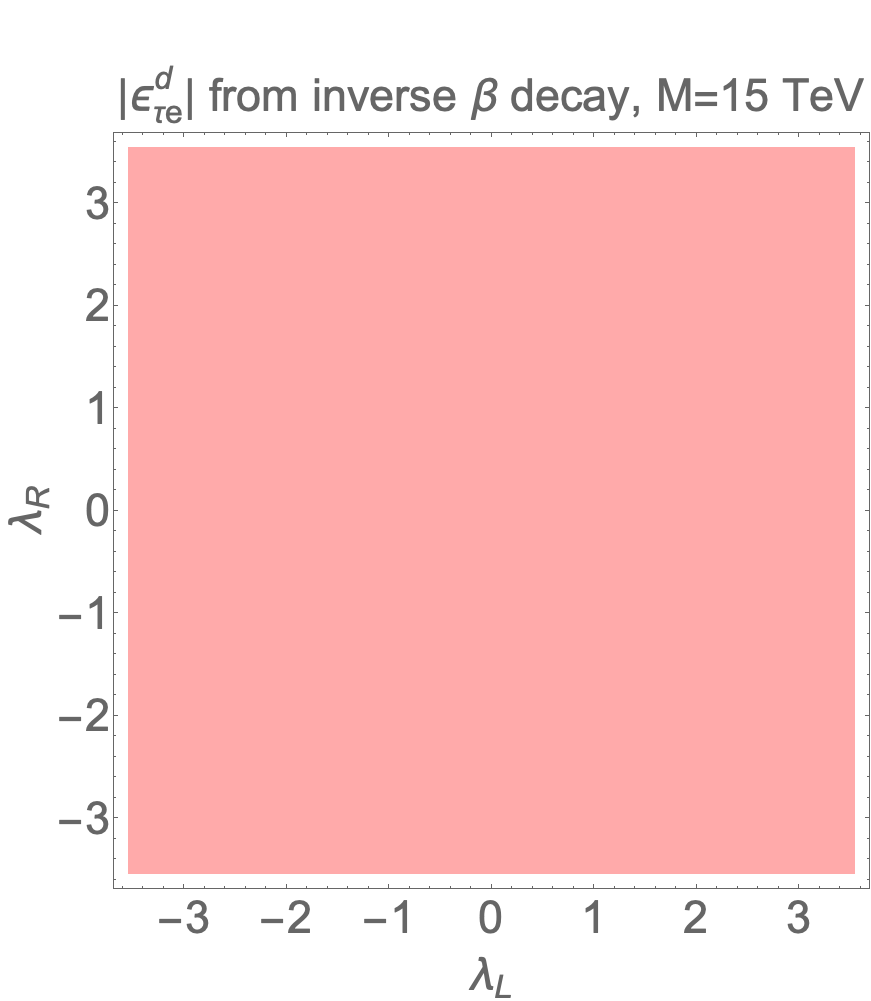}
\end{tabular}
  \end{adjustbox}}
\caption{Constraints on the simplified leptoquark model parameter space from current constraints on $|\epsilon^d_{\ell e}|$ ($\ell=e,\mu,\tau$) from inverse beta decay summarized in table\,\ref{NSIbounds}. The pink region(s) in each figure is (are) still allowed at 95\% CL, obtained with fixed leptoquark mass $M$.}\label{app:NSIRegionLQPart22}
\end{figure}

\begin{figure}[t]
\centering{
  \begin{adjustbox}{max width = \textwidth}
\begin{tabular}{cccc}
\includegraphics[scale=0.25]{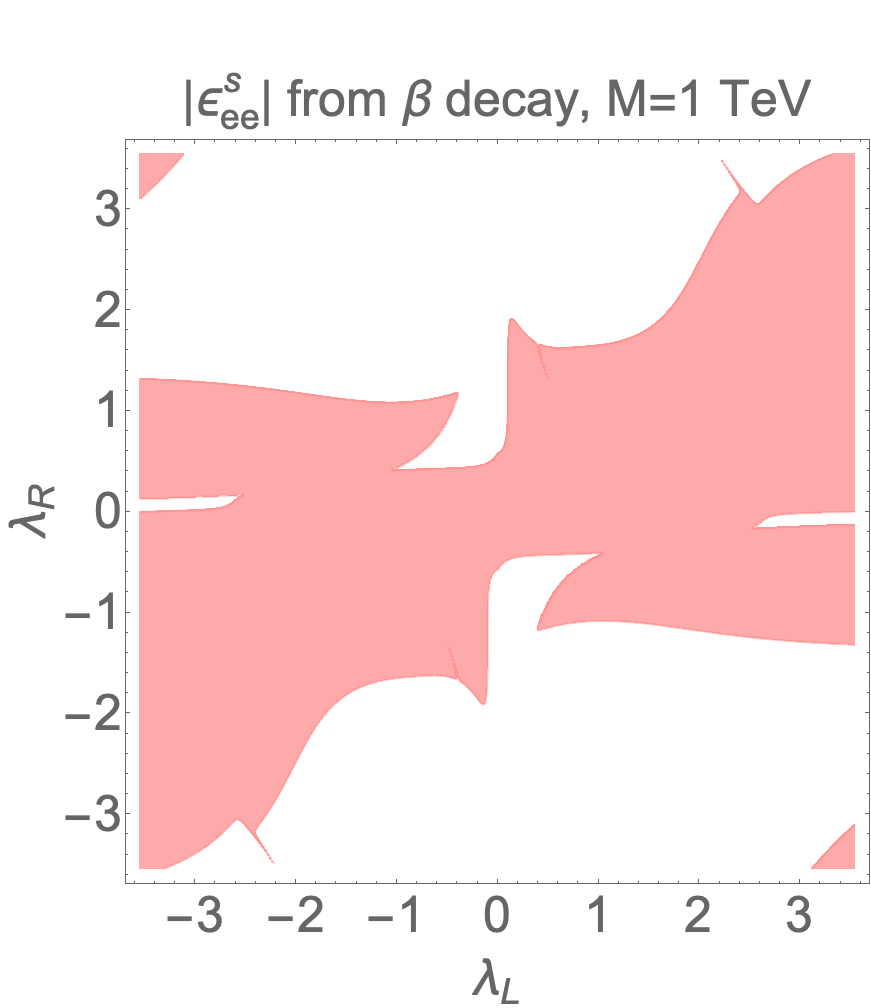} ~& ~ \includegraphics[scale=0.25]{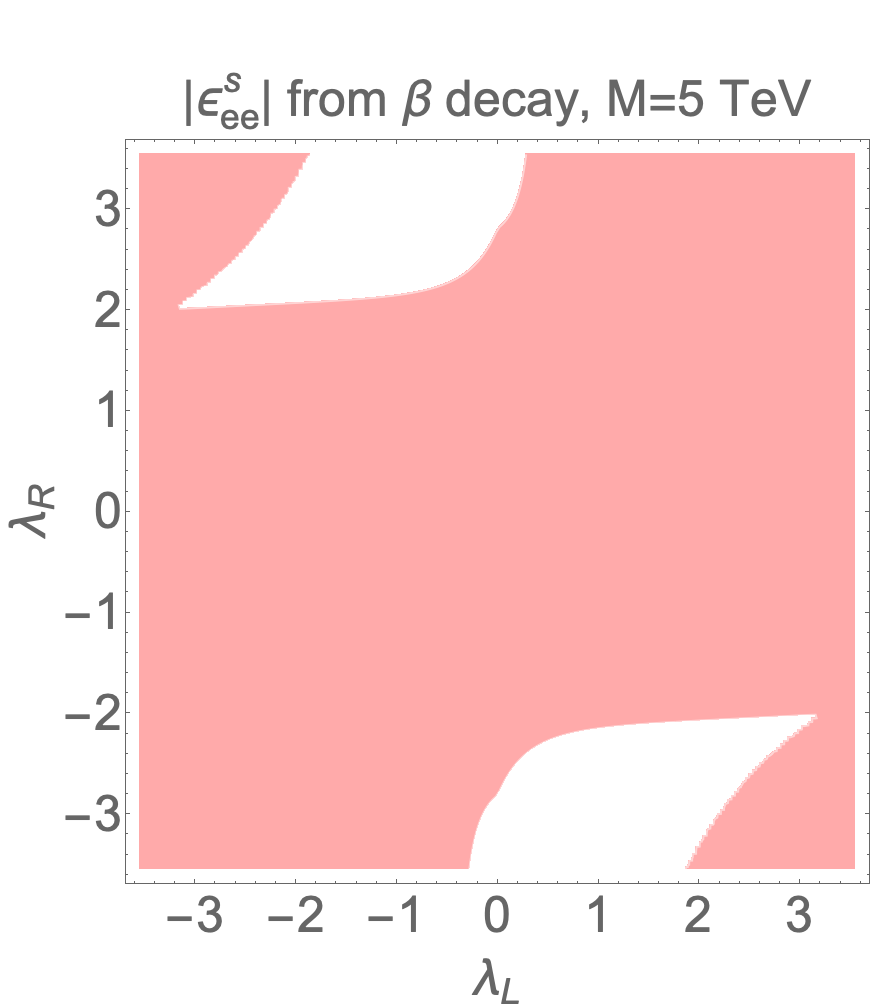}~&~\includegraphics[scale=0.25]{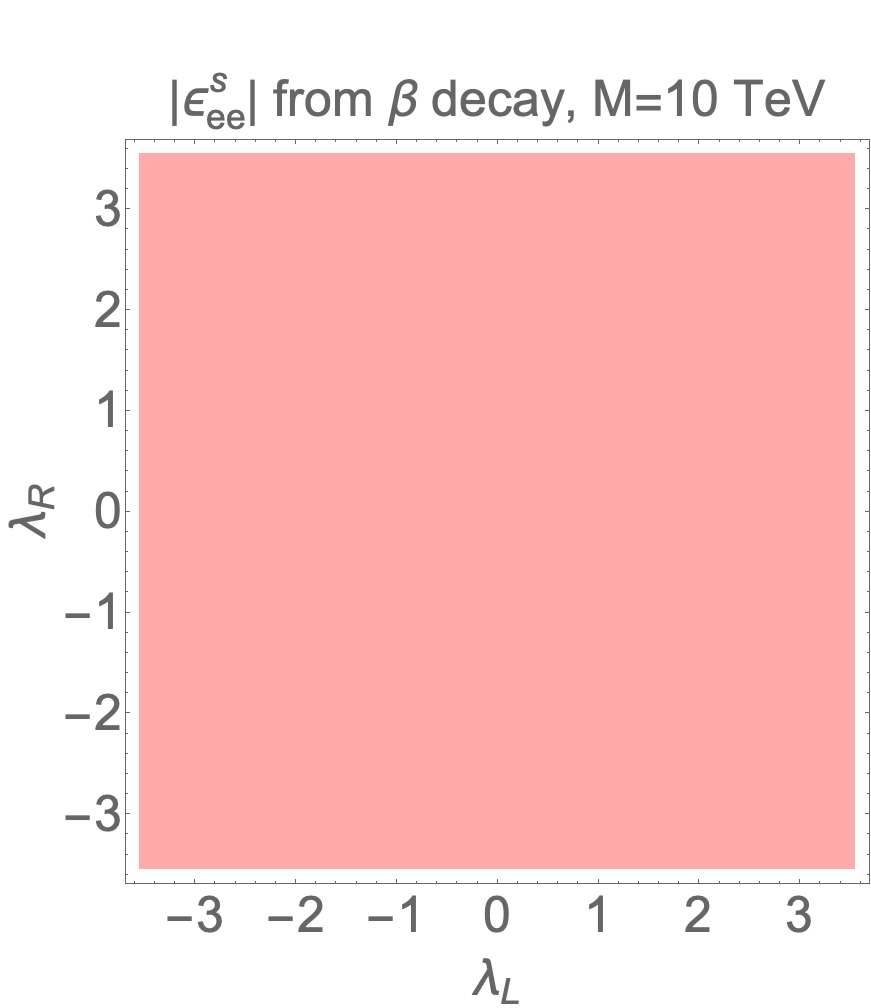}~&~\includegraphics[scale=0.25]{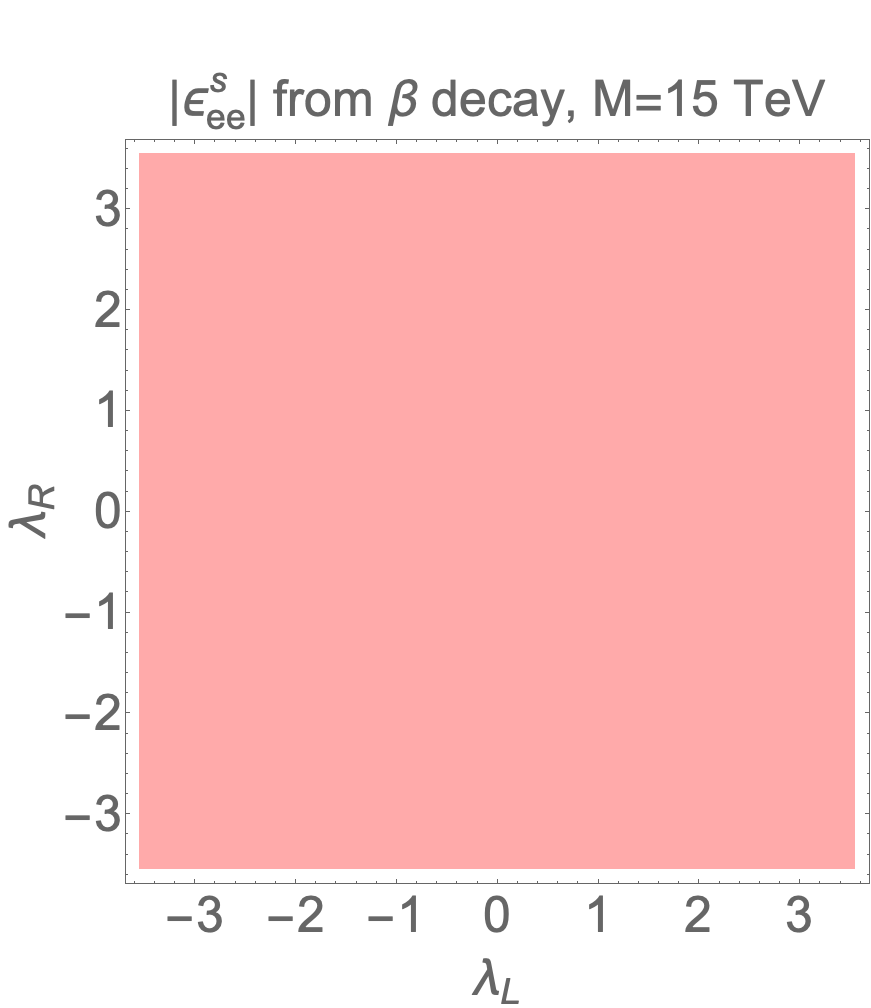}\\
\includegraphics[scale=0.25]{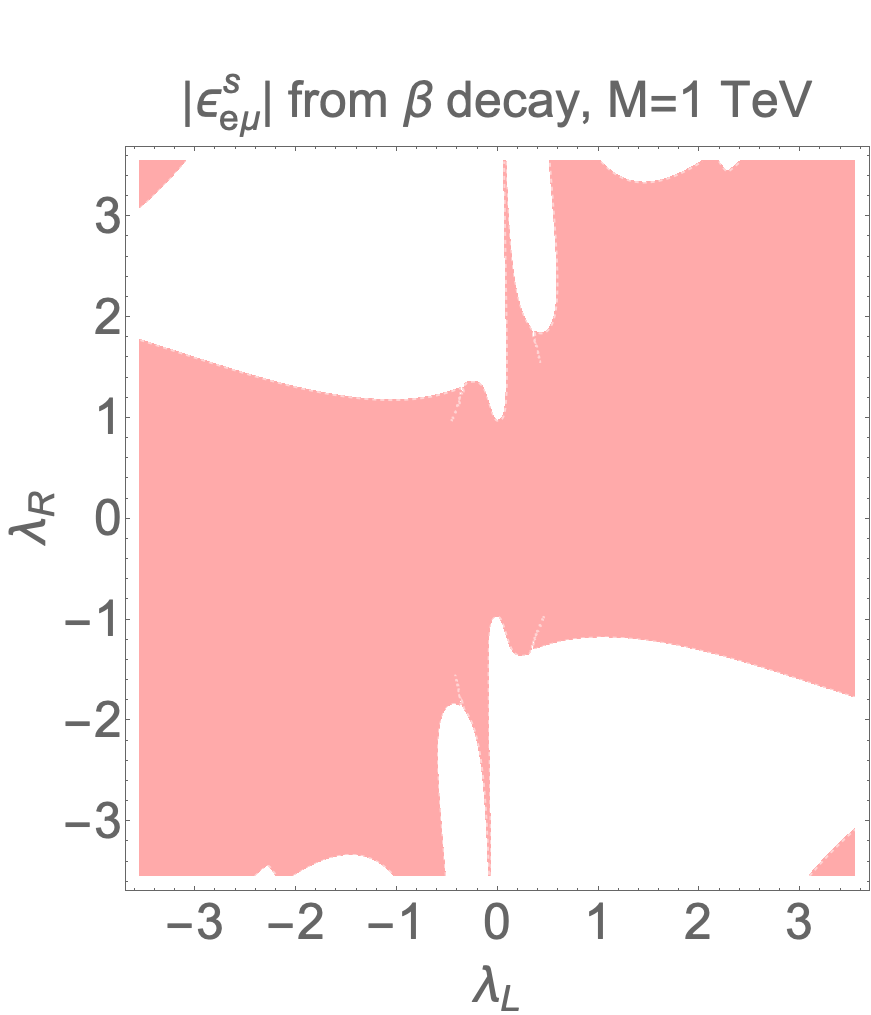} ~& ~ \includegraphics[scale=0.25]{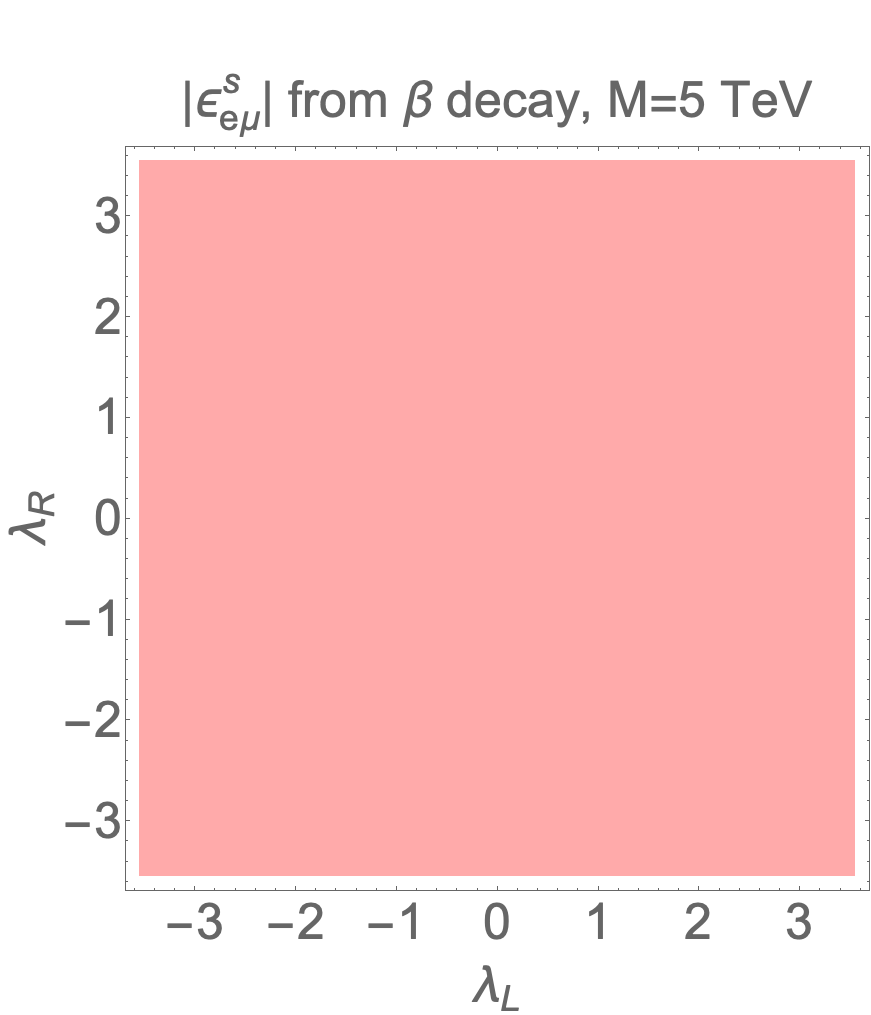}~&~\includegraphics[scale=0.25]{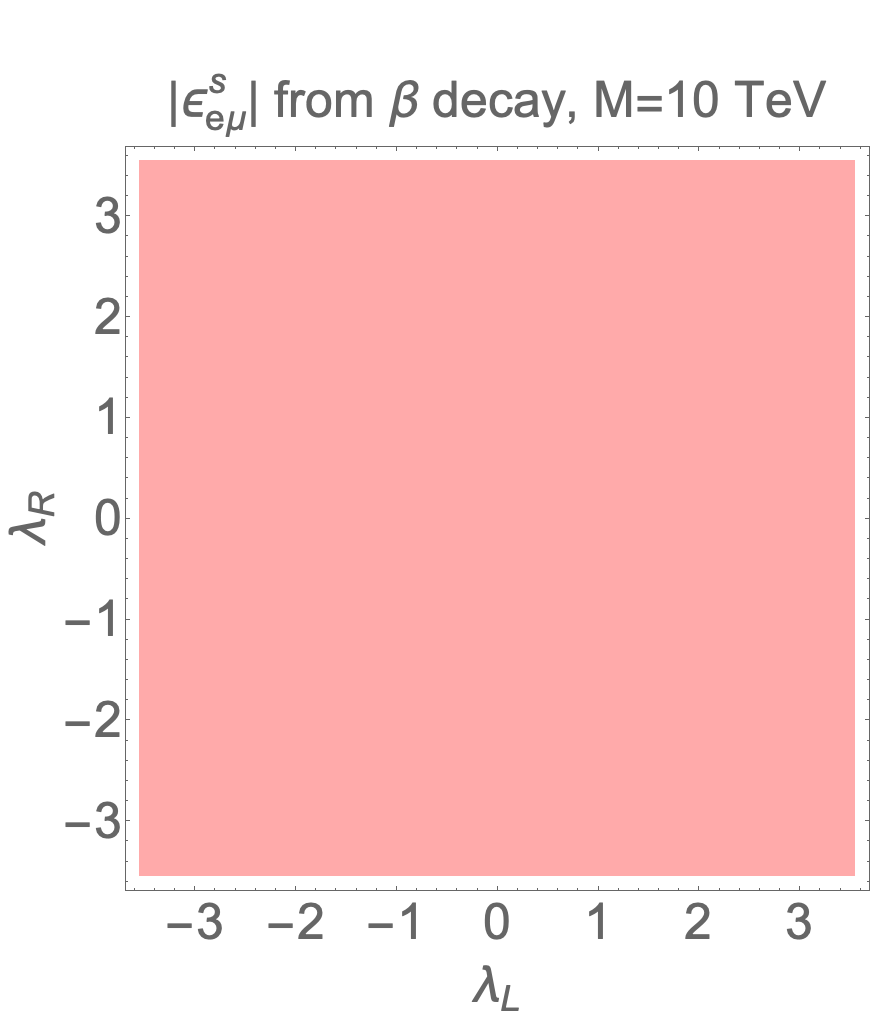}~&~\includegraphics[scale=0.25]{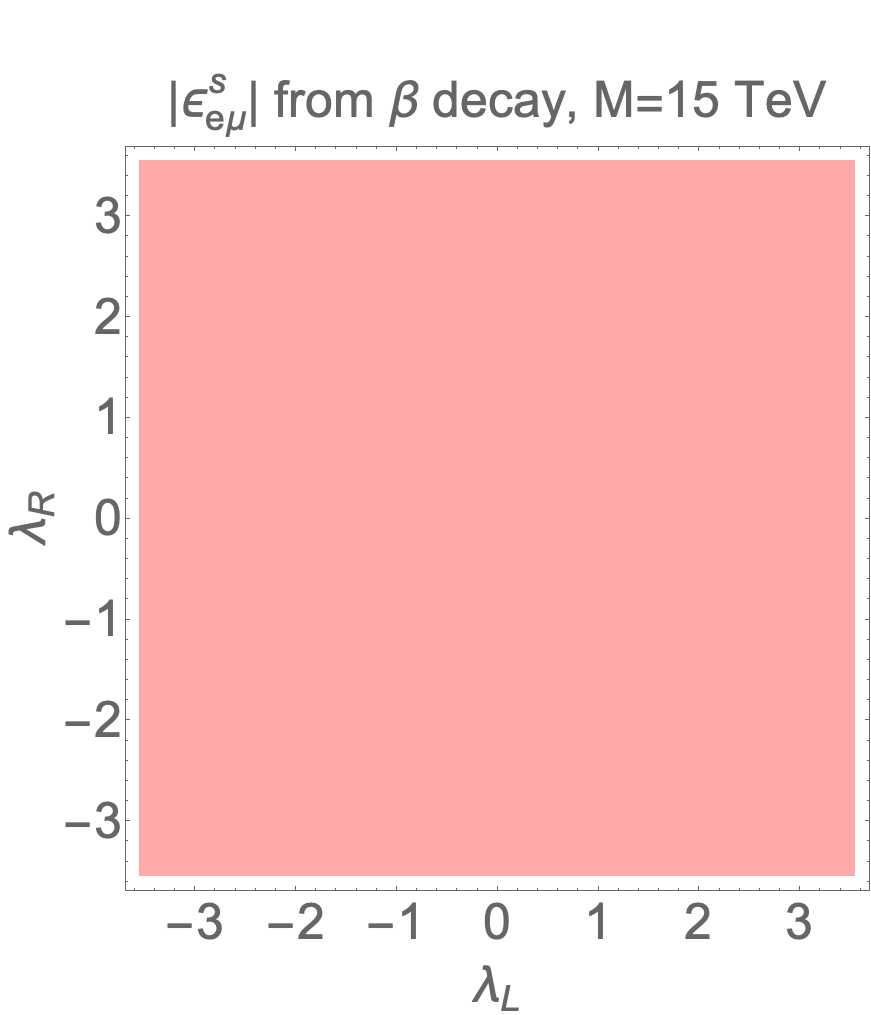}
\\
\includegraphics[scale=0.25]{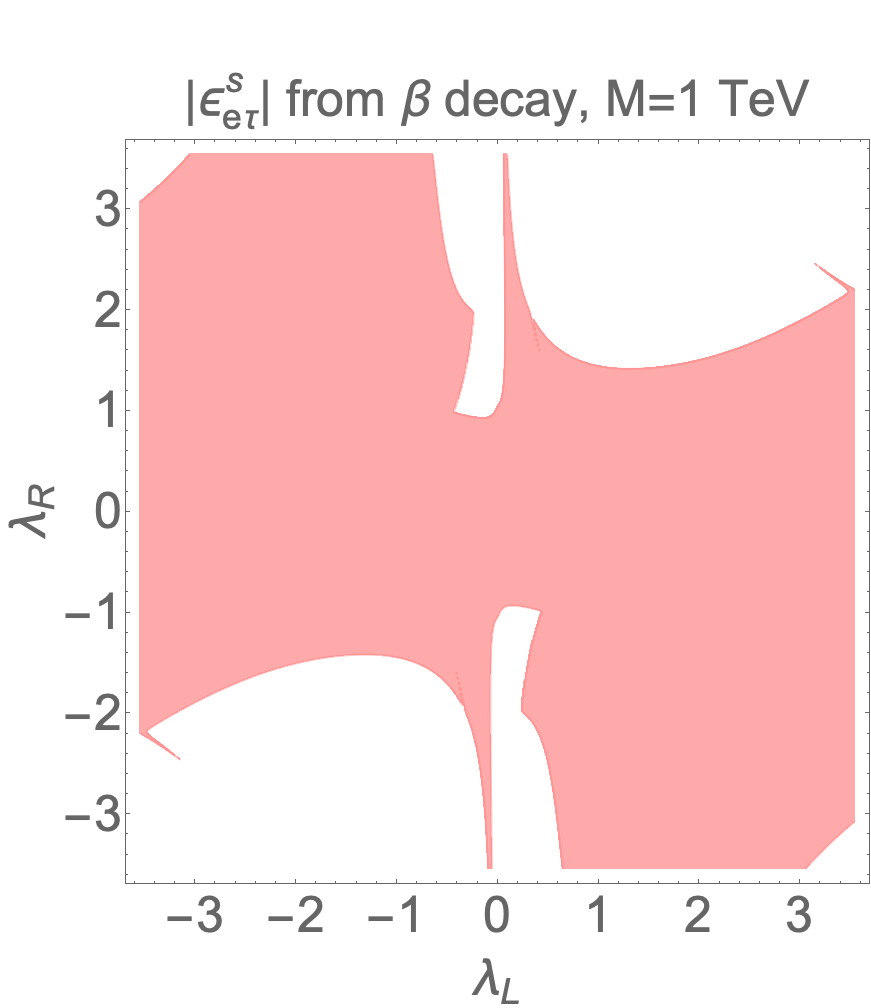} ~& ~ \includegraphics[scale=0.25]{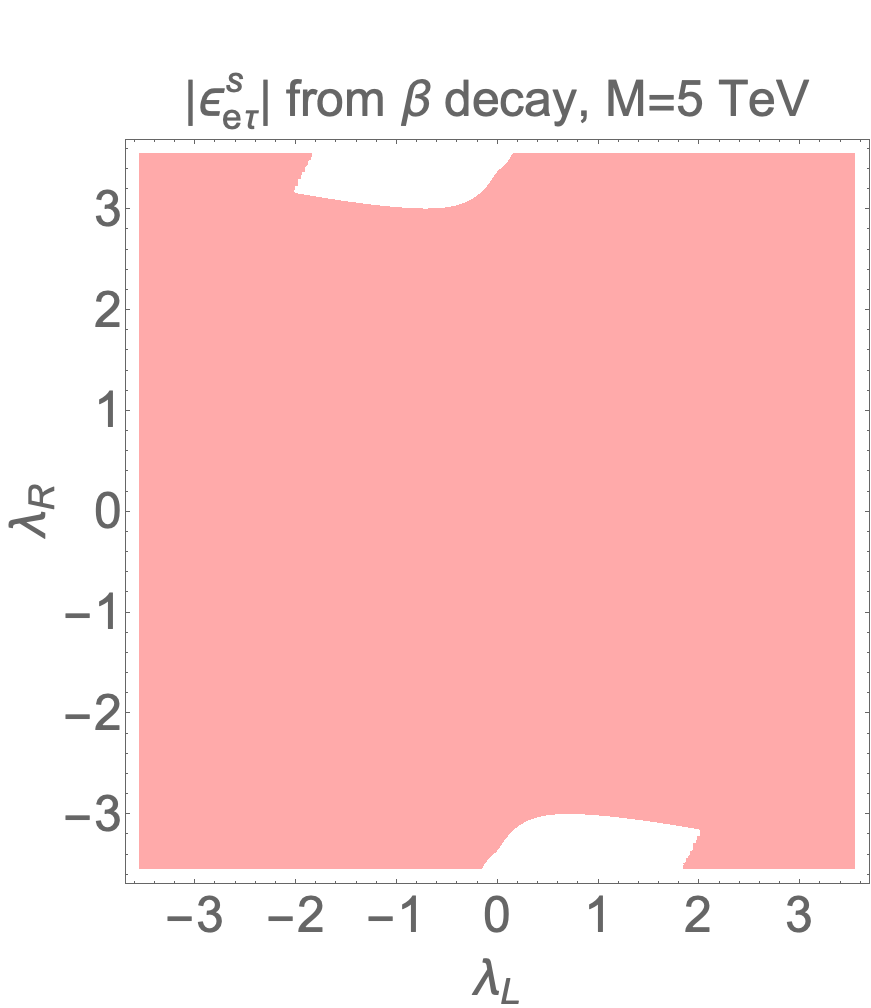}~&~\includegraphics[scale=0.25]{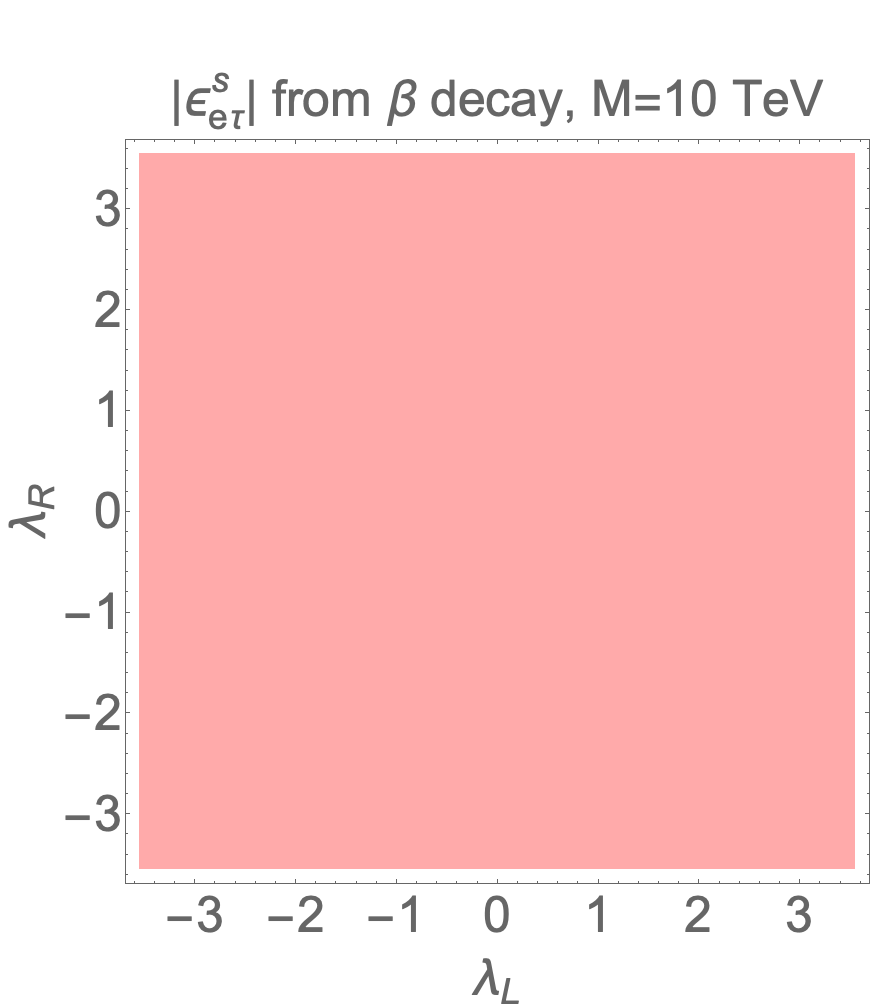}~&~\includegraphics[scale=0.25]{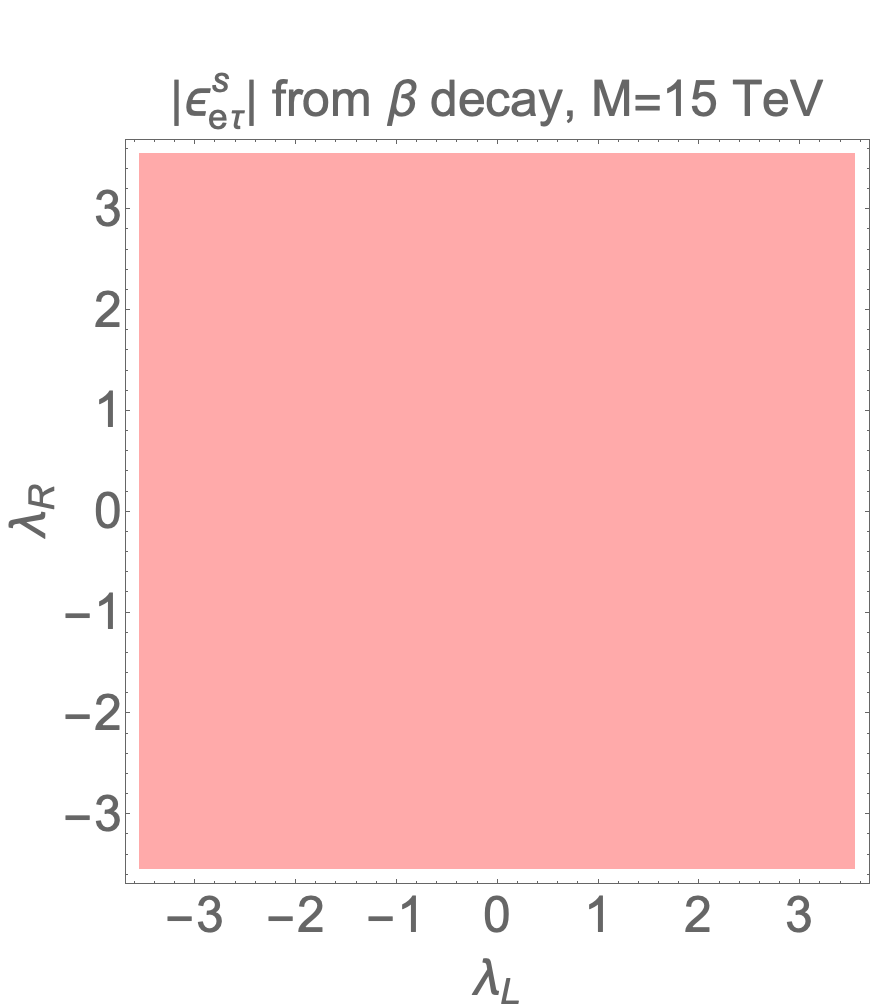}
\end{tabular}
  \end{adjustbox}}
\caption{Constraints on the simplified leptoquark model parameter space from current constraints on $|\epsilon^s_{e\ell}|$ ($\ell=e,\mu,\tau$) from beta decay summarized in table\,\ref{NSIbounds}. The pink region(s) in each figure is (are) still allowed at 95\% CL, obtained with fixed leptoquark mass $M$.}\label{app:NSIRegionLQPart23}
\end{figure}

\yong{Clearly, when $\lambda_{L}$ and/or $\lambda_{R}$ are tiny, contributions to the NSI parameters from this simplified model are suppressed for any fixed $M$. Alternatively, equivalent suppression can be achieved by increasing $M$, as a result, moderate or even large $\lambda_{L,R}$ are allowed as can be seen from the last three columns of figure\,\ref{NSIRegionLQ}. However, we point out that, even when the leptoquark mass is large of $\mathcal{O}(10\rm\,TeV)$, $\lambda_L$ and $\lambda_R$ can not simultaneously become large to survive from the current constraint on $|\epsilon_{\mu e}^{s}|$, shown in the last plot in the first row of figure\,\ref{NSIRegionLQ}. Constraints from other NSI parameters turn out to be relatively weaker than those shown in figure\,\ref{NSIRegionLQ}{, as one can see clearly from figures\,\ref{app:NSIRegionLQPart22} and \ref{app:NSIRegionLQPart23}}}.

\iffalse
%%%%%%%%%%%%%%%%%%
\subsection{\label{sec:collider}Constraints from a global fitting of neutrino oscillation experiments}
%%%%%%%%%%%%%%%%%%

{\color{black}The contents of this part has been absorbed into figure\,\ref{NSIRegionLQ}. We can remove this subsection.}

We studied the sensitivities to the leptoquark properties in LBL and reactor neutrino experiments. The results are presented in figure\,\ref{fig:leptoquark}. At 95\% CL, the upper bound of 1.8$\times$10$^{-7}$ GeV$^{-2}$ is obtained on $\lambda^2/M^2$ from LBL experiments T2K and NO$\nu$A, and 1.2$\times$10$^{-7}$ GeV$^{-2}$ from reactor experiments Daya Bay, Double Chooz and RENO, respectively.

  \begin{figure}[!htb]
        \center{\includegraphics[width=0.9\textwidth]
        {plots/Leptoquark}}
        \caption{\label{fig:leptoquark}\yong{This plot will be combined with other plots as shown above.} Exclusion limits to the leptoquark simplified model in neutrino oscillation experiments. The statistical significance for ruling out the simplified model is shown as function of $\lambda^2/M^2$ in GeV$^{-2}$ in the LBL experiments T2K and NO$\nu$A (black line) and in the reactor experiments Daya Bay, Double Chooz and RENO (red line). The statistical limit corresponding to 95\% CL is shown with gray dashed line.}
      \end{figure}
      
{\color{red} (1) Add comparison between this result and collider results; (2) Add comparison between this result and Yong's result.}
\fi

%%%%%%%%%%%%%%%%%%%%%%%%%%%%%%%%%%%%%%%%%%%%%
\section{Constraints on the dimension-6 SMEFT operators from neutrino oscillation experiments}
\label{sec:pheno}
%%%%%%%%%%%%%%%%%%%%%%%%%%%%%%%%%%%%%%%%%%%%%

\yong{In the last section, the robustness of using neutrino NSI parameters to study \sampsa{new physics at high energies} is illustrated by the simplified scalar leptoquark model. In this section, following figure\,\ref{fig:workflow}, we generalize our study to all dimension-6 SMEFT operators and present our final results \sampsa{as lower constraints} on the UV scale $\Lambda$ as well as \sampsa{upper constraints} on the Wilson coefficients \sampsa{pertaining to individual operators}. Due to the large number of dimension-6 SMEFT operators, we only present the results for operators that \sampsa{are within the reach of the neutrino experiments considered in this work.} To that end, we first take the bottom-up EFT approach where one assumes all the dimension-6 SMEFT operators are independent at the UV scale $\Lambda$ and then derives the experimental constraints on $\Lambda$ and the Wilson coefficient of each operator. The results are presented in subsection\,\ref{sec:singleop} and achieved by simulating the \sampsa{neutrino oscillation data in} the LBL neutrino experiments \sampsa{T2K and NO$\nu$A} and the reactor antineutrino experiments \sampsa{Daya Bay, Double Chooz and RENO}.

On the other hand, as discussed in section\,\ref{sec:SMEFTtoNSI}, the correlation among different operators will in general get lost in the bottom-up EFT approach. However, as already can be seen in the simplified scalar leptoquark model case presented in section\,\ref{sec:leptoquark}, effects due to the correlation among the dimension-6 SMEFT operators already show up at the UV scale $\Lambda$. As a result, neutrino NSI parameters at the 2\,GeV scale will also be affected by the correlation. We present our results demonstrating the correlation among different operators in subsection\,\ref{sec:multiop}.}

\subsection{\label{sec:singleop}Constraints on a single operator}

\begin{figure}[t]
\centering{
\begin{tabular}{c}
\includegraphics[width = \textwidth]{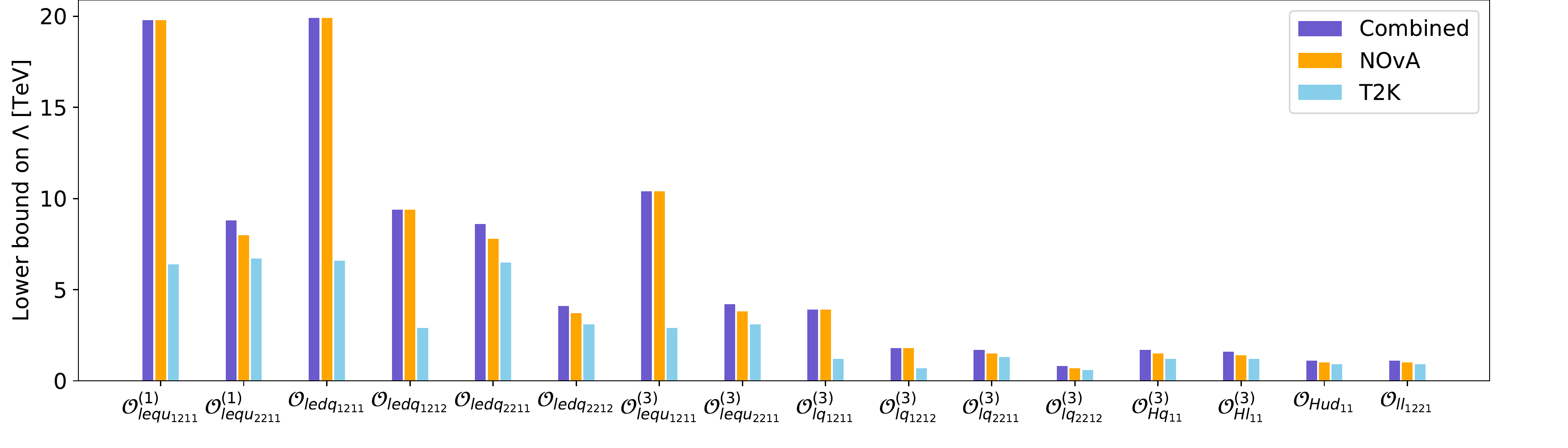}\\
\includegraphics[width = \textwidth]{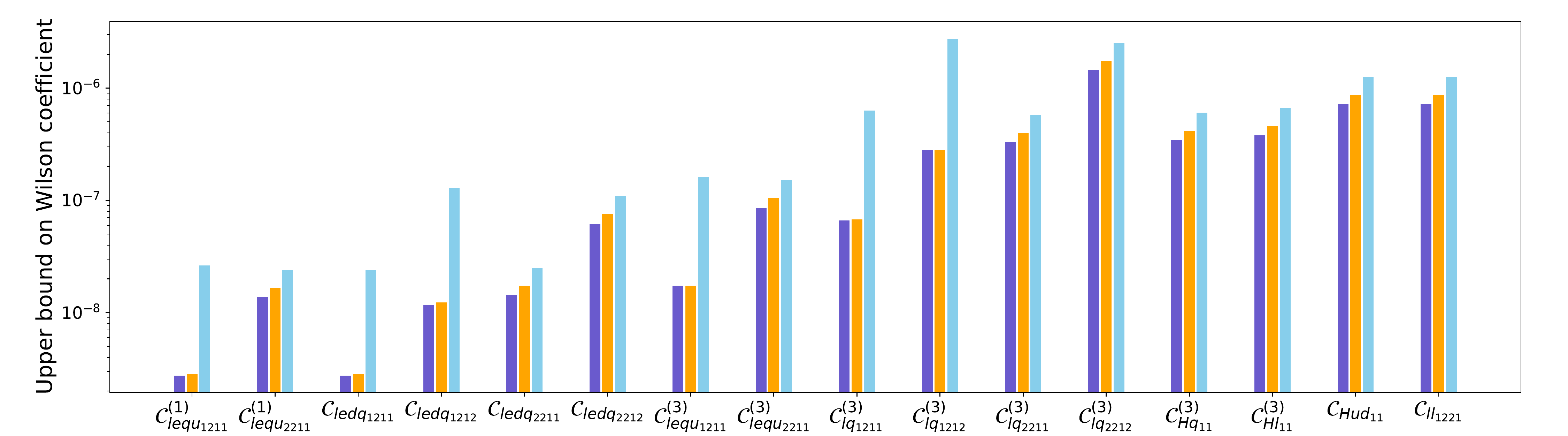}
\end{tabular}}
\caption{\label{fig:lblconstr} Constraints on a part of dimension-6 operators in tables~\ref{lambdabounds1} and \ref{wcbounds1} in LBL experiments T2K and NO$\nu$A at 95\% CL of statistical significance. \sampsa{The lower bounds are presented on the UV scale $\Lambda$ in each operator, when the corresponding Wilson coefficient is set to unity.} Correspondingly, the upper constraints on the respective Wilson coefficients \sampsa{are shown when the scale is set to $\Lambda =$ 1~TeV.}}
\end{figure}

{\color{black}Treating the Wilson coefficients and the UV scale $\Lambda$ as arbitrary parameters, the non-observation of new physics in the neutrino oscillation experiments can be used to derive constraints. In this section, we simulate the oscillation data that has recently been gathered in the LBL experiments T2K and NO$\nu$A and the reactor experiments Daya Bay, Double Chooz and RENO to extract the experimental sensitivities to each operator.}

{\color{black} We first present the constraints on the Wilson coefficient and the new physics scale $\Lambda$ for each operator individually in figure\,\ref{fig:lblconstr}. The constraints are provided at 95\% CL of statistical significance for 16 different operators using the neutrino oscillation data from the LBL experiments T2K and NO$\nu$A. The oscillations in these experiments are influenced through the pion decay process as discussed in section\,\ref{sec:NSIparam}. The upper panel shows the lower bound on the $\Lambda$ in the LBL experiments when the Wilson coefficient of the corresponding operator is set to unity. Correspondingly, the upper bounds on the Wilson coefficients shown in the lower panel were obtained by keeping the scale parameter at $\Lambda =$ 1~TeV. The results are presented for both experiments individually as well as for the combined bound of T2K and NO$\nu$A.} \sampsa{Owing mainly to the higher fiducial mass in the near detector, we remark that NO$\nu$A places the most restrictive constraint of the two experiments.}

\yong{Note that in figure\,\ref{fig:lblconstr}, the type-A dimension-6 SMEFT operators discussed in section\,\ref{sec:SMEFTtoNSI} are in general most stringently constrained from the neutrino NSI parameters. This is what one would expect since these operators contribute directly to the Wilson coefficients $\epsilon_{L,R,S,P,T}$ in the LEFT unless one gets a suppression factor from the off-diagonal elements of the CKM matrix as is the case for the ${\cal O}_{\substack{ledq \\ 2212}}$ operator. In contrast, the type-B operators are relatively less constrained since they contribute to neutrino NSI parameters indirectly through modifying the interacting strength between gauge bosons and fermions at the weak scale $\mu=m_W$. Furthermore, among all the type-A operators, ${\cal O}^{(1)}_{\substack{lequ \\ 1211}}$ and ${\cal O}^{(1)}_{\substack{ledq \\ 1211}}$ dominate over all \sampsa{other operators} for the following two reasons: (1) There is no off-diagonal CKM matrix element suppression in the quark sector; (2) For LBL experiments through pion decay, the dominant contributions to the neutrino NSI parameters come from the pseudo-scalar interacting term $\epsilon_P$ in the LEFT defined in eq.\,\eqref{eq:CC}, which can be clearly understood from eq.\,\eqref{eq:sNSI2} due to the enhancement factor ${m_{\pi}^{2}}/{[m_{\mu}\left(m_{u}+m_{d}\right)]}$.}

\begin{figure}
\centering{
\begin{tabular}{c}
\includegraphics[width = \textwidth]{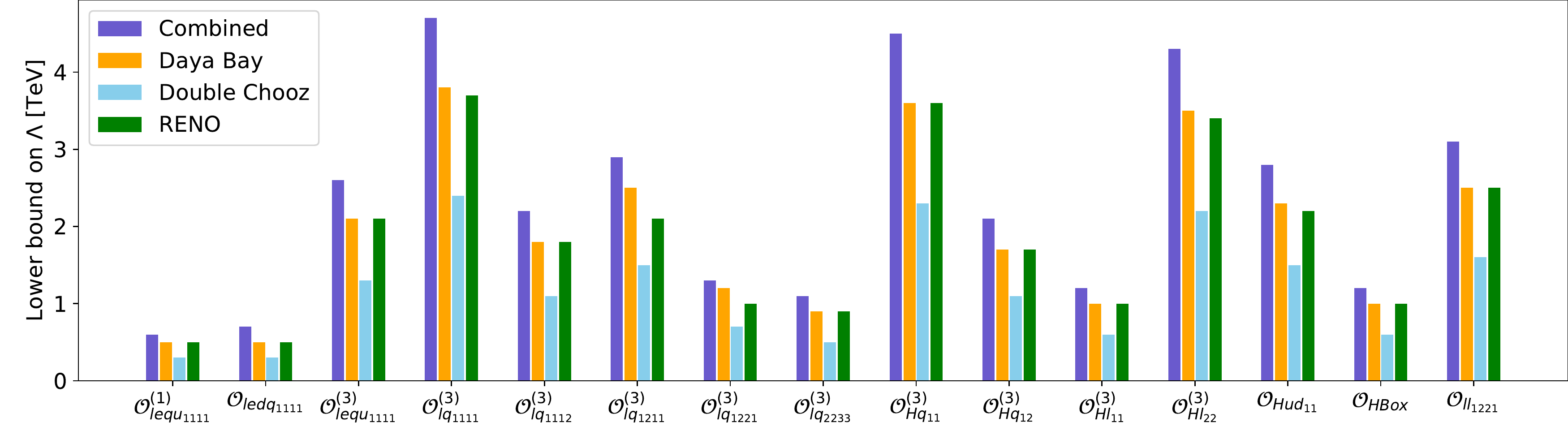}\\
\includegraphics[width = \textwidth]{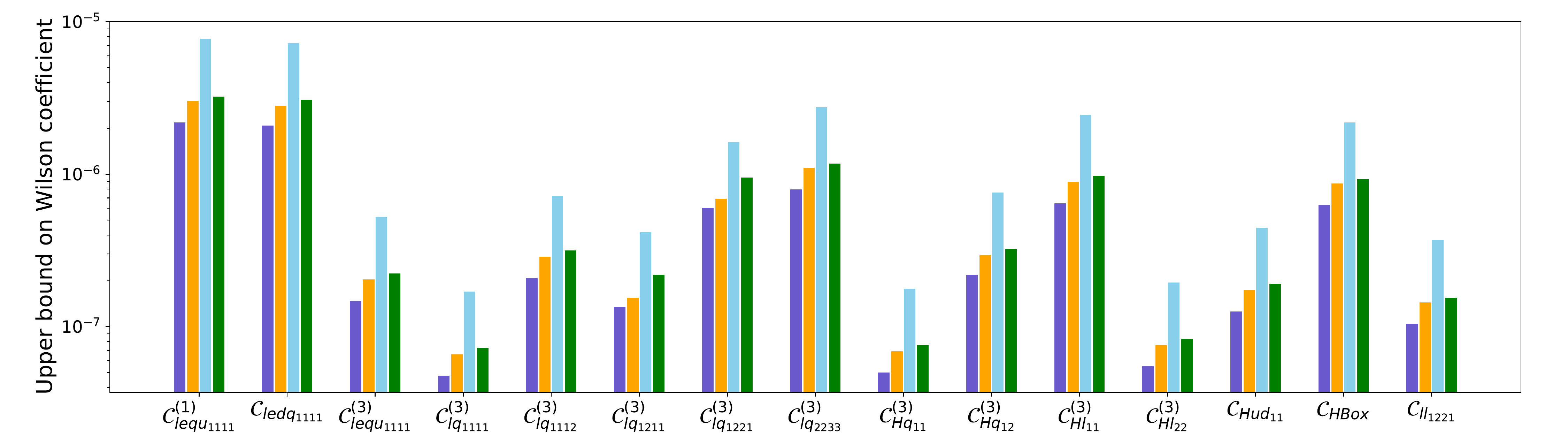}
\end{tabular}}
\caption{\label{fig:rconstr} \sampsa{Constraints on part of dimension-6 operators in tables~\ref{lambdabounds2} and \ref{wcbounds2} in reactor neutrino experiments Daya Bay, Double Chooz and RENO. The new physics scale $\Lambda$ associated with a specific operator is constrained from below and the corresponding Wilson coefficient from above when $\Lambda =$ 1~TeV. The results are obtained at 95\% CL.}}
\end{figure}

Similarly, the constraints from the reactor antineutrino experiments Daya Bay, Double Chooz and RENO are shown in figure\,\ref{fig:rconstr}, also presented at 95\% CL for each of the considered reactor experiments. A combined bound from all the three experiments is also provided for each operator. We implement the non-standard physics through the beta decay and inverse beta decay processes, which are both present in all the three reactor experiments. {\color{black} Of the three reactor experiments, Day Bay and RENO impose similar limits to the observables, while Double Chooz presents somewhat weaker contribution due to its smaller exposure.}

\yong{Different from the LBL neutrino experiments where the neutrino NSI parameters are sensitive to the pseudo-scalar interaction, the reactor neutrino experiments are dominated by the vector and axial-vector type interactions described by the $\epsilon_{L,R}$ terms in eq.\,\eqref{eq:CC}. This can also be understood by looking at the matching formulas in eqs.\,(\ref{eq:sNSI1}-\ref{inverseBetaDecayFor}) and explains why ${\cal O}^{(3)}_{\substack{lq \\ 1111}}$, ${\cal O}^{(3)}_{\substack{Hq \\ 11}}$, and ${\cal O}^{(3)}_{\substack{Hl \\ 22}}$ are more constrained than all the other operators listed in figure\,\ref{fig:rconstr}. Furthermore, (1) the type-A operator ${\cal O}^{(3)}_{\substack{lq \\ 1111}}$ dominates over the type-B operators ${\cal O}^{(3)}_{\substack{Hq \\ 11}}$ and ${\cal O}^{(3)}_{\substack{Hl \\ 22}}$ for the same reason as explained for the LBL neutrino experiments; (2) ${\cal O}^{(3)}_{\substack{Hl \\ 22}}$ is more constrained compared with ${\cal O}^{(3)}_{\substack{Hl \\ 11}}$ since the second generation charged lepton is heavier than the first one, thus its contribution to the corresponding NSI parameter is less suppressed than that from ${\cal O}^{(3)}_{\substack{Hl \\ 11}}$.}

{\color{black}Comparing figures\,\ref{fig:lblconstr} and \ref{fig:rconstr}, one notices that the results extracted from the LBL neutrino oscillation experiments are generally more restrictive compared to those obtained from the reactor experiments. While the strongest bound on the UV scale $\Lambda$ is nearly 20~TeV from the LBL experiments, the most restrictive bound from the reactor experiments is about 4~TeV. However, it should be emphasized that the results following from the LBL and the reactor experiments are mostly sensitive to different operators. This highlights the complementary nature that lies between the LBL and the reactor experiments.}

\begin{figure}[t]
        \center{\includegraphics[width=\textwidth]
        {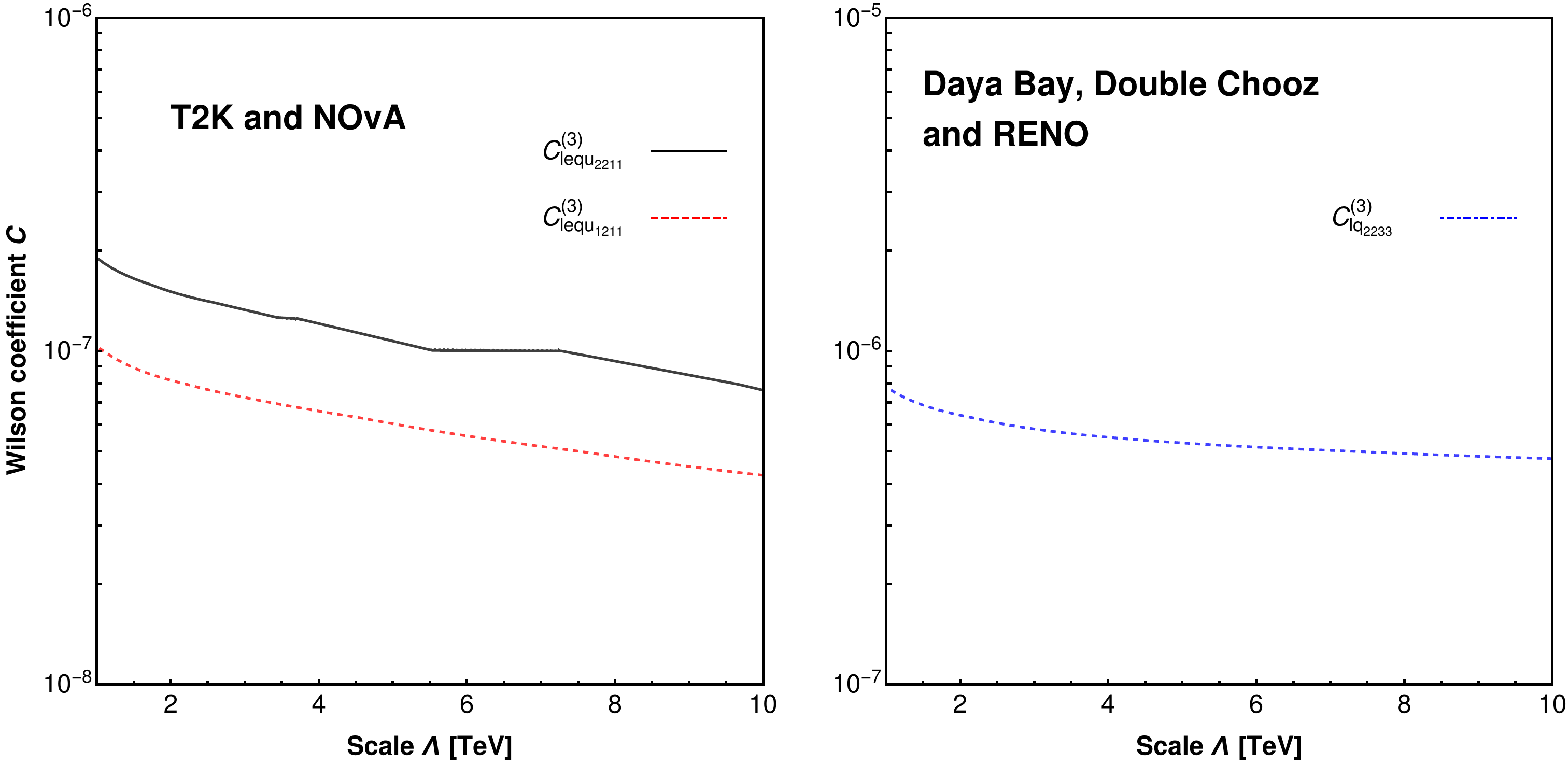}}
        \caption{\label{fig:scaledep} Upper constraints on Wilson coefficient as function of the scale $\Lambda$. The left panel shows the scale-dependence for operators in LBL experiments T2K and NO$\nu$A whilst the right panel represents the reactor experiments Daya Bay, Double Chooz and RENO. The upper constraints on the Wilson coefficients are obtained at 95\% CL.}
      \end{figure}

\yong{We also point out that, for certain operators, the upper bounds provided for the Wilson coefficients in figures\,\ref{fig:lblconstr} and \ref{fig:rconstr} are sensitive to the UV scale $\Lambda$. Among all the dimension-6 SMEFT operators that contribute dominantly to neutrino NSI parameters, we find that there are only three operators whose Wilson coefficients are sensitive to $\Lambda$. We \sampsa{show} them in the left and the right panels of figure\,\ref{fig:scaledep} for the LBL and the reactor neutrino experiments respectively for $\Lambda$ in the [1, 10]\,TeV range.\footnote{\sampsa{As one may see from the results shown in figure\,\ref{fig:scaledep} it still holds that the low-energy effects are approximately independent from the UV scale $\Lambda$.}} In addition, for these three operators, we also observe that their phases oscillate very rapidly as a function of $\Lambda$, implying {\sl CP}-violating NSI phases that we further explore in section\,\ref{sec:NSIphase}.}

\begin{figure}[t]
\centering{
  \begin{adjustbox}{max width = \textwidth}
\begin{tabular}{cc}
\includegraphics[scale=0.5]{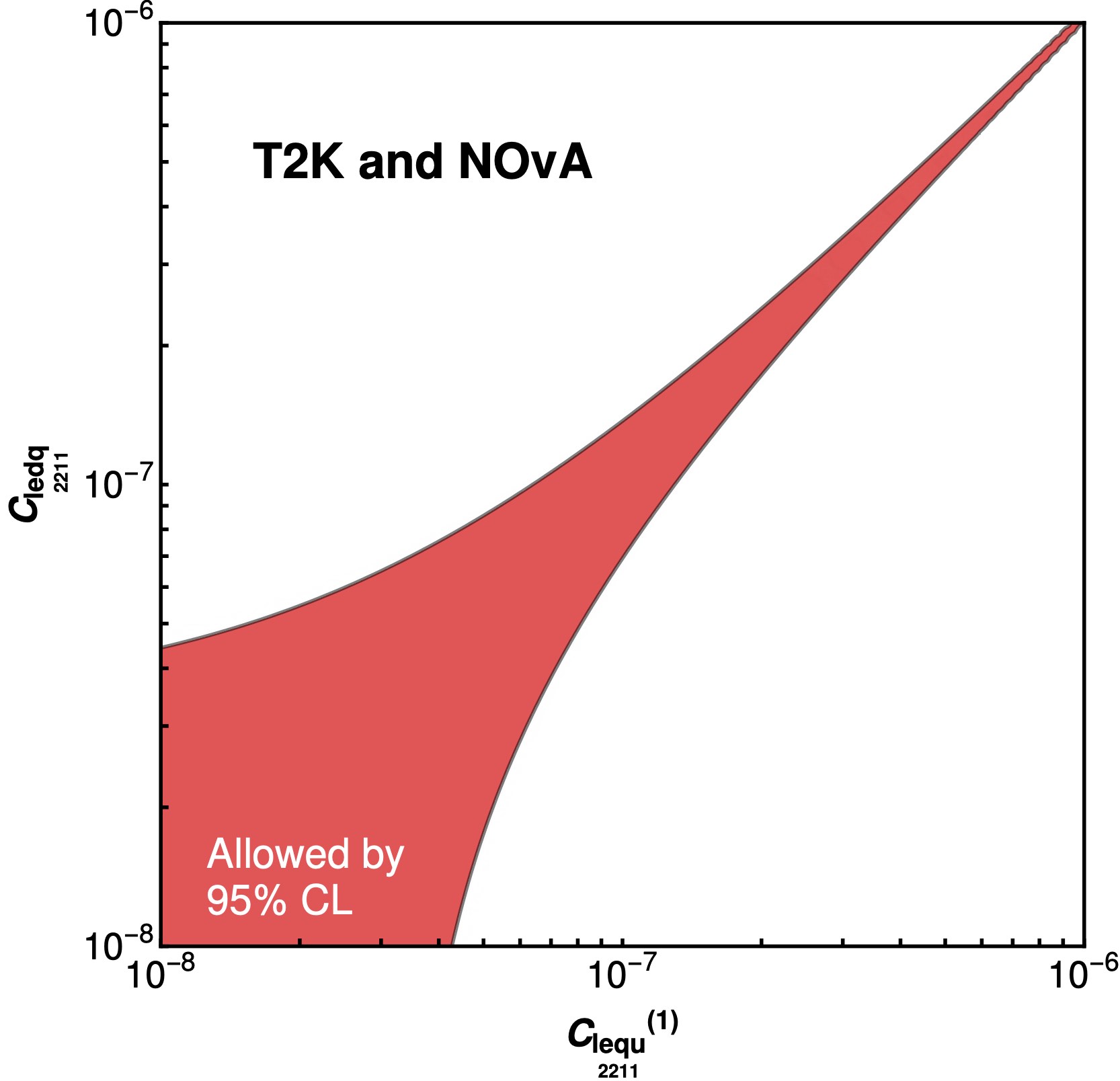} ~& ~ \includegraphics[scale=0.5]{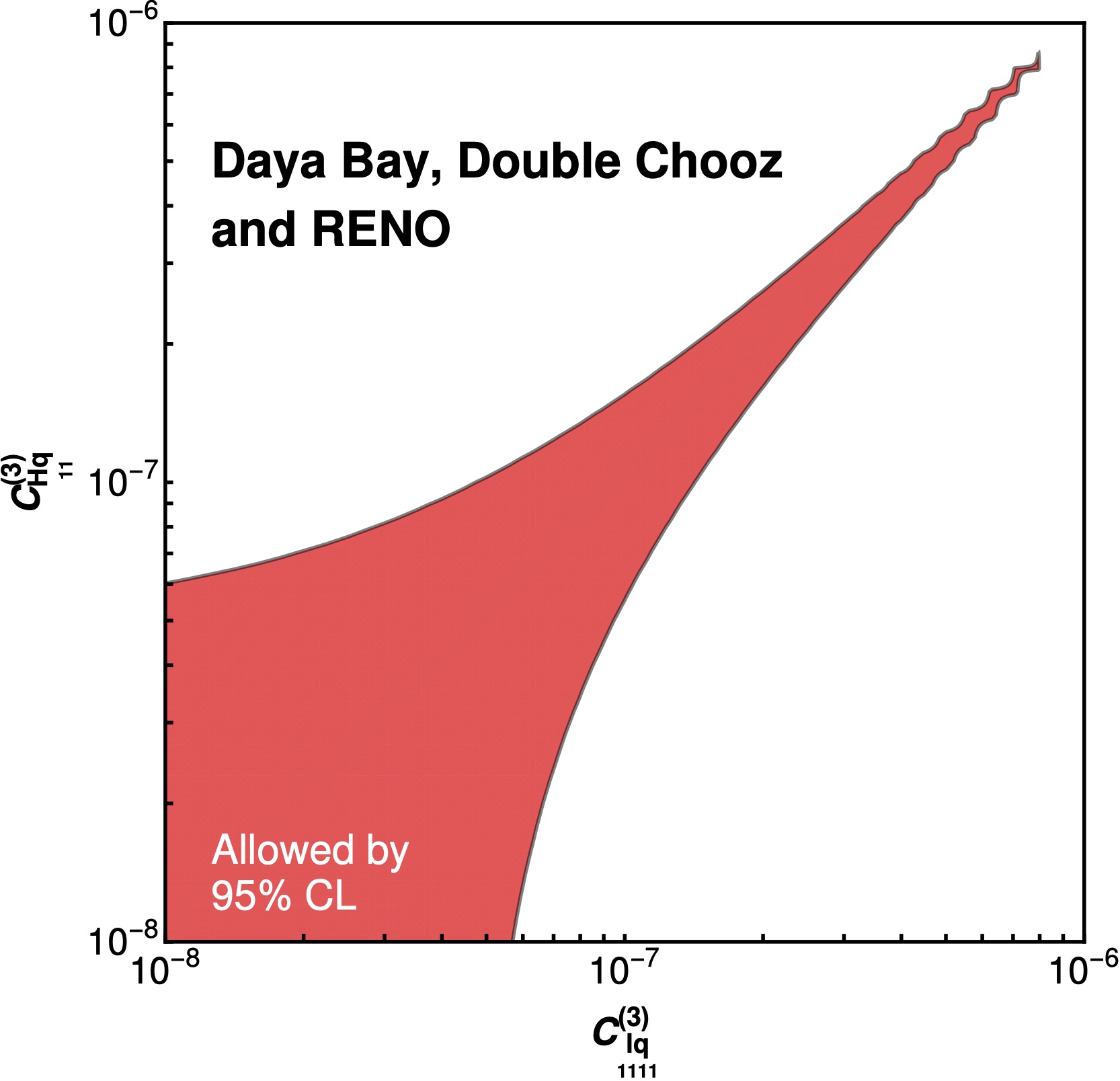}
\end{tabular}
  \end{adjustbox}}
\caption{{\color{black}Left panel: Upper constraints on the Wilson coefficients corresponding to the operators $(\bar{\ell}_2 e_2) \varepsilon_{jk} (\bar{q}_1 u_1)$ and $(\bar{\ell}_2 e_2) (\bar{d}_1 q_1)$ with $\Lambda =$ 1\,TeV. The constraints are obtained for the LBL experiments T2K and NO$\nu$A at 95\% CL significance. Right panel: Same as the left panel but with the application of Daya Bay, Double Chooz and RENO on operators $(\bar{\ell}_1 \sigma_{\mu \nu} e_1) \varepsilon_{jk} (\bar{q}_1 \sigma^{\mu \nu} u_1)$ and $(H^\dagger i \Vec{D}_\mu H) (\bar{q}_1 \tau^I \gamma^\mu q_1)$ instead.}}\label{fig:two-operatorsTot}
\end{figure}

%\iffalse
%%%%%%%%%%%%%%%%%%%%%%%%%%%%%%%%%%%%%%%%%%%%%%%%%%%%%%%%%%%%%%
% The numerical values of figures 4 and 5 are provided here: %
%%%%%%%%%%%%%%%%%%%%%%%%%%%%%%%%%%%%%%%%%%%%%%%%%%%%%%%%%%%%%%
\begin{table}[!t]
\caption{\label{lambdabounds1} Lower bounds on the UV scale $\Lambda$ as obtained from the long-baseline neutrino experiment data. The results are arranged from strongest to weakest. The constraints are provided in units of TeV at 95\% CL significance for T2K and NO$\nu$A, both individually and combined.}
\begin{center}
%\resizebox{\linewidth}{!}{%
\begin{tabular}{cccc}\hline\hline
Operator & T2K & NO$\nu$A & Combined \\ \hline
\rule{0pt}{3ex} ${\cal O}_{\substack{ledq \\ 1211}}$ & 6.4 & 19.8 & 19.8\\
\rule{0pt}{4ex} ${\cal O}^{(1)}_{\substack{lequ \\ 1211}}$ & 6.7 & 8.0 & 8.8\\
\rule{0pt}{4ex} ${\cal O}^{(1)}_{\substack{lequ \\ 2211}}$ & 6.6 & 8.0 & 8.8\\
\rule{0pt}{4ex} ${\cal O}_{\substack{ledq \\ 2211}}$ & 2.9 & 9.4 & 9.4\\
\rule{0pt}{4ex} ${\cal O}^{(3)}_{\substack{lequ \\ 1211}}$ & 6.5 & 7.8 & 8.6\\
\rule{0pt}{4ex} ${\cal O}_{\substack{ledq \\ 1212}}$ & 3.1 & 3.7 & 4.1\\
\rule{0pt}{4ex} ${\cal O}_{\substack{ledq \\ 2212}}$ & 2.9 & 9.4 & 9.4\\
\rule{0pt}{4ex} ${\cal O}^{(3)}_{\substack{lequ \\ 2211}}$ & 3.1 & 3.7 & 4.1\\
\rule{0pt}{4ex} ${\cal O}^{(3)}_{\substack{lq \\ 1211}}$ & 1.2 & 10.4 & 10.4\\
\rule{0pt}{4ex} ${\cal O}^{(3)}_{\substack{lq \\ 2211}}$ & 0.7 & 1.8 & 1.8\\
\rule{0pt}{4ex} ${\cal O}^{(3)}_{\substack{H q \\ 11}}$ & 1.3 & 1.5 & 1.7\\
\rule{0pt}{4ex} ${\cal O}^{(3)}_{\substack{H l \\ 11}}$ & 0.6 & 0.7 & 0.8\\
\rule{0pt}{4ex} ${\cal O}_{\substack{H ud \\ 11}}$ & 1.2 & 1.5 & 1.7\\
\rule{0pt}{4ex} ${\cal O}_{\substack{ll \\ 1221}}$ & 1.2 & 1.4 & 1.6\\
\rule{0pt}{4ex} ${\cal O}^{(3)}_{\substack{lq \\ 1212}}$ & 0.9 & 1.0 & 1.1\\
\rule{0pt}{4ex} ${\cal O}^{(3)}_{\substack{lq \\ 2212}}$ & 0.9 & 1.0 & 1.1\\ \hline\hline
\end{tabular}%}
\end{center}
\end{table}

\begin{table}[!t]
\caption{\label{wcbounds1} Upper bounds on the Wilson coefficient $\mathcal{C}$ as obtained from the long-baseline neutrino experiment data. The constraints are provided in units of TeV at 95\% CL significance for T2K and NO$\nu$A, both individually and combined.}
\begin{center}
%\resizebox{\linewidth}{!}{%
\begin{tabular}{cccc}\hline\hline
Operator & T2K & NO$\nu$A & Combined \\ \hline
\rule{0pt}{3ex} ${\cal C}_{\substack{ledq \\ 1211}}$ & 2.6$\times$10$^{-8}$ & 2.8$\times$10$^{-9}$ & 2.8$\times$10$^{-9}$\\
\rule{0pt}{4ex} ${\cal C}^{(1)}_{\substack{lequ \\ 1211}}$ & 2.4$\times$10$^{-8}$ & 8.0$\times$10$^{-8}$ & 8.8$\times$10$^{-8}$\\
\rule{0pt}{4ex} ${\cal C}^{(1)}_{\substack{lequ \\ 2211}}$ & 2.4$\times$10$^{-8}$ & 2.8$\times$10$^{-9}$ & 2.8$\times$10$^{-9}$\\
\rule{0pt}{4ex} ${\cal C}_{\substack{ledq \\ 2211}}$ & 1.3$\times$10$^{-7}$ & 1.2$\times$10$^{-8}$ & 1.2$\times$10$^{-8}$\\
\rule{0pt}{4ex} ${\cal C}^{(3)}_{\substack{lequ \\ 1211}}$ & 2.5$\times$10$^{-8}$ & 1.7$\times$10$^{-8}$ & 1.4$\times$10$^{-8}$\\
\rule{0pt}{4ex} ${\cal C}_{\substack{ledq \\ 1212}}$ & 1.1$\times$10$^{-7}$ & 7.6$\times$10$^{-8}$ & 6.2$\times$10$^{-8}$\\
\rule{0pt}{4ex} ${\cal C}_{\substack{ledq \\ 2212}}$ & 1.6$\times$10$^{-7}$ & 1.7$\times$10$^{-8}$ & 1.7$\times$10$^{-8}$\\
\rule{0pt}{4ex} ${\cal C}^{(3)}_{\substack{lequ \\ 2211}}$ & 1.5$\times$10$^{-7}$ & 1.0$\times$10$^{-7}$ & 8.5$\times$10$^{-8}$\\
\rule{0pt}{4ex} ${\cal C}^{(3)}_{\substack{lq \\ 1211}}$ & 6.3$\times$10$^{-7}$ & 6.8$\times$10$^{-8}$ & 6.6$\times$10$^{-8}$\\
\rule{0pt}{4ex} ${\cal C}^{(3)}_{\substack{lq \\ 2211}}$ & 2.8$\times$10$^{-6}$ & 2.8$\times$10$^{-7}$ & 2.8$\times$10$^{-7}$\\
\rule{0pt}{4ex} ${\cal C}^{(3)}_{\substack{H q \\ 11}}$ & 5.8$\times$10$^{-7}$ & 4.0$\times$10$^{-7}$ & 3.3$\times$10$^{-7}$\\
\rule{0pt}{4ex} ${\cal C}^{(3)}_{\substack{H l \\ 11}}$ & 2.5$\times$10$^{-6}$ & 1.7$\times$10$^{-6}$ & 1.4$\times$10$^{-6}$\\
\rule{0pt}{4ex} ${\cal C}_{\substack{H ud \\ 11}}$ & 6.0$\times$10$^{-7}$ & 4.2$\times$10$^{-7}$ & 3.5$\times$10$^{-7}$\\
\rule{0pt}{4ex} ${\cal C}_{\substack{ll \\ 1221}}$ & 6.6$\times$10$^{-7}$ & 4.6$\times$10$^{-7}$ & 3.8$\times$10$^{-7}$\\
\rule{0pt}{4ex} ${\cal C}^{(3)}_{\substack{lq \\ 1212}}$ & 1.3$\times$10$^{-6}$ & 8.7$\times$10$^{-7}$ & 7.2$\times$10$^{-7}$\\
\rule{0pt}{4ex} ${\cal C}^{(3)}_{\substack{lq \\ 2212}}$ & 1.3$\times$10$^{-6}$ & 8.7$\times$10$^{-7}$ & 7.2$\times$10$^{-7}$\\ \hline\hline
\end{tabular}%}
\end{center}
\end{table}

\begin{table}[!t]
\caption{\label{lambdabounds2} Lower bounds on the UV scale $\Lambda$ as obtained from the reactor neutrino experiment data. The constraints are provided in units of TeV at 95\% CL significance for Daya Bay, Double Chooz and RENO, both individually and combined.}
\begin{center}
%\resizebox{\linewidth}{!}{%
\begin{tabular}{ccccc}\hline\hline
Operator & Daya Bay & Double Chooz & RENO & Combined \\ \hline
\rule{0pt}{3ex} ${\cal O}^{(1)}_{\substack{lequ \\ 1111}}$ & 0.5 & 0.3 & 0.5 & 0.6\\
\rule{0pt}{4ex} ${\cal O}_{\substack{ledq \\ 1111}}$ & 0.5 & 0.3 & 0.5 & 0.7\\
\rule{0pt}{4ex} ${\cal O}^{(3)}_{\substack{lequ \\ 1111}}$ & 2.1 & 0.3 & 2.1 & 2.6\\
\rule{0pt}{4ex} ${\cal O}^{(3)}_{\substack{lq \\ 1111}}$ & 3.8 & 2.4 & 3.7 & 4.7\\
\rule{0pt}{4ex} ${\cal O}^{(3)}_{\substack{lq \\ 1112}}$ & 1.8 & 1.1 & 1.8 & 2.2\\
\rule{0pt}{4ex} ${\cal O}^{(3)}_{\substack{lq \\ 1211}}$ & 2.5 & 1.5 & 2.1 & 2.9\\
\rule{0pt}{4ex} ${\cal O}^{(3)}_{\substack{lq \\ 1221}}$ & 1.2 & 0.7 & 1.0 & 1.3\\
\rule{0pt}{4ex} ${\cal O}^{(3)}_{\substack{lq \\ 2233}}$ & 0.9 & 0.5 & 0.9 & 1.1\\
\rule{0pt}{4ex} ${\cal O}^{(3)}_{\substack{H q \\ 11}}$ & 3.6 & 2.3 & 3.6 & 4.5\\
\rule{0pt}{4ex} ${\cal O}^{(3)}_{\substack{H q \\ 12}}$ & 1.7 & 1.1 & 1.7 & 2.1\\
\rule{0pt}{4ex} ${\cal O}^{(3)}_{\substack{H l \\ 11}}$ & 1.0 & 0.6 & 1.0 & 1.2\\
\rule{0pt}{4ex} ${\cal O}^{(3)}_{\substack{H l \\ 22}}$ & 3.5 & 2.2 & 3.4 & 4.3\\
\rule{0pt}{4ex} ${\cal O}_{\substack{H ud \\ 11}}$ & 2.3 & 1.5 & 2.2 & 2.8\\
\rule{0pt}{4ex} ${\cal O}_{H \Box}$ & 1.0 & 0.6 & 1.0 & 1.2\\
\rule{0pt}{4ex} ${\cal O}_{\substack{ll \\ 1221}}$ & 2.5 & 1.6 & 2.5 & 3.1\\ \hline\hline
\end{tabular}%}
\end{center}
\end{table}

\begin{table}[!t]
\caption{\label{wcbounds2} Upper bounds on the Wilson coefficient $\mathcal{C}$ as obtained from the reactor neutrino experiment data. The constraints are provided in units of TeV at 95\% CL significance for Daya Bay, Double Chooz and RENO, both individually and combined.}
\begin{center}
%\resizebox{\linewidth}{!}{%
\begin{tabular}{ccccc}\hline\hline
Operator & Daya Bay & Double Chooz & RENO & Combined \\ \hline
\rule{0pt}{3ex} ${\cal C}^{(1)}_{\substack{lequ \\ 1111}}$ & 3.0$\times$10$^{-6}$ & 7.8$\times$10$^{-6}$ & 3.2$\times$10$^{-6}$ & 2.2$\times$10$^{-6}$\\
\rule{0pt}{4ex} ${\cal C}_{\substack{ledq \\ 1111}}$ & 2.8$\times$10$^{-6}$ & 7.2$\times$10$^{-6}$ & 3.1$\times$10$^{-6}$ & 2.1$\times$10$^{-6}$\\
\rule{0pt}{4ex} ${\cal C}^{(3)}_{\substack{lequ \\ 1111}}$ & 2.0$\times$10$^{-7}$ & 5.2$\times$10$^{-7}$ & 2.2$\times$10$^{-7}$ & 1.5$\times$10$^{-7}$\\
\rule{0pt}{4ex} ${\cal C}^{(3)}_{\substack{lq \\ 1111}}$ & 6.6$\times$10$^{-8}$ & 1.7$\times$10$^{-7}$ & 7.2$\times$10$^{-8}$ & 4.8$\times$10$^{-8}$\\
\rule{0pt}{4ex} ${\cal C}^{(3)}_{\substack{lq \\ 1112}}$ & 2.9$\times$10$^{-7}$ & 7.2$\times$10$^{-7}$ & 3.2$\times$10$^{-7}$ & 2.1$\times$10$^{-7}$\\
\rule{0pt}{4ex} ${\cal C}^{(3)}_{\substack{lq \\ 1211}}$ & 1.5$\times$10$^{-7}$ & 4.2$\times$10$^{-7}$ & 2.2$\times$10$^{-7}$ & 1.3$\times$10$^{-7}$\\
\rule{0pt}{4ex} ${\cal C}^{(3)}_{\substack{lq \\ 1221}}$ & 6.9$\times$10$^{-7}$ & 1.6$\times$10$^{-6}$ & 9.5$\times$10$^{-7}$ & 6.0$\times$10$^{-7}$\\
\rule{0pt}{4ex} ${\cal C}^{(3)}_{\substack{lq \\ 2233}}$ & 1.1$\times$10$^{-6}$ & 2.8$\times$10$^{-6}$ & 1.2$\times$10$^{-6}$ & 7.9$\times$10$^{-7}$\\
\rule{0pt}{4ex} ${\cal C}^{(3)}_{\substack{H q \\ 11}}$ & 6.9$\times$10$^{-8}$ & 1.8$\times$10$^{-7}$ & 7.6$\times$10$^{-8}$ & 5.0$\times$10$^{-8}$\\
\rule{0pt}{4ex} ${\cal C}^{(3)}_{\substack{H q \\ 12}}$ & 3.0$\times$10$^{-7}$ & 7.6$\times$10$^{-7}$ & 3.2$\times$10$^{-7}$ & 2.2$\times$10$^{-7}$\\
\rule{0pt}{4ex} ${\cal C}^{(3)}_{\substack{H l \\ 11}}$ & 8.9$\times$10$^{-7}$ & 2.5$\times$10$^{-6}$ & 9.8$\times$10$^{-7}$ & 6.5$\times$10$^{-7}$\\
\rule{0pt}{4ex} ${\cal C}^{(3)}_{\substack{H l \\ 22}}$ & 7.6$\times$10$^{-8}$ & 1.9$\times$10$^{-7}$ & 8.3$\times$10$^{-8}$ & 5.5$\times$10$^{-8}$\\
\rule{0pt}{4ex} ${\cal C}_{\substack{H ud \\ 11}}$ & 1.7$\times$10$^{-7}$ & 4.5$\times$10$^{-7}$ & 1.9$\times$10$^{-7}$ & 1.3$\times$10$^{-7}$\\
\rule{0pt}{4ex} ${\cal C}_{H \Box}$ & 8.7$\times$10$^{-7}$ & 2.2$\times$10$^{-6}$ & 9.3$\times$10$^{-7}$ & 6.3$\times$10$^{-7}$\\
\rule{0pt}{4ex} ${\cal C}_{\substack{ll \\ 1221}}$ & 1.4$\times$10$^{-7}$ & 3.7$\times$10$^{-7}$ & 1.5$\times$10$^{-7}$ & 1.0$\times$10$^{-7}$\\ \hline\hline
\end{tabular}%}
\end{center}
\end{table}
%\fi

\subsection{\label{sec:multiop}Constraints on multiple operators}

{\color{black}\yong{Constraints on the Wilson coefficients could change when more than one dimension-6 SMEFT operator is present at the UV scale, as is the case for the simplified scalar leptoquark model discussed in section\,\ref{sec:leptoquark}. The change is due to the operator mixing through the RGEs. As a result, the existence of multiple SMEFT operators may contribute simultaneously to the neutrino NSI parameters at the low energy scale, such that the sensitivities to individual Wilson coefficients also change as the other Wilson coefficients may produce a similar imprint on the neutrino oscillation data. The correlation among different dimension-6 SMEFT operator at the UV scale $\Lambda$ is absent in the case presented in last subsection, where we only consider one non-vanishing dimension-6 SMEFT operator at a time at the UV scale. Our discussion on the correlation among different operators at the UV scale shall not be confused with the study of a subset of SMEFT operators discussed in Refs.\,\cite{Elias-Miro:2013eta,Henning:2014wua}, where one only focuses on a closed subset of the full RGEs related to the observables under the study in the latter case. As we shall see later in this subsection, the correlation we study here is in general important and shall be considered for a relatively more rigorous study.}

Without losing generality, we assume that all the dimension-6 SMEFT operators are associated with the same high-energy scale $\Lambda$ and then present the interference of multiple operators in figure\,\ref{fig:two-operatorsTot}. The effects, resulting from the correlation among different operators, on the upper bounds of two Wilson coefficients are shown in the left (right) panel for LBL (reactor) experiments. The shaded areas indicate the allowed values of two Wilson coefficients at 95\% CL statistical significance when no other operators are present. The scale is set at $\Lambda =$ 1~TeV. The operators selected for this example represent both the type-A and the type-B classes, but they generate similar source and detection NSI parameter values. As one can see from the resulting sensitivities, \yong{adding} a second Wilson coefficient decreases the sensitivities to both Wilson coefficients as opposed to if they were assumed to be the single non-vanishing ones.}

\yong{To see the quantitative impact of extra operators on the upper bound of the Wilson coefficient, we can, for example, compare the difference between the upper bound for $C_{\substack{ledq\\2211}}$ when only one dimension-6 SMEFT operator is considered at the input scale and that when multiple operators exist at the UV scale. In the former case, as one can see from the lower panel of figure\,\ref{fig:lblconstr}, its upper bound from the combined analysis is $C_{\substack{ledq\\2211}}\lesssim 10^{-8}$; while in the latter case, according to the left panel of figure\,\ref{fig:two-operatorsTot}, it relaxes to $\lesssim 5\times10^{-8}$ when the Wilson coefficient $C^{(1)}_{\substack{lequ\\2211}}$ also has a non-zero input at the UV scale $\Lambda$. Moreover if $C_{\substack{ledq\\2211}}$ happens to be roughly equal to $C^{(1)}_{\substack{lequ\\2211}}$, then both of them can be as large as $\sim10^{-6}$ or even larger as indicated by the diagonal pin in the left panel of figure\,\ref{fig:two-operatorsTot}, the reason for this is that the two operators contributes opposite sign for the $\epsilon_P$ in the LEFT of which the effect is dominant in the LBL experiments.   Similar behavior repeats in the reactor experiments, as demonstrated for coefficients $C^{(3)}_{\substack{lq\\1111}}$ and $C^{(3)}_{\substack{Hq\\11}}$ in the right panel of figure\,\ref{fig:two-operatorsTot}.

\begin{figure}[t]
\centering{
  \begin{adjustbox}{max width = \textwidth}
\begin{tabular}{cc}
\includegraphics[scale=0.5]{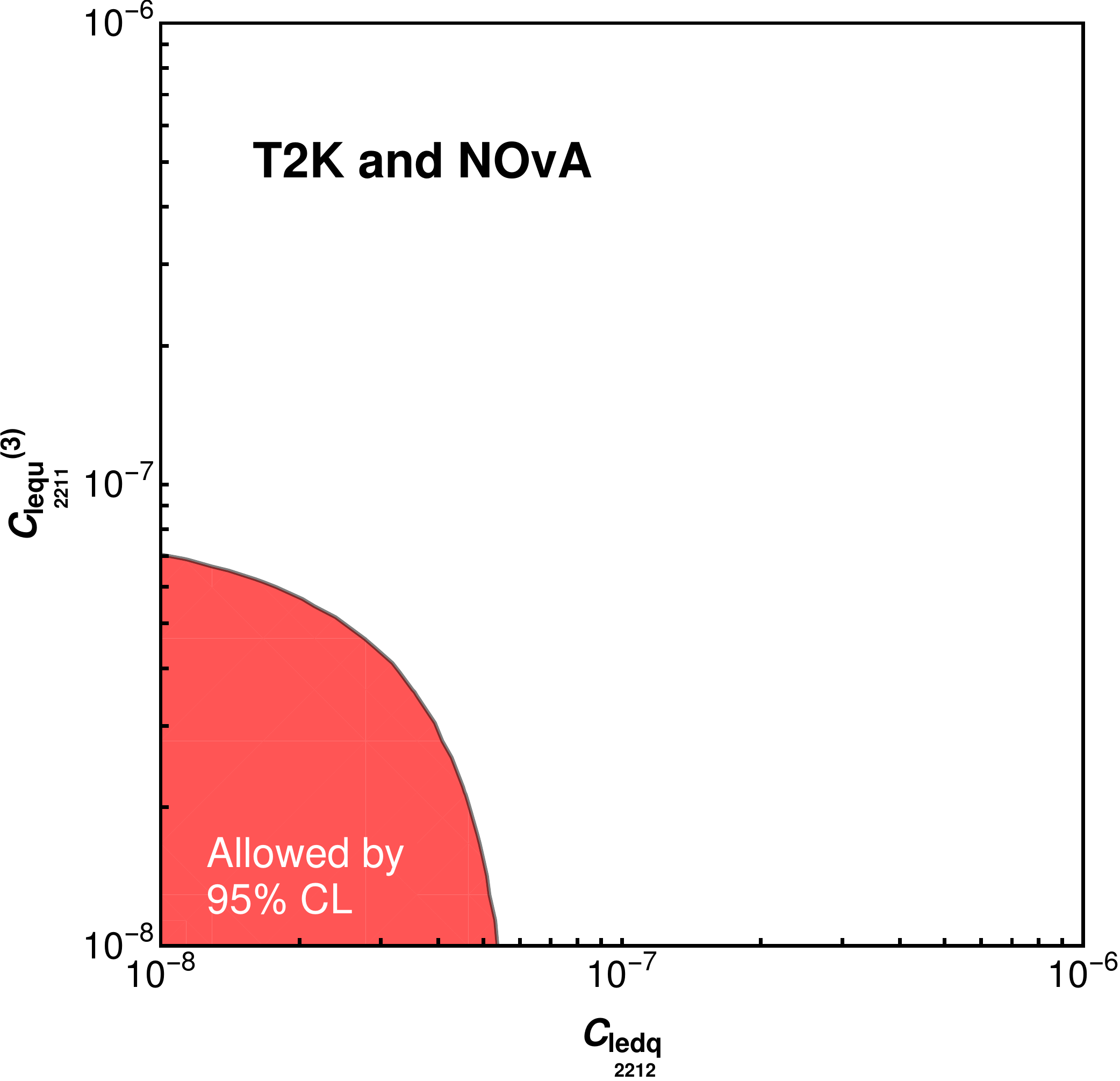} ~& ~ \includegraphics[scale=0.5]{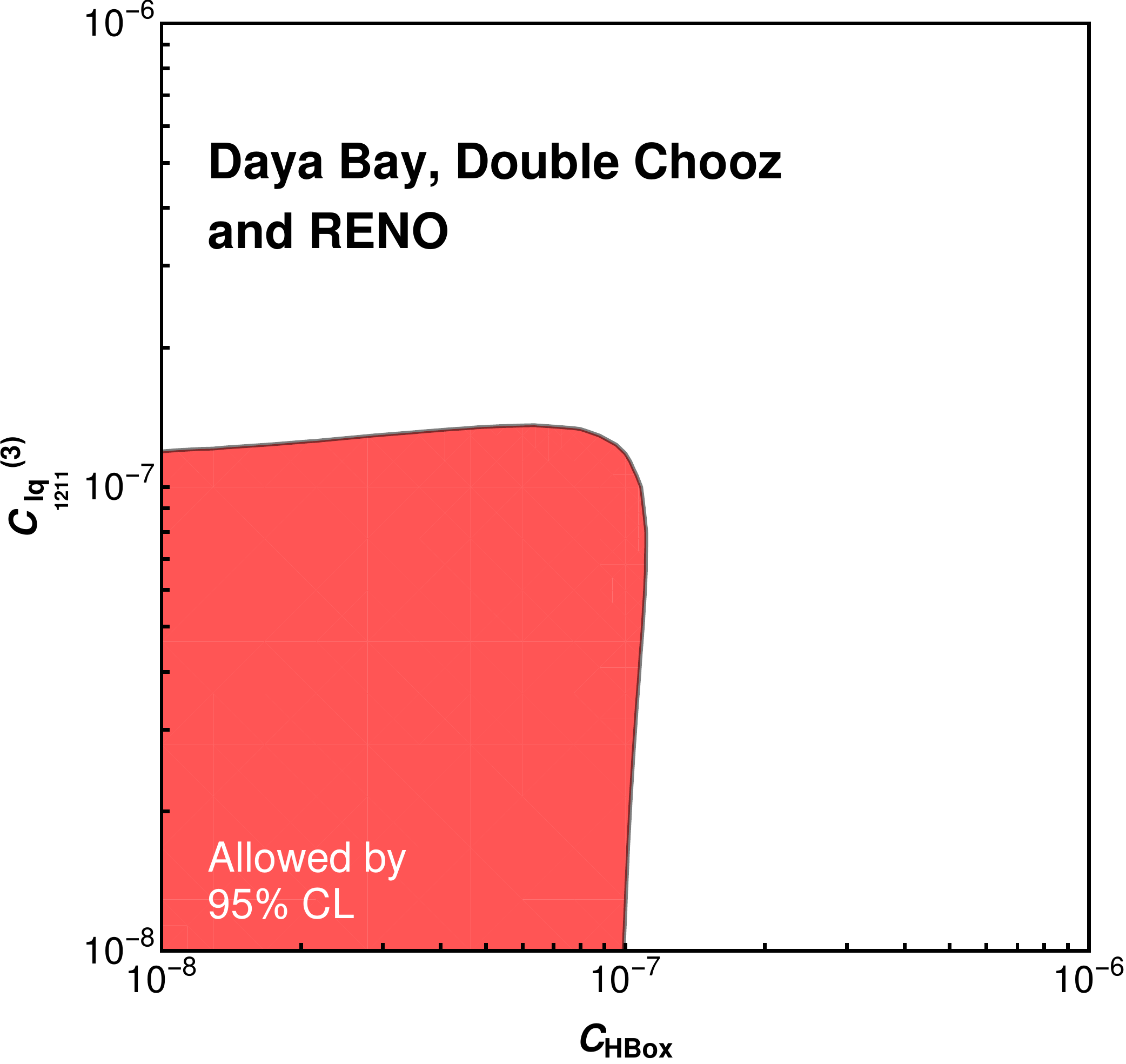}
\end{tabular}
  \end{adjustbox}}
\caption{{\color{black}Left panel: Upper constraints on the Wilson coefficients corresponding to the operators $(\bar{\ell}_2 e_2) \varepsilon_{jk} (\bar{q}_1 u_1)$ and $(\bar{\ell}_2 e_2) (\bar{d}_1 q_1)$ with $\Lambda =$ 1\,TeV. The constraints are obtained for the LBL experiments T2K and NO$\nu$A at 95\% CL significance. Right panel: Same as the left panel but with the application of the reactor experiments Daya Bay, Double Chooz and RENO with operators $(\phi^\dagger \phi)\Box(\phi^\dagger \phi)$ and $(\bar{\ell}_1^j \sigma_{\mu \nu} e_1) \varepsilon_{j k} (\bar{q}_1^k \sigma^{\mu \nu} u_1)$ instead.}}\label{fig:two-operatorsTotcounter}
\end{figure}

We point out that, due to the correlation of multiple operators at the UV scale $\Lambda$, relaxation on the upper (lower) bounds of the Wilson coefficients (UV scale $\Lambda$) turns out to be a common feature from our study. However, we are not claiming that inclusion of extra operators will always lead to stronger (weaker) constraints on the Wilson coefficients (UV scale). One counterexample is shown in figure\,\ref{fig:two-operatorsTotcounter}, where the left and the right panels are for LBL and reactor neutrino experiments respectively. Comparing with the upper bounds obtained when only one operator is considered at the UV scale $\Lambda$, inclusion of an extra operator leads to stronger constraints on all the operators shown in figure\,\ref{fig:two-operatorsTotcounter}.

Based on observation above, we claim that, even though it may be computationally challenging to include the correlation among all observable-related SMEFT operators at the UV scale in practice, one shall at least take into account the full correlation among different operators for specific UV models at the UV scale, as has been done for the simplified scalar leptoquark model in this work.}

\section{\label{sec:NSIphase}{\sl CP} violation from NSI Phases}

\begin{figure}[!t]
        \center{\includegraphics[width=\textwidth]
        {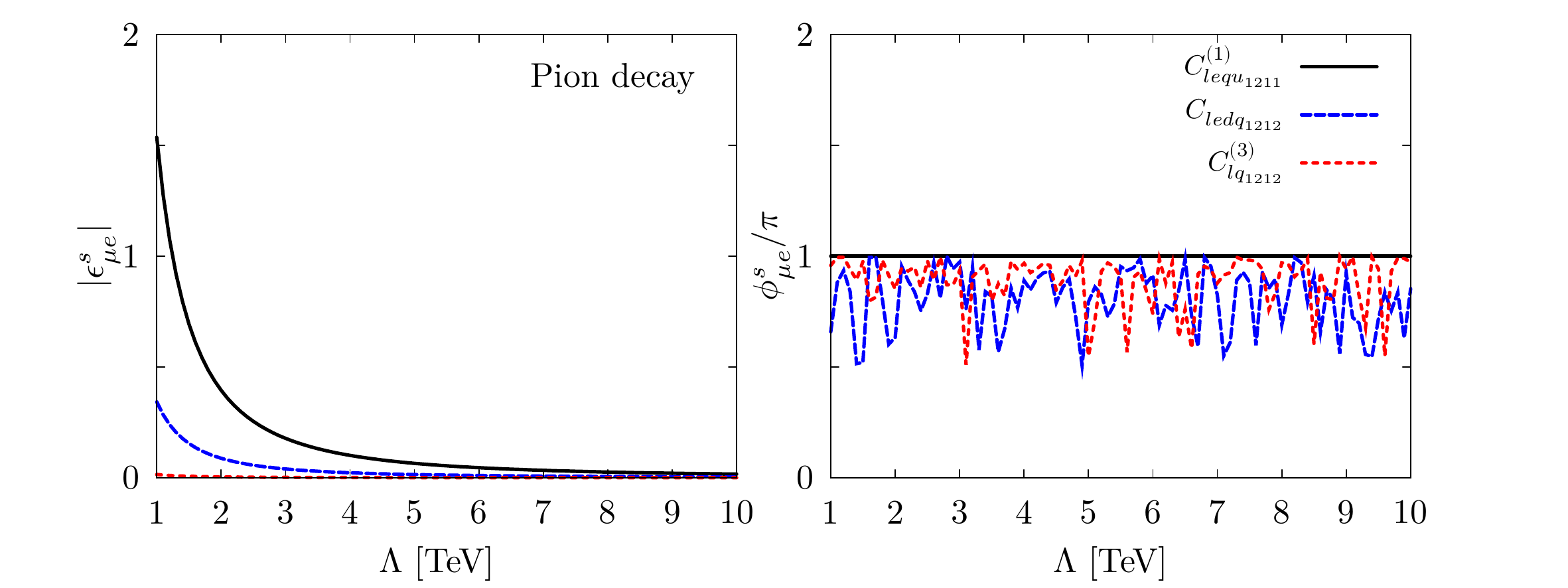}}
        \caption{\label{fig:NSIphase_superbeam} The magnitude and phase of the source NSI parameter $\epsilon_{\mu e}^s$ in pion-decay-driven neutrino oscillation experiments. The values obtained for the NSI parameter are shown for one {\sl CP}-conserving operator (solid black) and two {\sl CP}-violating operators (dashed blue and dotted red) as function of the scale $\Lambda$.}
      \end{figure}

\begin{figure}[!t]
        \center{\includegraphics[width=\textwidth]
        {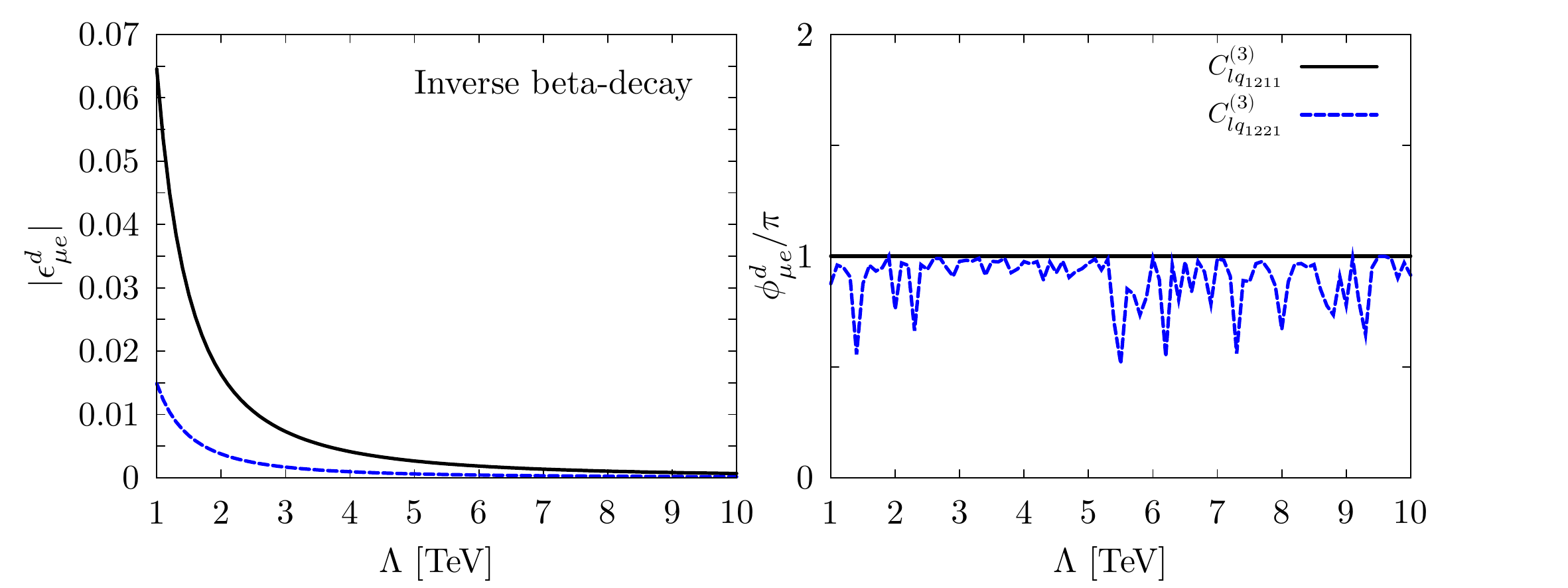}}
        \caption{\label{fig:NSIphase_reactor} The magnitude and phase of the detection NSI parameter $\epsilon_{\mu e}^d$ in reactor experiments, where detection is based on inverse beta decay. As an example, the values acquired for the NSI parameter are presented for one {\sl CP}-conserving operator (solid black) and one {\sl CP}-violating operator (dashed blue) as function of the UV-scale $\Lambda$.}
      \end{figure}

One of the primary objectives in the on-going and future neutrino oscillation experiments is to investigate whether there exists {\sl CP} violation in the leptonic sector. One possible source of {\sl CP} violation is the Dirac CP phase $\delta_{CP}$, which corresponds to either {\sl CP}-conserving when $\sin \delta_{CP} =$ 0 or {\sl CP}-violating when $\sin \delta_{CP} \neq$ 0. \sampsa{Such {\sl CP} violation may also emerge due to the neutrino NSIs in the source or the detector. The importance of studying {\sl CP} violation in the NSI parameters has been highlighted in Ref.~\cite{GonzalezGarcia:2001mp}. It has been shown that {\sl CP}-violating NSI phases may resolve tensions in experiments~\cite{Denton:2020uda,Chatterjee:2020kkm} as well as disrupt future experimental programs~\cite{Agarwalla:2016xlg}. In this section, we set focus on the new NSI phases that could be generated by dimension-6 operators in SMEFT.

Using the {\tt Wilson} package and the matching method discussed in sections~\ref{sec:NSIparam} and \ref{sec:SMEFTtoNSI}, we are able to find two operators that could generate a {\sl CP}-violating NSI phase in LBL experiments such as T2K and NO$\nu$A, where neutrino production is based on pion decay, and one in reactor experiments like Daya Bay, Double Chooz and RENO, where both processes based on beta decays and inverse beta decays occur.} We calculate the values that would be acquired in the source and detection NSI parameters $\epsilon^s_{\mu e}$ and $\epsilon^d_{\mu e}$ in the event where operators ${\cal O}_{\substack{ledq \\ 1212}}$ and ${\cal O}^{(3)}_{\substack{lq \\ 1212}}$ are considered at the UV scale $\Lambda$ in the LBL neutrino experiments and operator ${\cal O}^{(3)}_{\substack{lq \\ 1221}}$ in the reactor neutrino experiments, respectively\footnote{We also obtain similar values for $\epsilon_{e \mu}^s$ in beta decay.}. We also study the NSI parameter values obtained for the {\sl CP}-conserving operators ${\cal O}^{(1)}_{\substack{lequ \\ 1211}}$ and ${\cal O}^{(3)}_{\substack{lq \\ 1211}}$. In each case we consider only one operator at a time. The source and detection NSI computed from individual operators are presented for LBL experiments in figure\,\ref{fig:NSIphase_superbeam} and reactor experiments in figure\,\ref{fig:NSIphase_reactor}, respectively.

It can be seen in the results presented in figure\,\ref{fig:NSIphase_superbeam} that the operators ${\cal O}_{\substack{ledq \\ 1212}}$ and ${\cal O}^{(3)}_{\substack{lq \\ 1212}}$ lead to a sizeable value in the {\sl CP}-violating NSI phase $\phi^s_{\mu e}$ in LBL experiments where neutrinos are produced via pion decay. On the other hand, it is evident in figure\,\ref{fig:NSIphase_reactor} that rather large {\sl CP}-violating NSI phase $\phi^d_{\mu e}$ can be generated from operator ${\cal O}^{(3)}_{\substack{lq \\ 1211}}$ in reactor neutrino experiments, where detection via inverse beta decay is prevalent. As an example of the {\sl CP}-conserving operators, which we find all other operators in our consideration to fall into, are also present the NSI generated from operators ${\cal O}^{(1)}_{\substack{lequ \\ 1211}}$ and ${\cal O}^{(3)}_{\substack{lq \\ 1211}}$ in LBL and reactor experiments, respectively. The exact value of the NSI phase depends largely on the scale $\Lambda$, which we consider to be in the [1, 10]\,TeV range.

One may ask whether the NSI generated from operators ${\cal O}_{\substack{ledq \\ 1212}}$ and ${\cal O}^{(3)}_{\substack{lq \\ 1212}}$ can induce a noticeable signal in the neutrino oscillation experiments we consider in this work. Although the NSI phases pertaining to these operators can be {\sl CP}-violating, the relatively small magnitude generated for these operators are below the scope of the present data available in T2K and NO$\nu$A. Respectively, the data obtained from the reactor neutrino experiments Daya Bay, Double Chooz and RENO suggests that the {\sl CP}-violating NSI phase from operator ${\cal O}^{(3)}_{\substack{lq \\ 1221}}$ is also below the present limit.

%%%%%%%%%%%%%%%%%%%%%%%%%%%%%%%%%%%%%%%%%%%%%
\section{\label{sec:concl}Conclusions}
%%%%%%%%%%%%%%%%%%%%%%%%%%%%%%%%%%%%%%%%%%%%%
Ongoing neutrino oscillation experiments are providing more and more precise measurements on neutrino properties, which, on one hand, deepens our understanding of neutrinos, and on the other hand, opens the possibility to study neutrino NSIs more closely. SMEFT provides a systematic and model-independent method for this purpose. In this work, we study constraints on neutrino NSIs from neutrino experiments T2K, NO$\nu$A, Daya Bay, Double Chooz and RENO in the SMEFT framework. \sampsa{The SMEFT operators} are naturally defined at a UV scale $\Lambda$ that is larger than the scale where neutrino oscillation experiments \sampsa{operate}. As a result, matching of different EFTs and the renormalization group running need to be taken into account at different scales. A schematic workflow of this procedure is depicted in figure\,\ref{fig:workflow}, together with the dimension-6 operators that contribute to neutrino NSIs under the corresponding EFT names.

A simplified scalar leptoquark model is adopted for the illustration of our workflow in the top-down EFT approach. At tree level, there are about 300 SMEFT operators that are related to neutrino NSIs  at the UV scale. At a different scale, the SMEFT operators mix through the anomalous dimension matrix, and as a consequence, all SMEFT operators could contribute to the neutrino NSI parameters at the 2\,GeV scale. Using \sampsa{presently available} upper bounds on \sampsa{the} NSI parameters from neutrino oscillation experiments, we investigate how these experiments can be used to constrain this simplified model. We find that the current upper bound on $|\epsilon_{\mu e}^s|$ from pion decay at \sampsa{the LBL experiments T2K and NO$\nu$A lead} to the most stringent constraint on \sampsa{the} simplified model as shown in figures\,\ref{LQModelMassScan} and \ref{NSIRegionLQ}. Furthermore, we conclude that constraints from the \sampsa{charged current} neutrino NSI parameters are much stronger than those obtained from collider studies at the ATLAS and the CMS, implying complementarity between low and high-energy experiments in searching for new physics.

We also generalize our study to obtain \sampsa{the constraints} on individual dimension-6 SMEFT operators from neutrino oscillation experiments in the bottom-up approach, where only one dimension-6 SMEFT operator is assumed to have a non-vanishing value at the UV scale. The results are shown in figure\,\ref{fig:lblconstr} and figure\,\ref{fig:rconstr} for LBL and reactor neutrino experiments respectively. We find that LBL neutrino experiments \sampsa{T2K and NO$\nu$A} are already sensitive to new physics around 20\,TeV by fixing the Wilson coefficients at unity, as indicated by the $\mathcal{O}^{(1)}_{\substack{lequ\\1211}}$ and the $\mathcal{O}_{\substack{lequ\\1211}}^{(1)}$ operators in the upper panel of figure\,\ref{fig:lblconstr}. In contrast, reactor neutrino experiments \sampsa{Daya Bay, Double Chooz and RENO} are now reaching the 5\,TeV range as implied by the $\mathcal{O}^{(3)}_{\substack{lq\\1111}}$, $\mathcal{O}^{(3)}_{\substack{Hq\\12}}$ and $\mathcal{O}^{(3)}_{\substack{Hl\\22}}$ operators in the upper panel of figure\,\ref{fig:rconstr}. \sampsa{Though LBL neutrino experiments impose generally} much stronger constraints on the SMEFT operators than the reactor neutrino experiments, \sampsa{the two experiment groups are mainly} sensitive to different sets of SMEFT operators, which highlights the complementarity \sampsa{between} LBL and reactor neutrino experiments.

\sampsa{The correlation among the different dimension-6 operators becomes important when more than one operator is studied at the same time.} In section\,\ref{sec:multiop}, effects from the correlation among multiple operators are studied. The results are obtained by considering two non-vanishing dimension-6 SMEFT operators at the same time at the UV scale $\Lambda$. The effect from this correlation is presented in figures\,\ref{fig:two-operatorsTot} and \ref{fig:two-operatorsTotcounter} for both LBL and reactor neutrino experiments. We find that the correlation among different operators at the UV scale could weaken the constraints on SMEFT operators by several orders of magnitude as shown in figure\,\ref{fig:two-operatorsTot}. We point out that it is also possible that stronger constraints on the SMEFT operators could be obtained with the inclusion of multiple operator correlation, as is shown in figure\,\ref{fig:two-operatorsTotcounter}. \sampsa{Basing on this observation, we conclude our work with the following remark: Though} it may be technically challenging to include the correlation among all observable-related SMEFT operators at the UV scale, one shall at least take the correlation among all operators that are induced by a specific UV model into account as is studied for the simplified scalar leptoquark model in section\,\ref{sec:leptoquark}.

{We also examine whether new sources of {\sl CP} violation can be found at the UV scale and identify several operators for which considerable NSI phases could be generated, with three such operators showcased in figures\,\ref{fig:NSIphase_superbeam} and \ref{fig:NSIphase_reactor}. The magnitude of the relevant neutrino NSI is however found to fall below the sensitivities of the present data.}

\acknowledgments{We thank Adam Falkowski and Gang Li for helpful discussion, the HPC Cluster of ITP-CAS for the computing support and TseChun Wang for his contribution in the early stage of this project. YD, HLL and JHY were supported by the National Science Foundation of China (NSFC) under Grants No. 11875003 and No. 12022514. JT and SV were supported in part by National Natural Science Foundation of China under Grant Nos. 12075326, 11505301 and 11881240247, Guangdong Basic and Applied Basic Research Foundation under Grant No. 2019A1515012216. SV was also supported by China Postdoctoral Science Foundation under Grant No.  2020M672930. YD was also supported in part under U.S. Department of Energy contract DE-SC0011095. JHY was also supported by the National Science Foundation of China (NSFC) under Grants No. 11947302. JT also acknowledges the support from the CAS Center for Excellence in Particle Physics (CCEPP).}

\appendix

\section{\label{sec:muonLEFT}LEFT for muon decay and the matching between QFT and QM formalisms}
In this section, we present an example to show the details of the matching between the QFT and the QM formalisms of neutrino NSIs following the procedure in Ref.\,\cite{Falkowski:2019kfn}. Since only one neutrino is involved in pion decay, as well as in beta decay and inverse beta decay, the matching procedure is relatively straightforward and we reproduced the matching formulae presented in Ref.\,\cite{Falkowski:2019kfn}. This procedure is less trivial in muon decay as the muon neutrino and electron antineutrino are produced simultaneously. We thus use muon decay as our example here for illustration.\footnote{We thank Mart\'in Gonz\'alez-Alonso for bringing their published version of Ref.\,\cite{Falkowski:2019kfn} to our attention, where they also discussed the matching for muon decay. We also thank Mart\'in for pointing out the disagreement between our previous results presented in this section, which motivates us to show the details of our calculation here.}

To start, we write the most general low-energy EFT describing muon decay as follows, with an introduction of the right-handed partner of neutrinos\,\cite{Mursula:1984zb,Scheck:1977yg,Scheck:1984md},\footnote{To obtain the results in SM, one may simply set the right-handed neutrino terms to zero.}
\eqal{\mathcal{L}_\mu=\frac{G_{\rm F}}{\sqrt{2}}\{ &h_{11}(s+p)_{e \nu_{e}}(s+p)_{\nu_{\mu} \mu}+h_{12}(s+p)(s-p)+h_{21}(s-p)(s+p)+h_{22}(s-p)(s-p)\nb
&+g_{11}\left(v^{\alpha}+a^{\alpha}\right)_{e \nu_{e}}\left(v_{\alpha}+a_{\alpha}\right)_{\nu_{\mu} \mu}+g_{12}\left(v^{\alpha}+a^{\alpha}\right)\left(v_{\alpha}-a_{\alpha}\right)+g_{21}\left(v^{\alpha}-a^{\alpha}\right)\left(v_{\alpha}+a_{\alpha}\right)\nb
&+{(1+g_{22})\left(v^{\alpha}-a^{\alpha}\right)\left(v_{\alpha}-a_{\alpha}\right)}+f_{11}\left(t^{\alpha \beta}+u^{\alpha \beta}\right)_{e \nu_{e}}\left(t_{\alpha \beta}+u_{\alpha \beta}\right)_{\nu_{\mu} \mu}\nb
&+f_{22}\left(t^{\alpha \beta}-u^{ \alpha \beta}\right)\left(t_{\alpha \beta}-u_{\alpha \beta}\right)+\rm{h.c.} \},\label{Lagmuon}}
where
\eqal{s_{ij}\equiv\bar{u}_i u_j, p_{ij}\equiv\bar{u}_i\gamma_5u_j, v_{ij}^\mu\equiv\bar{u}_i\gamma^\mu u_j,
a_{ij}^\mu\equiv\bar{u}_i\gamma^\mu\gamma_5u_j, t_{ij}^{\alpha\beta}\equiv\bar{u}_i\sigma^{\mu\nu}u_j, u_{ij}^{\alpha\beta}\equiv\bar{u}_i\sigma^{\mu\nu}\gamma_5u_i.}
For comparison, the relation between our notations and those in Ref.\,\cite{Falkowski:2019kfn}, denoted as ``FGT'' in the following, takes the form
\eqal{g_{22} = \rho_L^{\rm FGT},\quad h_{21}=-2\rho_R^{\rm FGT}\label{notationRel}
}

We denote the muon decay process as $\mu^-(p_2)\to\nu_\mu(p_1)+e^-(p_3)+\bar{\nu}_e(p_4)$, with their four-momenta specified explicitly. To obtain the matching between the Wilson coefficients in the LEFT above and the NSI parameters $\epsilon^{s,d}$, we follow the notations in Ref.\,\cite{Falkowski:2019kfn}. At linear level in the Wilson coefficients in $\mathcal{L}_\mu$, the master formula relevant for our discussion here can be written as\,\cite{Falkowski:2019kfn}
\eqal{\epsilon^s_{\alpha\beta}=\sum\limits_{X}p_{XL}[\epsilon_X]^*_{\alpha\beta},}
with the production coefficients $p_{XL}$ defined as
\eqal{p_{XL}\equiv\frac{\int d\Pi_{P'}A_X^P\bar{A}_L^P}{\int d\Pi_{P'}A_L^P\bar{A}_L^P},\quad d\Pi_{P'}\equiv\frac{d\Pi_P}{dE_\nu},\label{ProdCoeff}}
and $d\Pi_P\equiv\prod\limits_{i}d^3p_i/[(2\pi)^32E_i](2\pi)^4\delta^{(4)}(\sum\limits_{\rm in}p_-\sum\limits_{\rm out}P)$ the usual definition of phase space and $E_{\nu}$ the energy of the (anti)neutrino produced in the decay process. In our setup, we find the only non-vanishing terms contributing to the production coefficients $p_{XL}$ are the $g_{22}$ and $h_{21}$ terms in eq.\,\eqref{Lagmuon}, whereas all the others vanish due to chiral flip. As a result, one only needs to calculate the following two integrals:
\eqal{I_1\equiv\int d\Pi_{P'}A_L^P\bar{A}_L^P, \quad I_2\equiv\int d\Pi_{P'}A_{h21}^P\bar{A}_L^P,}
where $A_{h21}$ corresponds to the amplitude from the $h_{21}$ term in eq.\,\eqref{Lagmuon}. We find
\eqal{I_1=\mathcal{F}\int d\Pi_{P'}256(p_1\cdot p_3) (p_2\cdot p_4), \quad I_2=\mathcal{F}\int d\Pi_{P'}64m_em_\mu (p_1\cdot p_4),} 
where $\mathcal{F}$ is an overall factor that will cancel eventually when one calculates the production coefficients. Therefore, the problem boils down to a three-body phase space integral, for which we will do in the following general form and start with the full phase space integral $d\Pi_{P}$ instead of $d\Pi_{P'}$:
\eqal{
I\equiv\int d\Pi_p\langle\mathcal{M}^2\rangle,\quad {\text{with }} \langle\mathcal{M}^2\rangle = 
\left\{\begin{array}{cc}
256(p_1\cdot p_3) (p_2\cdot p_4), & {\text{for }} I_1 \\ 
64m_em_\mu (p_1\cdot p_4), & {\text{for }} I_2
\end{array} 
\right.\label{M2Results}
}

Writing explicitly,
\eqal{
I=&\int \frac{d^3p_1}{(2\pi)^32E_1}\frac{d^3p_4}{(2\pi)^32E_4}\frac{d^3p_3}{(2\pi)^32E_3} (2\pi)^4\delta^{(4)}(p_2-p_1-p_3-p_4) \langle\mathcal{M}^2\rangle\nb
=&\int \frac{d^3p_1}{(2\pi)^32E_1}\frac{d^3p_4}{(2\pi)^32E_4}\frac{d^4p_3}{(2\pi)^3}\Theta(E_3)\delta(p_3^2-m_e^2) (2\pi)^4\delta^{(4)}(p_2-p_1-p_3-p_4) \langle\mathcal{M}^2\rangle\nb
=&\frac{1}{(2\pi)^5}\frac{1}{2^2}\int\frac{d^3p_1}{E1}\frac{d^3p_4}{E4}\Theta(E_2-E_1-E_4)\left.\langle\mathcal{M}^2\rangle\right|_{p_3\to p_2-p_1-p_4}\nb
&\quad\quad\quad\quad\times\delta(m_\mu^2-m_e^2-2p_2\cdot p_1-2p_2\cdot p_4+2p_1\cdot p_4)\nb
}
Working in the rest frame of the decaying muon and performing the angular integrals, we arrive at
\eqal{I=\frac{1}{32\pi^3}&\int dE_1\int dE_4\Theta(m_\mu-E_1-E_4)\nb
&\times\left.\left(\left.\langle\mathcal{M}^2\rangle\right|_{p_3\to p_2-p_1-p_4}\right)\right|_{\cos\theta=\frac{m_\mu^2-m_e^2-2m_\mu(E_1+E_4)+2E_1E_4}{2E_1E_4}},}
where $\theta$ is the angle between the three-momenta $\vec{p}_1$ and $\vec{p}_4$, and $\Theta$ is the Heaviside function that ensures the positiveness of the electron's energy in the final state. Constraints on $E_{1,4}$ can be obtained by requiring 
\eqal{m_\mu-E_1-E_4>0,\quad -1\le\cos\theta\le1,}
leading to
\eqal{
\frac{m_\mu^2-m_e^2}{2m_\mu}-E_4\le &E_1\le \frac{m_\mu}{2}-\frac{m_e^2}{2(m_\mu-2E_4)}\\
\frac{m_\mu^2-m_e^2}{2m_\mu}-E_1\le &E_4\le \frac{m_\mu}{2}-\frac{m_e^2}{2(m_\mu-2E_1)}
}

Therefore, for neutrino production, one finally has
\eqal{I_{1,2}^\nu=\frac{dI}{dE_1} = \frac{1}{32\pi^3}&\int_{E_4^{\rm min}}^{E_4^{\rm max}} dE_4\left.\left(\left.\langle\mathcal{M}^2\rangle\right|_{p_3\to p_2-p_1-p_4}\right)\right|_{\cos\theta=\frac{m_\mu^2-m_e^2-2m_\mu(E_1+E_4)+2E_1E_4}{2E_1E_4}},}
and, for antineutrino production,
\eqal{I_{1,2}^{\bar{\nu}}=\frac{dI}{dE_4} = \frac{1}{32\pi^3}&\int_{E_1^{\rm min}}^{E_1^{\rm max}} dE_1\left.\left(\left.\langle\mathcal{M}^2\rangle\right|_{p_3\to p_2-p_1-p_4}\right)\right|_{\cos\theta=\frac{m_\mu^2-m_e^2-2m_\mu(E_1+E_4)+2E_1E_4}{2E_1E_4}},}
where $I_{1,2}^{\nu,\bar{\nu}}$ are obtained with appropriate $\langle\mathcal{M}^2\rangle$ given in eq.\,\eqref{M2Results}.

Finally, using eq.\,\eqref{ProdCoeff} and $I_{1,2}^{\nu,\bar{\nu}}$ derived above, we obtain\footnote{We use $E_{\nu(\bar{\nu})}$ to replace $E_{1(4)}$ here to specify the general fact here that the production coefficients are (anti)neutrino-energy dependent. The difference in the matching formulae for neutrinos and antineutrinos results from the variable being integrated over or equivalently the feature of the amplitudes in eq.\,\eqref{M2Results}, as was also explained in Ref.\,\cite{Falkowski:2019kfn}.}
\eqal{
&p_{LL} = 1, \quad p_{h_{21}L} = \frac{3m_em_\mu(m_\mu-2E_\nu)}{16m_\mu E_\nu^2+6m_\mu(m_\mu^2+m_e^2)-4E_\nu(5m_\mu^2+m_e^2)},\\
&\bar{p}_{LL} = 1, \quad \bar{p}_{h_{21}L} = \frac{m_e}{4(m_\mu-2E_{\bar{\nu}})},
}
for neutrino and antineutrino production respectively, and thus the matching formulae can be written as
\eqal{\epsilon_{\mu\beta}^s &= \left[ g_{22} + \frac{3m_em_\mu(m_\mu-2E_\nu)}{16m_\mu E_\nu^2+6m_\mu(m_\mu^2+m_e^2)-4E_\nu(5m_\mu^2+m_e^2)}h_{21} \right]_{\mu\beta}^*,\\
\epsilon_{e\beta}^s &= \left[ g_{22} + \frac{m_e}{4(m_\mu-2E_{\bar{\nu}})}h_{21} \right]_{e\beta}^*.
}
We checked that these results agree with those in Ref.\,\cite{Falkowski:2019kfn} after performing a small $m_e$ expansion and using the notation relations specified in eq.\,\eqref{notationRel}.\footnote{Note that we have ignored the other flavor indices above compared with the results in Ref.\,\cite{Falkowski:2019kfn}.}

\clearpage

\bibliographystyle{JHEP}
\bibliography{ref}

\end{document}